\documentclass[aps,prd,twocolumn,superscriptaddress,showpacs,preprintnumbers,nofootinbib,notitlepage,longbibliography]{revtex4-2}
\usepackage{graphicx}
\usepackage{bm}
\usepackage[utf8]{inputenc}
\usepackage[hyperfootnotes=false]{hyperref}
\usepackage{multirow}
\usepackage{amsmath}
\usepackage{amsfonts}
\usepackage{amssymb} 
\usepackage{color}
\usepackage{xcolor}
\usepackage{xspace}
\usepackage{braket}
\usepackage{units}
\usepackage{slashed}
\usepackage{multirow}
\usepackage[normalem]{ulem}

\graphicspath{ {plots/} }

\newcommand{\be}{\begin{equation}}
\newcommand{\ee}{\end{equation}}

\usepackage{bm} 

\renewcommand{\Re}{\ensuremath{\textrm{Re}}}
\renewcommand{\Im}{\ensuremath{\textrm{Im}}}

\newcommand{\mevnospace}{\ensuremath{{\mathrm{\,Me\kern -0.1em V}}}}
\newcommand{\gevnospace}{\ensuremath{{\mathrm{\,Ge\kern -0.1em V}}}}
\newcommand{\tevnospace}{\ensuremath{{\mathrm{\,Te\kern -0.1em V}}}}

\newcommand{\sig}
{\ensuremath{\sigma/f_0(500)}\xspace}

\newcommand{\pipi}
{\ensuremath{\pi \pi \to \pi \pi}\xspace}

\usepackage{mfirstuc} 
\newcommand{\addReviewer}[2]{
  \expandafter\newcommand\csname #1\endcsname[1]{{\bf \color{#2} \capitalisewords{#1}:\,##1}}
  \expandafter\newcommand\csname #1cor\endcsname[2]{{\color{#2} \capitalisewords{#1}:\,\st{##1}{\,\bf ##2}}}
  \expandafter\newcommand\csname #1color\endcsname{\,#2}
}

\usepackage{tikz, marginnote} 
\newcommand{\checkedby}[1]{
\ifdefined\CROSSCHECKS
  \marginnote{
    \begin{tikzpicture}
      \foreach \x [count=\xi] in {#1} {
         \node[shape=circle,inner sep=0mm,
         minimum size=2mm,
         fill=\csname \x color\endcsname] at (\xi*3mm,0) {};
       }
    \end{tikzpicture}
  }
\else
\fi
}

\usepackage{soul,color}
\definecolor{chromeyellow}{rgb}{1.0, 0.65, 0.0}
\definecolor{DodgeBlue}{rgb}{0.118, 0.565,1.000}
\definecolor{asparagus}{rgb}{0.53, 0.66, 0.42}
\definecolor{cadmiumgreen}{rgb}{0.0, 0.42, 0.24}

\addReviewer{AR}{asparagus}
\addReviewer{JR}{red}
\addReviewer{JRE}{blue}

\newcommand{\ucm}{Departamento de F\'isica Te\'orica and IPARCOS, 
Universidad Complutense de Madrid, 
E-28040 Madrid, Spain}

\begin{document}
\title{Dispersive determination of resonances from $\pi\pi$ scattering data}
\author{J.~R.~Pel\'aez}
\email{jrpelaez@ucm.es}\affiliation{\ucm}
\author{P. Rab\'an}
\email{praban@ucm.es}\affiliation{\ucm}
\author{J.~Ruiz de Elvira}
\email{jacobore@ucm.es}\affiliation{\ucm}
\preprint{IPARCOS-UCM-25-065}
\begin{abstract}
We provide a precise, model- and parametrization-independent dispersive determination of the $f_0(500)$, $\rho(770)$, $f_0(980)$, $f_2(1270)$, $f_0(1370)$, $\rho(1450)$, $f_0(1500)$, and $\rho_3(1690)$ resonance pole parameters. They are obtained from the analytic continuation, by means of continued fractions, of forward dispersion relations, whose input is a recent global dispersive analysis of $\pi \pi$ scattering data. From this dispersive study, we find no indications of other resonant poles below 1.7 GeV.
Beyond this energy, we also provide resonance pole parameters from the direct analytic continuation of Global Fits to the three existing incompatible datasets. Depending on the dataset we find poles for the $\rho(1700)$, $f_0(1710)$, $\rho(1900)$, $f_2(1950)$, and $f_0(2020)$ resonances.
We also present the Argand diagrams of these Global Fits and illustrate that each resonance does not necessarily have to trace a full circle in the diagram.
\end{abstract}

\maketitle

\section{Introduction} 

The model- and process-independent definition of a resonance is given by the position of the associated $T$-matrix pole in the complex $s$ plane on an unphysical Riemann sheet, where $s$ denotes the energy squared in the center-of-mass frame. When poles are well isolated from other singularities, the resonance mass $M$ and width $\Gamma$ are related to the pole position by $\sqrt{s_\text{pole}}=M-i\Gamma/2$. Additionally, the pole residue is associated with the coupling of the resonance to the states involved in a specific process. The Review of Particle Physics (RPP) has adopted the pole definition~\cite{ParticleDataGroup:2024cfk}, and it has been gradually implemented in the light-meson sector over the last decade (see also~\cite{Pelaez:2025wma} for a recent review on light meson resonances).
A reliable determination of resonance pole parameters is notoriously challenging as it requires precise control over the analytic continuation from the physical region into the complex plane. Nevertheless, hadronic resonances are often extracted using models such as Breit-Wigner (BW)-like approximations or $K$-matrix parametrizations, which are not applicable when a resonance lies deep in the complex plane or the resonance is not well isolated from other singularities (associated with other poles or thresholds). This ``model problem" partly explains the current state of the low-energy meson spectrum, where many resonance states in the RPP are still classified as ``need confirmation". In contrast, others have only ``educated guesses" for their parameters. The ``data-problem" explains the other part of the situation, since very often different experiments lead to conflicting datasets.

In particular, many determinations of resonances below 2 GeV are based on low-energy meson-meson scattering data. As these data are extracted from indirect measurements using models, they are often inconsistent between and within experiments, as well as with fundamental principles. For instance, in $\pi \pi$ scattering, data is obtained indirectly from the process $\pi N \to \pi \pi N'$ under the assumption of one-pion exchange dominance~\cite{Hyams:1973zf,Durusoy:1973aj,Losty:1973et,Cohen:1973yx,Protopopescu:1973sh,Grayer:1974cr,Hyams:1975mc,Hoogland:1977kt,Kaminski:1996da}. Such extractions often suffer from large systematic uncertainties, inconsistencies between different datasets, and multiple conflicting solutions. In addition, many of them do not satisfy several dispersion relations or sum rules \cite{Pelaez:2004vs}. As a consequence,
meson-meson scattering data and resonance determinations based on them are particularly prone to both the model and data problems (see the discussion in~\cite{Pelaez:2021dak}).

Causality imposes strong analyticity constraints on scattering amplitudes. These 
constraints can be formulated as integral equations, known as dispersion relations, which can also implement Lorentz invariance, crossing symmetry, and unitarity. As a result, dispersion relations serve as powerful tools to address both the ``data problem" and the ``model problem". On the one hand, they help discriminate among different scattering datasets, thereby mitigating the systematic model dependence inherent in their extraction.
On the other hand, as they are based on analyticity, dispersion relations provide a model-independent continuation into the complex plane, allowing for a rigorous and precise determination of resonance pole parameters.

For more than a decade, our group has employed dispersion relations to constrain parametrizations of $\pi\pi$ scattering data up to 1.4 GeV. Using partial-wave dispersion relations, valid up to 1.1 GeV, we also rigorously obtained precise pole parameters of resonances appearing in the elastic region, i.e., up to $\sim$1~GeV. Beyond that region, we still determined resonances precisely, but less rigorously, from data parametrizations. Nevertheless, we have recently proposed a method \cite{Pelaez:2022qby}, which is also valid in the inelastic regime, to determine pole parameters from forward dispersion relations. Even more recently \cite{Pelaez:2024uav}, we have extended some forward dispersion relations up to 1.6 GeV.

Our aim in this work is to use these last two recent pieces of progress to provide a dispersive determination of the pole parameters---pole position and coupling to $\pi\pi$---of all resonances appearing in $\pi\pi$ scattering\footnote{As long as they couple strongly enough to $\pi\pi$.} below 1.7 GeV. We begin by briefly presenting the current status of dispersive $\pi\pi$ scattering analyses and resonance determination. In Sec.~\ref{sec:method}, we review the recently proposed method to combine forward dispersion relations and continued fractions. Then, in Sec.~\ref{sec:results}, we present our dispersive results.
These are compared and cross-checked with resonance determinations obtained directly from parametrizations in Sec.~\ref{sec:parametrizations}. In that section, we also discuss the Argand diagrams for the resonances we find from the partial-wave parametrizations and clarify some common misconceptions.
Finally, in Sec.~\ref{sec:discussion}, we discuss and summarize our findings.

\subsection{Dispersive determinations of resonances from $\pi\pi$ scattering: Current status}\label{sec:masintro}

Let us review the present situation of dispersive determinations of resonances from $\pi\pi$ scattering. We discuss the advantages and disadvantages of previous treatments, and what can be achieved with our new method, using our latest parametrizations in~\cite{Pelaez:2024uav} as input. This section is conceptual; explicit equations and notation details will be given in Sec.~\ref{sec:method} only for our method, following closely~\cite{Pelaez:2024uav}. 

\subsubsection{Partial-wave versus forward dispersion relations}
Let us recall that two-body scattering amplitudes $F(s,t)$ depend on two independent variables, which intuitively correspond to the energy and the scattering angle. In the relativistic context, these are usually recast in terms of the usual $s$ and $t$ Mandelstam variables. The redundant variable $u$ is customarily used to make crossing symmetry explicit. 

Dispersion relations can be subtracted, i.e., the input divided by energy monomials, to improve the convergence of the integrals. In such cases, dispersion relations constrain the amplitude up to an energy polynomial of the same order as the number of subtractions, whose coefficients are called subtraction constants. Subtractions apart, the most frequently used dispersion relations for $\pi\pi$ scattering are of two types. The first are dispersion relations for fixed $t$; the special case $t=0$ is known as forward dispersion relations (FDRs). In our work, we employ FDRs for the $\pi^0\pi^0$, $\pi^0\pi^+$ scattering amplitudes, denoted as $F^{00}(s)\equiv F^{00}(s,0)$ and $F^{0+}(s)\equiv F^{0+}(s,0)$, respectively; they are $s\leftrightarrow u$ symmetric, once subtracted, and satisfy ``positivity", i.e., the imaginary part of the amplitude that appears in the integrand is the sum of positive terms, which makes them very accurate. The isospin basis is completed with the amplitude $F^{I_t=1}(s)$, which exchanges one unit of isospin $I$ in the $t$ channel. This amplitude is antisymmetric, and we implement it without subtractions, but it has no positivity. Note that FDRs constrain the amplitudes in the forward direction, without requiring partial-wave projection, and are formally valid at all energies.

The second type are dispersion relations for partial waves of definite angular momentum $\ell$. Imposing $s\leftrightarrow u$ crossing symmetry, these lead to a system of infinite coupled partial-wave integral equations generically known as Roy-like equations.
Since partial waves are suppressed at low energies with increasing $\ell$, Roy-like equations are customarily truncated to include low partial waves only, and the rest are considered input. Roy equations (RE) were originally derived~\cite{Roy:1971tc} with two subtractions to ensure the convergence of dispersive integrals. GKPY equations~\cite{Pelaez:2008ry,Kaminski:2008fu,GarciaMartin:2011cn} are similar but with one subtraction. For the same input, GKPY are more precise than RE in the resonance region, and vice versa near the $\pi\pi$ threshold. In practice, RE and GKPY equations are limited to energies below roughly 1.1 GeV, due to the requirements of convergence for the partial-wave expansion in their derivation.\footnote{Their applicability range can be extended with more cumbersome integral kernels~\cite{Mahoux:1974ej,Auberson:1974in,Roy:1975mq,Ananthanarayan:1998hj,EliasMiro:2025rqo}, which have not been used in practice.}

\subsubsection{Dispersive $\pi\pi$ data analyses}
In a series of works~\cite{Pelaez:2004vs,Kaminski:2006yv,Kaminski:2006qe,GarciaMartin:2011cn,Perez:2015pea}, our group has addressed the $\pi\pi$-scattering ``data problem", by providing dispersively constrained fits to data (CFD) 
in terms of partial-wave parametrizations. 
These fits were constrained to satisfy, within uncertainties, the RE and GKPY equations up to 1.1 GeV,
and FDRs up to $\sim$1.4~GeV. Above this energy, an average description of data, parametrized using Regge theory, was considered as input. Our ``data-driven" approach yields CFD parametrizations fairly compatible with other results where RE were solved below 0.8 GeV~\cite{Ananthanarayan:2000ht,Colangelo:2001df,Caprini:2011ky} and 1.1 GeV~\cite{Moussallam:2011zg}. For this ``solution" approach, the input consists of S and P waves above those energies, higher partial waves, a Regge description above 1.6 GeV, and, in some cases, low-energy input from chiral perturbation theory.

The CFD parametrizations in~\cite{GarciaMartin:2011cn}, together with the solutions~\cite{Ananthanarayan:2000ht,Colangelo:2001df,Caprini:2011ky,Moussallam:2011zg}, yielded a description of $\pi\pi$ scattering S and P partial waves below $\sim$1~GeV with unprecedented rigor and accuracy, far exceeding the experimental precision; an unusual case in the context of strong interactions. 

Note that, although the FDRs provide stringent constraints on the CFD up to 1.4 GeV, they do not constrain each partial wave individually.

\subsubsection{Extending dispersive $\pi\pi$ data analyses: Global Fits}
The CFD parametrizations in~\cite{GarciaMartin:2011cn} focused on the low-energy region---primarily on the $\pi\pi$ scattering threshold parameters and elastic resonances---thus precluding the study of the inelastic $\pi\pi$ spectrum above 1 GeV. 

As a first step toward extending this study to higher energies, in~\cite{Pelaez:2019eqa} we extended the description of the S0 and P partial waves above 1.4 GeV by matching phenomenological data parametrizations to the dispersive output below. The partial waves in these global parametrizations are individually consistent within uncertainties with both their RE and GKPY output in the real axis below $\sim$1.1~GeV, as well as in their domain of validity in the complex plane~\cite{Lehmann:1958}, and therefore describe the resonance poles in this region. In addition, they satisfy FDRs up to 1.4 GeV. Above 1.4 GeV, they were purely phenomenological fits to the three different $\pi\pi$ scattering datasets from the CERN-Munich experiment~\cite{Hyams:1973zf,Hyams:1975mc,Grayer:1974cr,Kaminski:1996da}. Therefore, we provided three ``Global Fits", I, II, and III, for the S0 and P waves, which were very compatible up to 1.4 GeV but clearly different above that energy.

Very recently~\cite{Pelaez:2024uav}, we have extended this global approach to $\pi\pi$ scattering data up to $\sim$1.8~GeV for the S2, D, F, and G partial waves, which are now much more accurate above 1 GeV than they were in~\cite{GarciaMartin:2011cn}.
In addition, this analysis refines the high‑energy behavior of the P wave by correctly incorporating the onset of inelasticity at the $\pi\omega$ threshold, as suggested in~\cite{Colangelo:2018mtw}. Another relevant improvement is the new choice of matching point with the Regge regime for the $F^{0+}$ and $F^{I_t=1}$ FDRs calculation, which are now satisfied up to 1.6 GeV. We will use these latest Global Fits as input for this work; the fact that the $F^{0+}$ FDR is well satisfied up to 1.6 GeV will play a crucial role.

\subsubsection{Dispersive pole determinations from $\pi\pi$ data}
Once a description of $\pi\pi$ data is consistent with dispersion relations, the very dispersive integrals provide a rigorous analytic continuation to the complex plane in the first Riemann sheet, removing parametrization and model dependencies. However,  poles are to be found in unphysical sheets. In particular, the most relevant poles lie on the contiguous Riemann sheet (also called the adjacent or proximal). Obtaining the amplitude in the contiguous sheet requires an additional step, which is straightforward in the elastic regime but not so much in the inelastic one.

In particular, RE or GKPY dispersive integrals provide a rigorous analytic continuation of partial waves to the first Riemann sheet within a region~\cite{Lehmann:1958,Caprini:2005zr} around the real axis up to 1.1 GeV.  
This range of energies covers the elastic regime, which, depending on the partial wave, extends up to 0.9 GeV, the $\pi\omega$ threshold, or $K\bar K$ threshold. For elastic amplitudes, there is only one unphysical sheet, for which the $S$-matrix is just the inverse of the $S$-matrix in the first sheet. This approach provides a straightforward and rigorous determination of
the $f_0(500)$, $\rho(770)$, and $f_0(980)$ poles. The consistency of the pole parameters obtained either from the RE solutions~\cite{Caprini:2005zr,Moussallam:2011zg} or the data-driven GKPY analysis~\cite{GarciaMartin:2011jx}, confirmed the existence of the long-debated $f_0(500)$ resonance and triggered its change of name and a substantial revision of its parameters in the RPP 2012 edition~\cite{Beringer:2012zz} (see \cite{Pelaez:2015qba} for a review).
The $f_0(980)$ suffered a minor revision as well, and in its 2021 edition, the RPP decided to start listing the $T$-matrix pole of the $\rho(770)$ meson too. At first, the RPP used only the parameters from~\cite{GarciaMartin:2011jx} and another approximate and less precise value from \cite{Pelaez:2003dy}. Nowadays the RPP $\rho(770)$ $T$-matrix parameter estimates also include Roy-like calculations from~\cite{Colangelo:2001df,Hoferichter:2023mgy}.

However, two problems appear when 
studying resonances in the inelastic regime using dispersion theory. First, above the first relevant inelastic threshold, the contiguous sheet is not obtained from the inverse of the $S$-matrix in the first sheet; instead, it must be built explicitly. Second, all such resonances lie beyond the applicability limit of the Roy-like equations used in the literature.

The first problem of analytically continuing an amplitude from its values in a real segment to the complex plane, without employing a specific parametrization or model, can be solved with several well-established techniques: conformal expansions~\cite{Yndurain:2007qm,Caprini:2008fc}, Laurent-Pietarinen expansions~\cite{Svarc:2013laa,Svarc:2014sqa,Svarc:2014aga}, sequences of Pad\'e approximants~\cite{Masjuan:2013jha,Masjuan:2014psa,Caprini:2016uxy,Pelaez:2016klv} or continued fractions~\cite{Schlessinger:1968,Tripolt:2016cya, Binosi:2019ecz,Binosi:2022ydc,Pelaez:2022qby}. Regardless of which inelastic threshold generates the corresponding discontinuity, these methods typically provide access to the first Riemann sheet in the upper-half $s$ plane and to the contiguous Riemann sheet in the lower half-plane.
To overcome the second problem, we proposed to make use of the output of FDRs in~\cite{Pelaez:2022qby}. Once again, the integral representation in the FDRs provides a rigorous analytic continuation to the first sheet, which allowed us to test the Padé approximant and continued fraction methods. Padé approximants require the numerical evaluation of high-order derivatives and proved somewhat unstable in practice, whereas continued fractions were found to be remarkably stable and accurate.
The only drawback of this approach, which we will call ``FDR$_{C_N}$" from now on, is that FDRs mix different partial waves. The isospin can be determined using the full basis of FDRs, but the spin must be determined by other means.

As a direct application of this method, and using the S0 Global Fits up to 1.4 GeV as input for the FDRs, Ref.~\cite{Pelaez:2022qby} provided the first precise and model-independent dispersive determination of the $f_0(1370)$ pole parameters---mass, width, and coupling to $\pi\pi$---from $\pi\pi$ scattering data. 
Remarkably, such a pole was absent in the original model-dependent analysis of these data made by the experimentalists~\cite{Hyams:1973zf,Hyams:1975mc},
and in the CFD~\cite{GarciaMartin:2011cn} analysis. In the study of Ref.~\cite{Pelaez:2022qby}, besides the controversial $f_0(1370)$, the $f_2(1270)$ and $f_0(1500)$ poles were also identified in the nearby region. Note that the latter could be reached by analytic continuation, even if the partial-wave input from the real axis only reached 1.4 GeV. It is also worth mentioning that, although the $f_{0}(1370)$ pole was not inserted by hand, it emerges naturally in the global parametrization in~\cite{Pelaez:2018qny} from the analyticity constraints built into it. However, this determination is inherently model dependent, since it relies on the chosen functional form. 

Once the FDR$_{C_N}$ approach has been shown to provide model-independent results for the inelastic isoscalar resonances that appear around 1.2 to 1.5 GeV in the $F^{00}$ FDR output, the natural step is to provide a model-independent dispersive study of all the resonances that appear in all FDRs in $\pi\pi$ scattering, which is the aim of this work. 
For this, we will use as input the Global Fits obtained in our recent dispersive analysis of $\pi\pi$ scattering data presented in~\cite{Pelaez:2024uav}, which, as explained above, extends to the other waves with $J\leq 4$, refines the P-wave analysis, and improves the matching with the Regge regime, thus enlarging the application of the $F^{0+}$ and $F^{I_t=1}$ FDRS to 1.6 GeV.

Before presenting our results, let us provide some notation and details of the FDR$_{C_N}$ approach.

\section{The FDR$_{C_N}$ method}
\label{sec:method}

The total amplitude $F^{(I)}(s,t)$ of definite isospin $I$ is  normalized as follows:
\begin{eqnarray}
F^{(I)}(s,t)&=&\frac{8}{\pi} \sum_{\ell=0}^{\infty}(2\ell+1)P_\ell(z(s,t))t^{(I)}_\ell(s)\;,\label{eq:FI}\\
t^{(I)}_\ell(s)&=&\frac{\pi}{16}\int_{-1}^{1} \text{d}z P_\ell(z) F^{(I)}(s, t(s,z))\;,\label{eq:pw}
\end{eqnarray}
where $t^{(I)}_\ell(s)$ are the partial waves of definite isospin $I$ and angular momentum $\ell$ and $z=\cos\theta$, with $\theta$ the scattering angle in the center-of-mass (c.m.) frame. The Mandelstam variables are the relativistic energy squared $s$ and  $t=-2k(s)^2\left(1-\cos\theta\right)$, with $k(s)^2=s/4 - m_\pi^2$ the c.m.-momentum squared. The third variable $u=4m_\pi^2-s-t$ is convenient to make explicit crossing symmetry, but it is not independent from the other two, and we will omit it for brevity.
In the literature, it is also common to use a different normalization for the total amplitude, $T^{(I)}(s,t)= 4\pi^2F^{(I)}(s,t)$. 

For our analysis, we apply FDRs to the next three amplitudes, which can be decomposed into $s$-channel isospin amplitudes as follows:
\begin{eqnarray}
F^{00}=\frac{1}{3}\left( F^{(0)}+2F^{(2)} \right),\;
F^{0+}=\frac{1}{2}\left( F^{(1)}+F^{(2)}\right), \label{eq:Fsym}\\
F^{I_t=1}=\frac{1}{6}\left(2F^{(0)}+3F^{(1)}-5F^{(2)}\right). \label{eq:Fanti}
\hspace{1cm}
\end{eqnarray}
The first two correspond to the $\pi^0\pi^0$ and $\pi^0\pi^+$ scattering amplitudes, respectively, and are symmetric under $s\leftrightarrow u$ crossing symmetry. Note that in the study of $F^{00}$ in~\cite{Pelaez:2022qby}, isovector resonances could not be determined, whereas here they can be accessed through the output of $F^{0+}$.
Since the $I=2$ channel is repulsive and nonresonant in the energy region of interest, all poles found in $F^{00}$ correspond to isoscalar resonances, while those in
$F^{0+}$  correspond to isovector ones.

With the above definitions, we can write a once-subtracted FDR, which reads:
\begin{eqnarray}
F^i(s,0)&=&F^i(4M^2_\pi,0)-\frac{s(s-4M^2_\pi)}{\pi} \label{eq:FDRsym}\nonumber\\
&\times& \int_{4M^2_\pi}^\infty \frac{(2s'-4M^2_\pi)\,\Im F^i(s',0)\,ds'}{s'(s'-s)(s'-4M^2_\pi)(s'+s-4M^2_\pi)}\;,\nonumber\\
\end{eqnarray}
where $i=00$ or $0+$.
When $s$ is in the physical region of the real axis, a principal value must be taken in the integral, and thus, the FDR only provides $\Re F^i(s,0)$ on the left-hand side (see~\cite{Pelaez:2024uav} for the explicit expression).

The input for these integrals will be the partial waves $t^{(I)}_\ell$ with $\ell\leq4$ in the Global Fits of~\cite{Pelaez:2024uav} and an average Regge parametrization used to describe the asymptotic region. Since the imaginary parts of $t^{(I)}_\ell$ are positive, all the contributions to the numerator in the integrals are positive. This positivity prevents cancellations and makes these two FDRs very accurate.

For the $F^{I_t=1}$ amplitude, it is possible to write an unsubtracted FDR, namely:
\begin{eqnarray}
 F^{I_t=1}(s,0)&=&\frac{2s-4M^2_\pi}{\pi}\nonumber
\\
&\times& \int_{4M^2_\pi}^\infty \frac{\Im F^{I_t=1}(s',0)\,ds'}{(s'-s)(s'+s-4M^2_\pi)}\;. \label{eq:FDRanti}
\end{eqnarray}
Once again, when $s$ lies in the real axis above the $\pi\pi$ threshold, a principal value must be taken on the integral, which only provides $\Re F^{I_t=1}(s,0)$.
Since this FDR is not positive defined, it leads to larger uncertainties in general. 
Moreover, in Eq.~\eqref{eq:Fanti} we see that the $F^{I_t=1}$ amplitude has contributions from all isospin amplitudes. Since both the isoscalar and isovector amplitudes contain resonances, their poles overlap more often in the FDR output, hindering their identification. Thus, although we have also studied the analytic continuation of $F^{I_t=1}$, it does not yield any relevant resonant information beyond the one already present in the $F^{00}$ and $F^{0+}$ FDRs. 

In~\cite{Pelaez:2024uav}, we made sure that the Global Fits satisfied the three FDRs within uncertainties for physical values of $s$ from the $\pi\pi$ threshold up to $\sqrt{s}=1.4\,$GeV for $F^{00}$, and 1.6~GeV for $F^{0+}$ and $F^{I_t=1}$. However, the FDR dispersive integrals also provide a rigorous analytic continuation to the first Riemann sheet of the complex plane. To reach an unphysical Riemann sheet, we will use continued fractions. This combination is what we refer to as the FDR$_{C_N}$ method. Nevertheless, the FDR output in the complex plane will be used below to check the goodness of the continued-fraction method.

Thus, following~\cite{Pelaez:2022qby}, from the FDR output and in a given real segment of $s$, we calculate the continued fraction of order $N$ as
\begin{equation}
C_{N}(s)=F\left(s_{1}\right)\Big/\left( \vcenter{\hbox{$ 1+\frac{\displaystyle a_{1}\left(s-s_{1}\right)}{\displaystyle 1+\frac{\displaystyle a_{2}\left(s-s_{2}\right)}{\displaystyle \ddots a_{N-1}\left(s-s_{N-1}\right)}}$}}\right)\;. 
\end{equation}
The $a_i$ coefficients are calculated recursively so that $F(s_i,0)=F(s_i)=C_N(s_i)$, i.e., so that the continued fraction reproduces the FDR output at the $N$ sampling points. Here, $F$ stands for any of the amplitudes $F^{00}$, $F^{0+}$ or $F^{I_t=1}$, and the points $\{s_i\}$ are taken to be $N$ equally spaced points (in $\sqrt{s}$) in the real segment of interest. Note that, being meromorphic, the $C_N$ functions have no cut; therefore, they continue $F(s)$ analytically to its first Riemann sheet in the upper half plane, but to the contiguous sheet in the lower-half plane, where we can search for resonance poles. Indeed, each $C_N$ is a different parametrization, i.e., a Pad\'e approximant of order $((N-1)/2, (N-1)/2)$ (we choose odd values of $N$ for technical reasons), which contains $(N-1)/2$ poles. Consequently, each of them provides a different approximation to the analytic continuation of the $F(s,0)$ amplitude in the complex plane.

Concerning uncertainties, for each region under study, we have sampled the segment size and Global-Fit parameters of our dispersive analysis to obtain a systematic uncertainty for each $N$.
To ensure model independence, we have also studied $N$ values from $N=11$ to $N=51$. With all these variations, we have collected tens of thousands of samples. Of course, only a few of the poles we find lie close to the real axis, and their parameters are stable within the sampling. These correspond to resonances, and we will see that their parameters are remarkably stable against variations of $N$.  Generically, the rest of the poles in $C_N$ appear far from the region of interest. However, given the large number of parameters and samples, as well as the nonlinearity of $C_N$, numerical calculations sometimes yield spurious artifacts, and our numerical pole-search algorithm does not find the pole of interest. This occurs in less than 1\%  of the time in all cases except two, where it happens in approximately 5\% or 10\% of the samples. These numerical artifacts disappear for small variations of the parameters and are not consistently found throughout the sampling, from which they can be safely removed. Thus, only the poles that appear consistently are considered resonances.

Regarding the pole residues, we make no prior assumption that they are tied to the pole position in any specific way. In addition, given that we only have access to poles from the $\pi\pi$ scattering amplitudes, these residues can be related to the coupling of the resonance to two pions, following the definitions in~\cite{GarciaMartin:2011jx}.

As already mentioned, one caveat of this method is that we cannot directly identify the spin of a given resonance from the FDRs. However, as we will see later, the partial-wave parametrizations enable us to identify the spin for the resonances extracted from the continuation of the FDR outputs.

Let us then discuss our results.

\begin{figure*}[t]
\centering
\includegraphics[width=\textwidth]{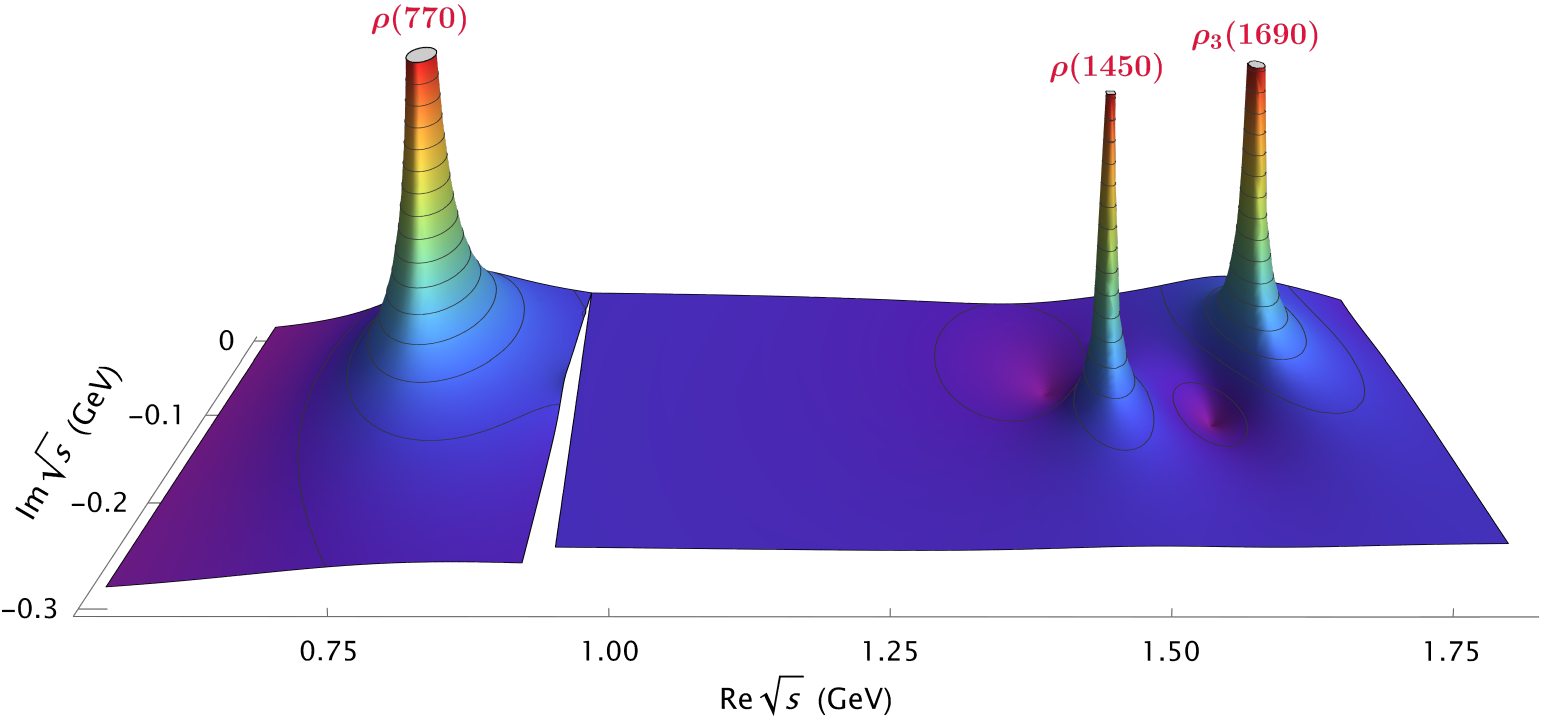}
\caption{ \small \label{fig:f0p-rep} 
Analytic continuation of the $F^{0+}$ FDR output, for the central values of the Global Fit~I input, and a representative $N$.
The amplitude is continued from the intervals [0.65, 0.85] GeV and [1.35, 1.55] GeV in the elastic and inelastic regions, respectively.
Each interval gives access to different contiguous sheets, which is illustrated with a discontinuity attached to the $\pi\omega$ threshold, $\sim$0.922 GeV.
Note the presence of the $\rho(770)$, $\rho(1450)$, and $\rho_3(1690)$ resonance poles.}
\end{figure*}

\section{Results}
\label{sec:results}

\vspace{-2mm}

We will present first the resonances obtained from the $F^{0+}$ FDR output, which is a completely new result. Later on, we will review the $F^{00}$ resonances above 1 GeV already obtained in~\cite{Pelaez:2022qby}. However, in this work, we will also demonstrate that the method yields results consistent with RE and GKPY determinations, for resonances in the elastic regime, which we had not previously studied using this method. We will only comment briefly on the $F^{I_t=1}$ FDR results, which are much less interesting due to their much larger uncertainties.
For the plots, we will use the results obtained from the Global Fit~I as input.
The other two Global Fits yield similar figures. In the tables, we will provide pole positions $\sqrt{s_\text{pole}}=M-i\Gamma/2$ and couplings to $\pi\pi$, defined as $g_{\pi\pi}=|g_{\pi\pi}|e^{i\phi}$, for all of them. Couplings are defined from partial waves as follows:
\begin{equation}
    g_{\pi\pi}^2=-16\pi\lim_{s\to s_\text{pole}}(s- s_\text{pole})t_\ell^{(I)}(s)(2\ell+1)/(2k)^{2\ell}.
\end{equation}
To extract the coupling from the pole residues in the
analytic continuation of the $F^{00}, F^{0+}$, and $F^{I_t=1}$ amplitudes, the corresponding normalization is obtained from Eqs.~\eqref{eq:FI},~\eqref{eq:Fsym}, and~\eqref{eq:Fanti}. 
The modulus $\vert g_{\pi\pi}\vert$ measures the coupling strength of the resonance to the $\pi\pi$ channel with the quantum numbers of the partial wave. In the case of 
narrow BW–like states, isolated from other resonances or thresholds, $\vert g_{\pi\pi}\vert$ is directly related to the partial width/branching fraction (see e.g., the RPP~\cite{ParticleDataGroup:2024cfk}); analogous relations hold for other simple parametrizations such as Flatté-type forms~\cite{Flatte:1976xu} (see~\cite{Burkert:2022bqo} for a broader discussion). 
The phase $\phi$ (defined modulo $\pi$) controls the interference of the pole contribution with the rest of the amplitude. 
For a strict BW formula arising from the tree-level exchange of a resonant state in a Lagrangian, one would expect the phase to be zero. Thus, the closer to this description, the smaller the phase.
In simple modifications of the single-channel narrow BW resonance formula plus a regular background, if the pole is well isolated from other singularities, the phase of the coupling seems to be well approximated  (see \cite{Ceci:2016pdn,Ceci:2025gsm,Ceci:2025yas}) by the sum of two angles in the $\sqrt{s}$-plane. The first one is the negative angle subtended at threshold by the positive real axis and the segment from the threshold to the pole. The second one is the angle subtended by the normal to the real axis that starts at the pole and the segment that joins the pole to the real energy where the phase shift reaches $\pi/2$.  Most frequently, the second does not compensate for the first, and the resulting phase comes out negative.
However, far from those simple scenarios, there is no clear geometric interpretation for $\phi$. 
In a multichannel $a\to b$ case, the pole's residue $r_{ab}\propto g_a g_b$, so that the residue phase encodes the relative phases between channels and is relevant for model building. This is of particular interest for meson-baryon phenomenology ~\cite{Anisovich:2011fc,Svarc:2014zja,Arndt:2006bf,Ronchen:2022hqk,Hoferichter:2023mgy}, and the reason why, in that case, such phases are listed in the RPP~\cite{ParticleDataGroup:2024cfk}. Although less usual for meson resonances, we provide them here, since we have been previously asked for the $\rho(770)$ and $K^*(892)$ coupling phases, of interest for studies of radiative transitions~\cite{Hoferichter:2017ftn,Dax:2020dzg,Niehus:2021iin}. In general, we provide coupling phases in the $[-90^\circ,90^\circ]$ determination, but, whenever it is useful for comparison with other Global Fit results, we use the $[0^\circ,180^\circ]$, or $[-180^\circ,0^\circ]$ determinations.

\subsection{Resonances in $F^{0+}$ }

In Fig.~\ref{fig:f0p-rep}, we show the analytic continuation to the lower half-plane of the $F^{0+}$ FDR in the contiguous sheet from the $2\pi$ threshold to 1.85 GeV. To make the figure more intuitive, we plot the modulus of the amplitude against $\Re \sqrt{s}$ and $\Im \sqrt{s}$ instead of $s$. Of course, the amplitude is analytic in the $s$ variable, except for physical singularities (cuts for thresholds and poles in unphysical sheets for resonances), so, for the continued fractions, we use the variable $s$. For this figure, we have used the FDR output from the central values of the Global Fit~I parameters for a representative choice of $N$ and different segments along the real axis from which the analytic continuation is calculated. Note that the analytic continuations above and below the $ \pi\omega$ threshold at $\sim$922~MeV, i.e., the first inelastic threshold in this amplitude, do not match continuously. This is because the $\rho(770)$ resonance dominates the elastic region and the continuation to the contiguous sheet below $\pi\omega$ threshold (the so-called second sheet); in contrast, it has much less influence on the contiguous sheet above the $\pi\omega$ threshold. 
One might worry about similar effects at other thresholds. However, such cases do not correspond to two-body thresholds of almost stable particles with such a dominant resonance nearby on one side. For example, both $4\pi$ and $K\bar K$ couple weakly to the $\rho(770)$ compared to $\pi\pi$ or $\pi\omega$. Consequently, except for the $\pi\omega$ threshold, the analytic continuations from different sheets become indistinguishable within the uncertainties of the data and the fits. Hence, in practice, the analytic continuation looks smooth in the two regions below and above $\pi\omega$.

\begin{figure}
\centering
\includegraphics[width=0.45\textwidth]{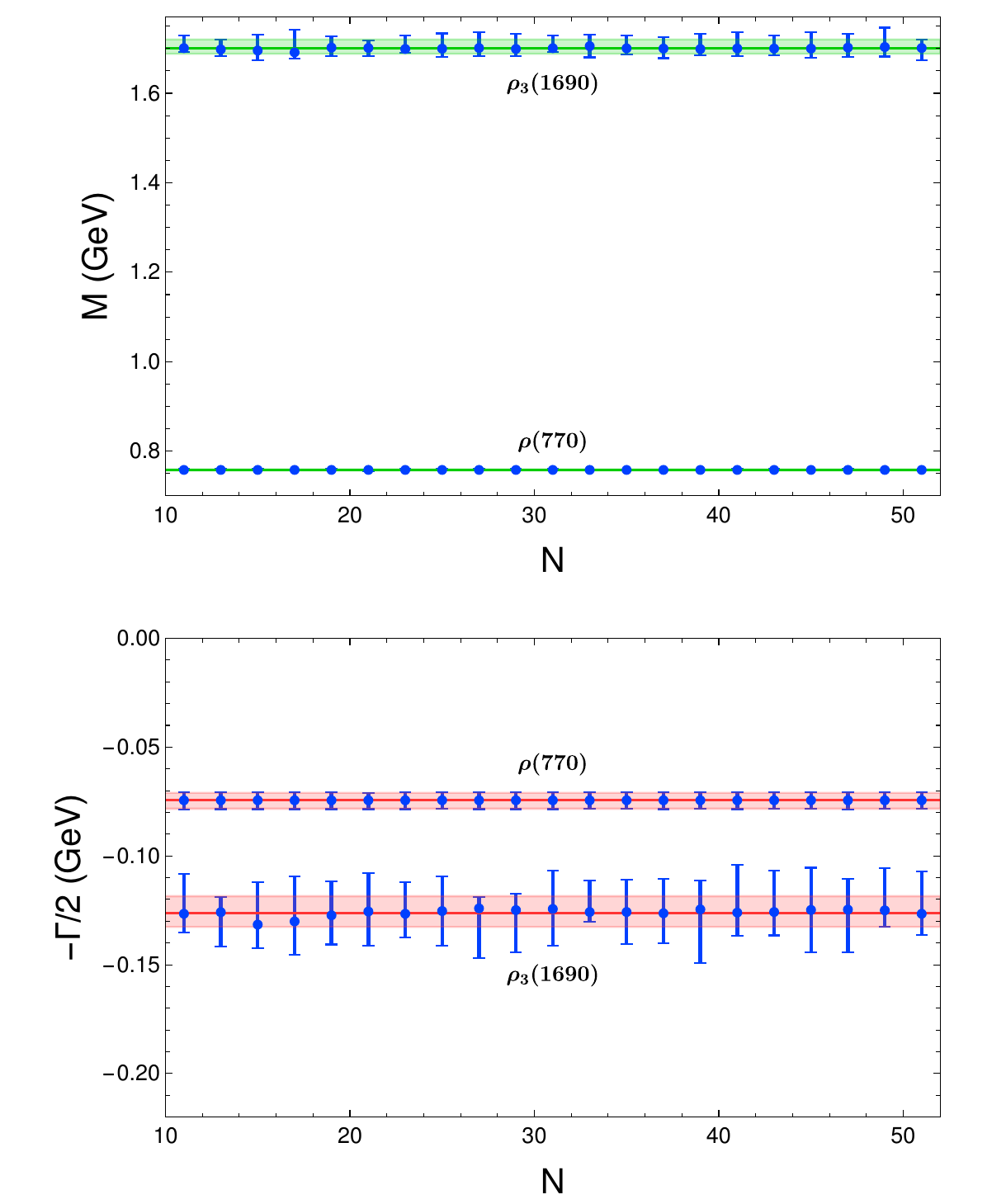}
\vspace{-2mm}
\caption{ \small \label{fig:resonances-f0p-1} 
Pole masses (top) and half-widths (bottom) for the $\rho(770)$ and $\rho_3(1690)$ resonances, obtained from the analytic continuations of the $F^{0+}$ FDR output using continued fractions $C_N$.  Results are remarkably stable against variations of $N$.}
\vspace{-4mm}
\end{figure}

\begin{figure}
\centering
\includegraphics[width=0.45\textwidth]{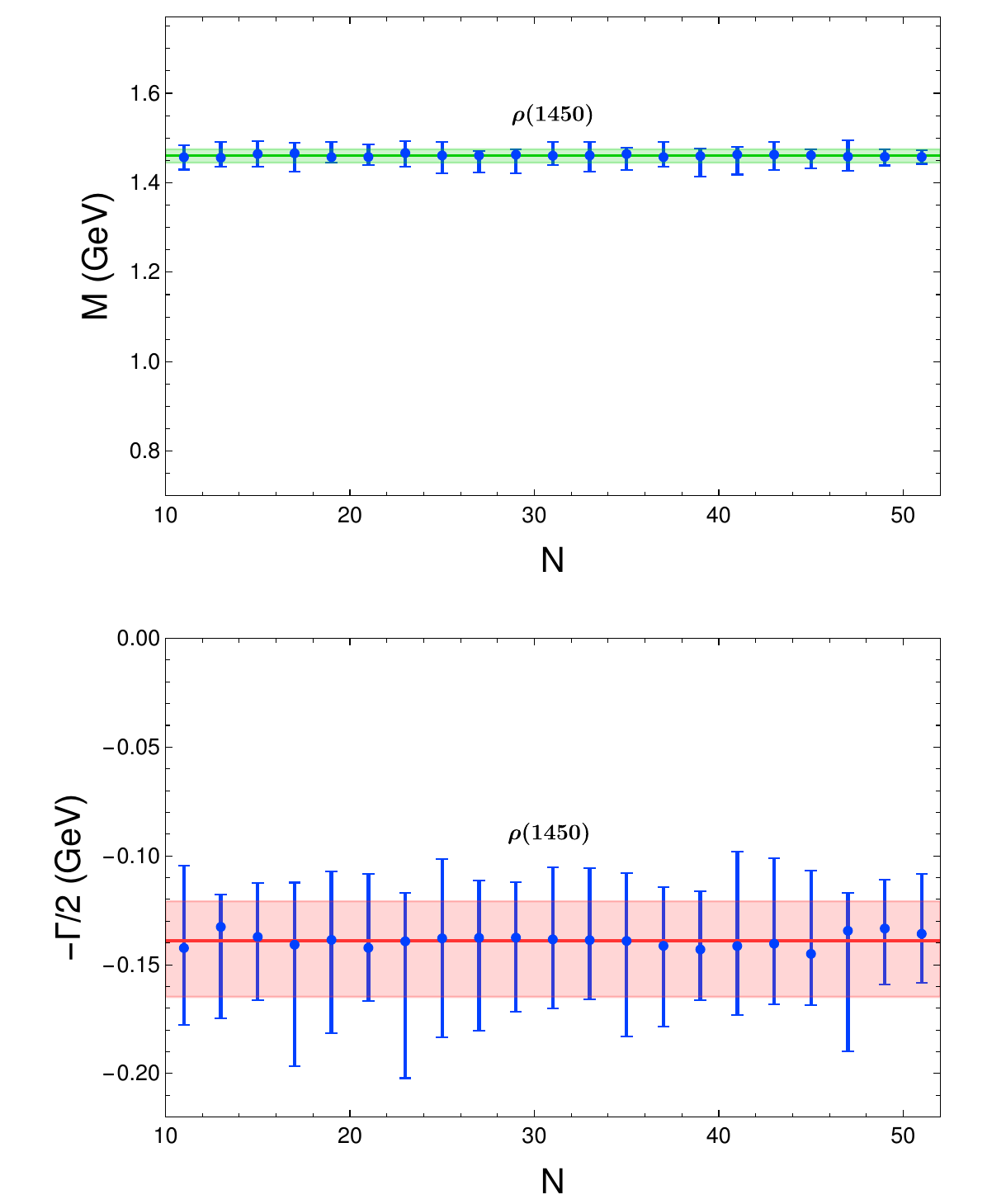}
\vspace{-2mm}
\caption{ \small \label{fig:resonances-f0p-2} 
Pole masses (top) and half-widths (bottom) for the $\rho(1450)$ resonance, obtained from the analytic continuations of the $F^{0+}$ FDR output using continued fractions $C_N$. Results are remarkably stable against variations of $N$.}
\end{figure}

The three poles seen in Fig.~\ref{fig:f0p-rep} can be identified with the $\rho(700)$, $\rho(1450)$, and $\rho_3(1690)$ resonances. In Figs.~\ref{fig:resonances-f0p-1} and~\ref{fig:resonances-f0p-2} we illustrate the remarkable stability of these poles against the choice of $N$. 
The error bars for each $N$ are calculated from the uncertainties in the Global Fit parameters used as input in the FDR and the variation of the boundaries of the segment of interest. These variations correspond to considering subsegments of the original intervals, where we vary one of the boundaries by multiples of 2 MeV up to a maximum variation of 24 MeV.

At this point, it is worth remarking that we do not find any other stable pole in the 1 to 1.7 GeV region coupling strongly to $\pi\pi$. In particular, the present edition of the RPP~\cite{ParticleDataGroup:2024cfk} lists in the particle listings (though not in the summary tables) another isovector-vector candidate: the $\rho(1570)$, which we do not find. This absence is not surprising since this state has only been seen to decay into three different states: $e^+e^-$, $\omega\pi$, and $\phi\pi$, but not into $\pi\pi$. Actually, the RPP suggests that this putative resonance might be just a misidentified Okubo–Zweig–Iizuka (OZI)-violating decay mode of the $\rho(1700)$.

There is yet another vector-isovector state first proposed in the 1970s, called the $\rho(1250)$,  which appeared in the RPP until 1986, when it was removed in favor of just one $\rho'$ state: the $\rho(1450)$. Still, in the 1990s, two more experimental claims of such a resonance with a mass around 1250 MeV were made by the LASS collaboration in the proceedings~\cite{Aston:1990wg}, and by the OBELIX Collaboration in~\cite{OBELIX:1997zla}. About a decade later, the P-wave $\pi\pi$ phase shifts and
inelasticities up to 1.9 GeV, measured in~\cite{Hyams:1973zf,Hyams:1975mc},  were studied within 
a  multichannel unitarity model~\cite{Surovtsev:2008zza}, later refined in~\cite{Surovtsev:2010cjf}, and also found such a resonance together with the $\rho(1450)$.
The subject has gained renewed interest after the very recent claim of strong evidence for the $\rho(1250)$ in a further improvement~\cite{Hammoud:2020aqi} of the P-wave $\pi\pi$ scattering analyses of~\cite{Surovtsev:2008zza,Surovtsev:2010cjf}.
In contrast, although we have explicitly looked for it, in this work we find no hint of a $\rho(1250)$ resonance pole in the analytic continuation of our dispersive $\pi\pi$ analysis to the contiguous Riemann sheet, where the corresponding pole should appear according to~\cite{Hammoud:2020aqi}.
If this resonance existed, our result would require a very small coupling to $\pi\pi$, in line with the findings of~\cite{henner:1985}, so that, in practice, it would be concealed in the uncertainty when using $\pi\pi$ scattering data alone. 

Let us then discuss separately the parameters of the three poles that we identify below $\sim$1700 MeV based on the analytic continuation of the $F^{0+}$ FDR output.

\subsubsection{$\rho(700)$}
\label{sssec:rho}

It is the lightest isovector resonance, observed in many processes, and known for a very long time with great accuracy. Concerning its $T$-matrix pole position, the RPP~\cite{ParticleDataGroup:2022pth} estimate is at $\sqrt{s_{\rho(770)}}=(761–765) - i (71–74)$ MeV, in view of several determinations from Roy-like equations~\cite{Colangelo:2001df,GarciaMartin:2011jx,Hoferichter:2023mgy}. 
Being so well-determined from robust and model-independent dispersive techniques, we can use this resonance to calibrate the goodness of our approach by comparing with the results from GKPY equations using the same input.

Thus, in Table~\ref{tab:rho770} we first show the $\rho(770)$ pole parameters obtained in~\cite{GarciaMartin:2011jx} from
GKPY equations, using as input their constrained fits to data (CFD).
Next, we present the results obtained here from the $F^{0+}$ FDR output using as input each of the three Global Fits from~\cite{Pelaez:2024uav} in the [0.65,~0.85]~GeV segment. For comparison, we also include the pole parameters found by solving the GKPY equations with the same Global Fit inputs, where the second-sheet amplitude is obtained from the inverse of the first-sheet S matrix.

The first observation is that our FDR$_{C_N}$ results from the three Global Fits are remarkably compatible among themselves and also with the GKPY results. 
Concerning uncertainties, those for the mass and coupling are somewhat smaller for our FDR$_{C_N}$ method, whereas they are rather similar for the width. All our new GKPY results are consistent with those of~\cite{GarciaMartin:2011jx}; our uncertainties are somewhat larger, particularly for the width, since Global Fits are designed to cover a much larger energy range, whereas the CFD focuses mainly on the elastic region for precision.

Since the results for each method are consistent across the three Global Fits, we have combined them into a single estimated range for each method, also provided in Table~\ref{tab:rho770}.

\begin{table}[h]
\footnotesize
\resizebox{.48\textwidth}{!}{
\renewcommand{\arraystretch}{1.8}
	      \begin{tabular}{ccccc} \hline\hline
		 Input & Method &  $\sqrt{s_{\rho(770)}}$ (MeV)& $|g_{\pi\pi}|$& $\phi$ ($^\circ$)\\ 
        \hline 
        CFD~\cite{GarciaMartin:2011jx} & GKPY& $\left(763.7^{+1.7}_{-1.5}\right)$$\,-\,$$i\left( 73.2^{+1.0}_{-1.1}\right)$ & $6.01^{+0.04}_{-0.07}$& \\
        \hline\hline
		\multirow{ 2}{*}{GFI} \hspace{0.2mm} & FDR$_{C_N}$ & $\left(758.1^{+1.2}_{-1.6}\right)$$\,-\,$$i\left(74^{+4}_{-3}\right)$  & $6.06^{+0.14}_{-0.12}$&  $-6.2^{+0.3}_{-0.4}$\\
         &GKPY& $\left(759 \pm 3\right)$$\,-\,$$i\left( 75\pm3\right)$ & $6.08^{+0.22}_{-0.19}$ & $-5.7\pm 1.0$ \\\hline
        \multirow{ 2}{*}{GFII} \hspace{0.2mm} & FDR$_{C_N}$ &$\left(758.5^{+1.6}_{-1.7}\right)$$\,-\,$$i\left(71\pm4\right)$ & $5.94^{+0.18}_{-0.14}$ & $-5.68^{+0.20}_{-0.22}$\\
         &GKPY&$\left(759^{+3}_{-5}\right)$$\,-\,$$i\left( 72^{+4}_{-6}\right)$& $5.97\pm0.17$ & $-5.2\pm 1.1$\\\hline
        \multirow{ 2}{*}{GFIII} \hspace{0.2mm} & FDR$_{C_N}$ &$\left(758.0^{+1.4}_{-1.8}\right)$$\,-\,$$i\left(72\pm4\right)$ & $5.98^{+0.16}_{-0.13}$ &  $-5.87^{+0.25}_{-0.21}$\\
         &GKPY&$\left(759 \pm 3\right)$$\,-\,$$i\left( 73\pm4\right)$&  $5.89^{+0.16}_{-0.14}$&$-5.2\pm 1.1$\\\hline
         \multirow{ 2}{*}{$\mathbf{\overline{\textbf{GF}}}$}        &   \textbf{FDR$\mathbf{_{\textbf{\textit{C}}_\textbf{\textit{N}}}}$} & $\mathbf{\left(758.2^{+1.9}_{-2.0}\right)-\textbf{\textit{i}}\left(73^{+5}_{-6}\right)}$  &   $\mathbf{6.00\pm0.20}$ & $\mathbf{-5.8^{+0.4}_{-0.8}}$ \\
              &   \textbf{GKPY} & $\mathbf{\left(759^{+3}_{-5}\right)-\textbf{\textit{i}}\left(74^{+4}_{-8}\right)}$  &   $\mathbf{5.96^{+0.34}_{-0.21}}$ & $\mathbf{-5.4\pm1.3}$ \\
      \hline
              \hline
	      \end{tabular}}
\caption{
Dispersive determinations of $\rho(770)$ pole parameters using as input in the dispersive integrals each of the three dispersively constrained 
Global Fits (GF in the table) from~\cite{Pelaez:2024uav}. We present results for our FDR$_{C_N}$ method as well as from the GKPY equations. Since the results for each method are consistent for the three Global Fits, we provide a single range in boldface for each method, covering the results of the three Global Fits.
The results are compatible with the values obtained using the GKPY equations with the CFD in~\cite{GarciaMartin:2011jx}, but with larger uncertainties.}
\label{tab:rho770}
\end{table}

The $\rho(770)$ case illustrates the reliability and competitiveness of the FDR$_{C_N}$ method. The reason for the similar or even better precision is twofold. On the one hand, the positivity of the $F^{0+}$ input. On the other hand, at low energies, the uncertainties come mostly from the S0 and P waves. GKPY equations include both waves as input for their integrals, whereas the $F^{0+}$ FDR does not require the S0 contribution. As we will see later, something similar happens for the $f_0(500)$ and $f_0(980)$ resonances from the $F^{00}$ FDR, which does not require the P contribution. 

\begin{figure}
\centering
\includegraphics[width=0.48\textwidth]{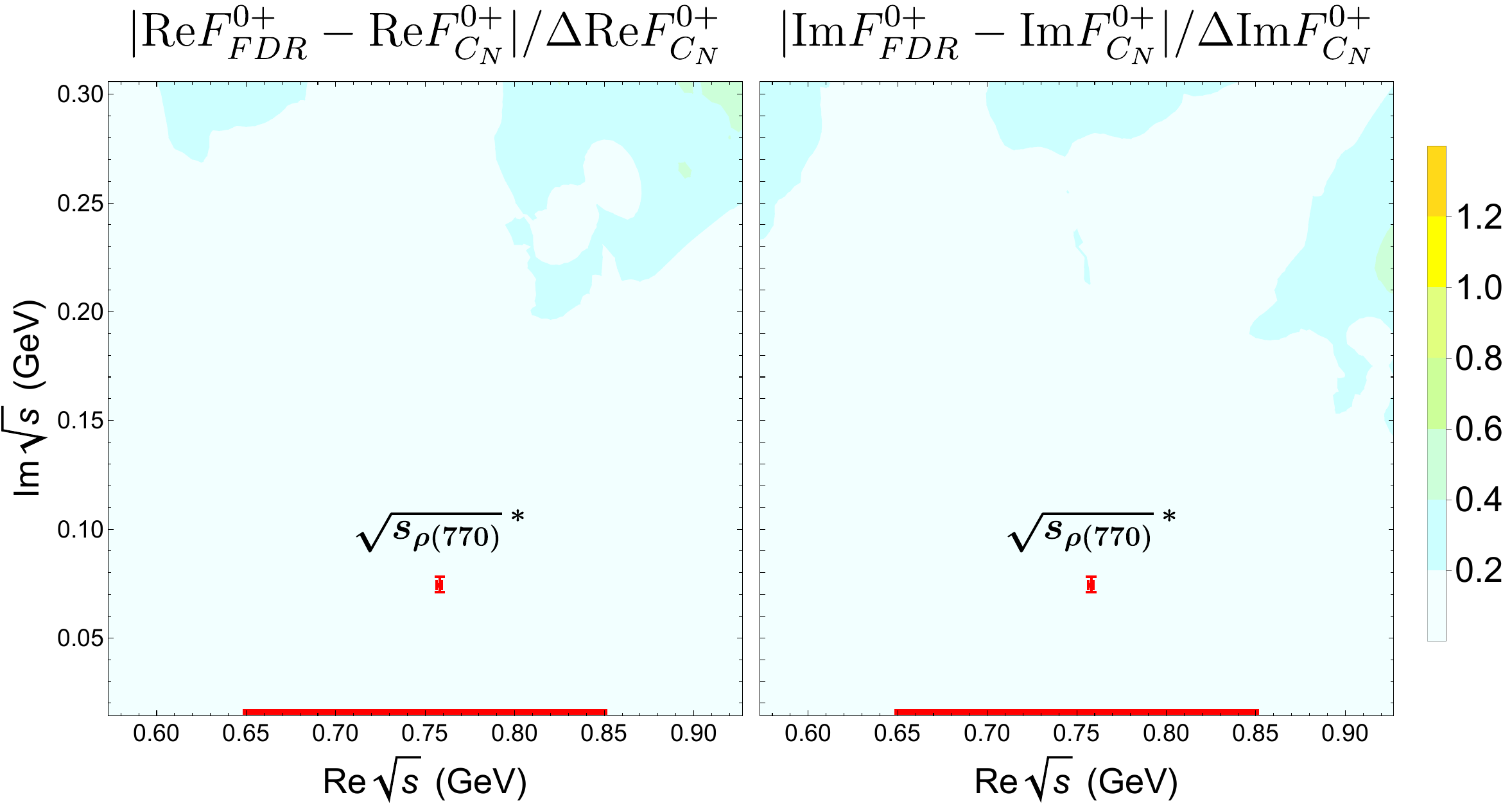}
\caption{ \small \label{fig:rhoplane} 
Reliability of the continued fraction analytic continuation method in the upper half $\sqrt s$-plane near the $\rho(770)$ region.
The first Riemann sheet value of the $F^{0+}$ FDR output, $F^{0+}_{FDR}$,
lies well within the estimated uncertainty of the continuation made with continued fractions $F^{0+}_{C_N}$. We show the absolute value of the difference for the real parts (left) and for the imaginary parts (right), divided by the uncertainty of the $C_N$ calculation.
For reference, we show the conjugate position of the $\rho(770)$ pole that appears in the second sheet (red cross) and the continued segment (red line).
}
\end{figure}

Last, we might wonder how reliable the $C_N$ analytic continuation and our uncertainty estimate are. As seen in Eq.~\eqref{eq:FDRsym}, the FDR integral provides us with a model-independent dispersive determination of the $F^{0+}$ amplitude in the complex plane. These values correspond to the first Riemann sheet, which is also reached by the $C_N$ analytic continuation method in the upper-half plane. Hence, in this region, we can compare the two of them. Thus, in Fig.~\ref{fig:rhoplane} we show how well the purely dispersive output in the upper half plane around the $\rho(770)$ region is described when the $C_N$ method is used to continue the FDR output from the [0.65,~0.85]~GeV segment. We see that both the real (left) and imaginary parts (right) differ by far less than the uncertainty we have estimated for the $C_N$ calculation.  Thus, the figure illustrates that the $C_N$ method provides a very good description of the dispersive result in the upper half-plane.
For reference, we have plotted in red the conjugate position of the $\rho(770)$ pole, which appears in the second sheet, and, also in red, the central value of the segment that has been used for the analytic continuation. Of course, the actual pole that dominates the amplitude in the real axis lies in the lower half-plane, and in the second sheet, but the latter can only be reached with the $C_N$ continuation method. 

\subsubsection{$\rho(1450)$}
\label{sssec:rho1450}

\begin{figure}
\centering
\includegraphics[width=0.49\textwidth]{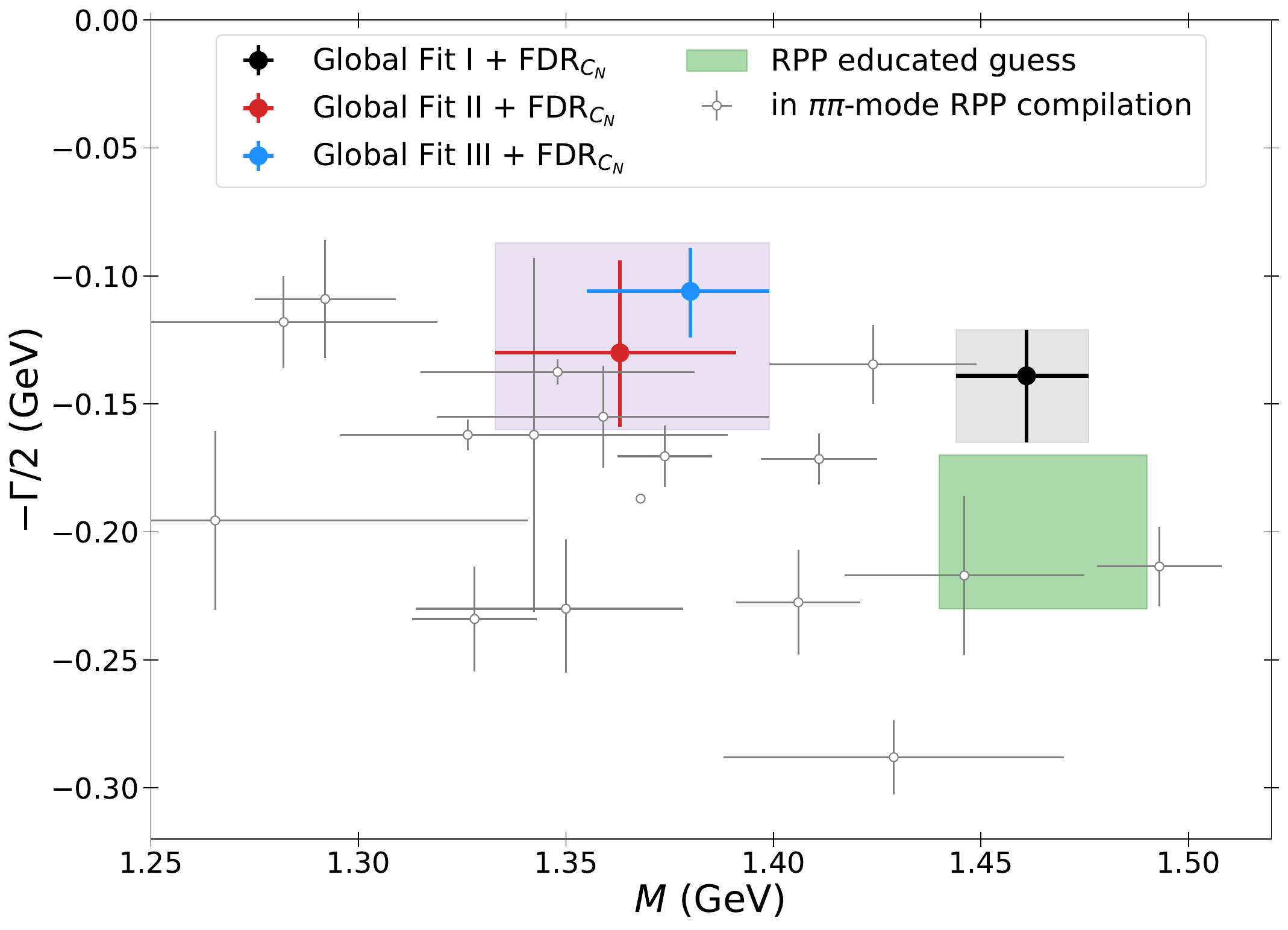}
\caption{ \small \label{fig:rho1450} 
$\rho(1450)$ poles obtained from the analytic continuation of the $F^{0+}$ FDR for the three Global Fits. We have provided a single range covering the results from Global Fits~II and III, as they are compatible.
For comparison, we also show the RPP educated guess (green area) and $ \pi\pi$ mode masses and widths collected there~\cite{ParticleDataGroup:2022pth}.
}
\end{figure}

This one is the next resonance with the quantum numbers of the $\rho(770)$, and because of that, it is sometimes referred to as $\rho'$. It is clearly visible in the $e^+e^-\to \text{hadrons}$ cross section, and it has 17 seen decay modes listed in the RPP, which collects tens of estimates from different groups and works. However, this resonance is not so well determined as the $\rho(770)$: data collected in the RPP are often incompatible among themselves (especially for the $\pi \pi$ mode), and there are no estimations of the branching ratios for any of the known decay modes. In addition, the RPP only provides an educated guess as an estimate for the mass: $M_{\rho(1450)}=1465\pm25\,$MeV,  which is incompatible with most of the data collected for the $\pi\pi$ mode (see Fig.~\ref{fig:rho1450}), which tend to be lighter (see, for instance, \cite{Bartos:2017ils}). Note, however, that even if these data contain a $\pi\pi$ in the final state, none of them come from $\pi\pi\to\pi\pi$ scattering. For the width, the PDG provides once again an educated guess of $\Gamma_{\rho(1450)}=400\pm60$ MeV. Remarkably,  the RPP does not provide a $T$-matrix pole estimate. Let us recall that Roy-like equations do not reach this far in energies, or into the inelastic region. Therefore, they cannot be used to determine this particular resonance pole. In contrast, our FDR$_{C_N}$ method can be applied in this region.

Thus, we have continued our FDR $F^{0+}$ output from a segment around the $\rho(1450)$, by means of continued fractions, using as input in the dispersive integrals the three Global Fits from~\cite{Pelaez:2024uav}.
At this point, it is convenient to recall that, as explained in~\cite{Pelaez:2024uav}, the three Global Fits have very similar phases up to 1.4 GeV. Still, they begin to behave differently above that energy, following the different datasets to which they are fitted.  Concerning the inelasticities, although they are incompatible within uncertainties, they have a similar behavior also up to 1.4 GeV. Above 1.6 GeV, even their qualitative behavior is rather different for both the phase and inelasticity, although Global Fits~II and III are more similar to each other. For these reasons, the pole parameters of the $\rho(1450)$ are going to differ when using different inputs.
The best-suited segment for the continuation of Global Fit~I is [1.35,~1.55]~GeV, but [1.3,~1.5]~GeV for Global Fits~II and III. As we will discuss at the end of this subsection, we have tried a wide variety of segments, also in search for other nearby resonances. However, we have not found any other pole in the $F^{0+}$ amplitude in this region.
Our results are collected in Table~\ref{tab:rho1450}. 
The width of the resonance is fairly compatible within errors when using any of the three Global Fits as input. However, from Global Fit~I, we obtain a much heavier resonance.
As expected, the results from Global Fits~II and III  are compatible since they come from similar datasets, and we have thus combined them into a single estimate $\overline{\text{GF}}_{\text{II,\,III}}$. Here, it is important to emphasize that these poles have not been imposed in the parametrizations, but appear as stable features of analytic continuations performed in terms of multiple continuous fractions.

Nevertheless, although the fulfillment of the $F^{0+}$ FDR up to 1.6 GeV is acceptable for Global Fits~II and III, it is much better for Global Fit~I. Moreover, as explained in~\cite{Pelaez:2024uav}, the imposition of FDR constraints on Global Fits~II and III changes them more from their original unconstrained fits than what happens with Global Fit~I. In addition, even after constraining them, Global Fits~II and III remain in some tension with the $F^{I_t=1}$ FDR in the [0.93,~1.06]~GeV and [1.46,~1.56]~GeV regions.
For these reasons, we have a slight preference for the dispersive outputs of Global Fit~I. However, the estimate coming from Global Fits~II and III is still acceptable and cannot be discarded.

\begin{table}
\renewcommand{\arraystretch}{1.6}
	      \begin{tabular}{ccccc} \hline\hline
		 FDR$_{C_N}$ &  $\sqrt{s_{\rho(1450)}}$ (MeV)& $|g_{\pi\pi}|$& $\phi$ ($^\circ$)\\ \hline
		\textbf{GFI}  & $\mathbf{\left(1461^{+15}_{-17}\right)-\textbf{\textit{i}}\left(139^{+26}_{-18}\right)}$  &   $\mathbf{1.8^{+0.6}_{-0.4}}$ & $\mathbf{-49^{+17}_{-16}}$\\ \hline
        GFII  &$\left(1363^{+28}_{-30}\right)$$\,-\,$$i\left(130^{+29}_{-36}\right)$  &   $0.8^{+0.4}_{-0.3}$&$-95^{+30}_{-28}$\\ 
        GFIII   &$\left(1380^{+19}_{-25}\right)$$\,-\,$$i\left(106^{+18}_{-17}\right)$  &  $0.8\pm0.3$&$-70^{+12}_{-13}$\\ 
        $\mathbf{\overline{\textbf{GF}}_{\textbf{II,\,III}}}$      & $\mathbf{\left(1374^{+25}_{-41}\right)-\textbf{\textit{i}}\left(112^{+47}_{-23}\right)}$  &   $\mathbf{0.8^{+0.4}_{-0.3}}$ & $\mathbf{-74^{+16}_{-49}}$ \\
          \hline
      \hline
	      \end{tabular}
\caption{Dispersive determinations of the $\rho(1450)$ pole parameters using as input in the dispersive integrals each one of the three dispersively constrained Global Fits (GF in the table) from~\cite{Pelaez:2024uav}. Global Fits~II and III are compatible and thus we have combined them into a single $\overline{\text{GF}}_{\text{II,\,III}}$ estimate (in boldface) covering both determinations. The fulfillment of the $F^{0+}$ and $F^{I_t=1}$ FDRs is better for Global Fit~I (also shown in boldface), and we consider it slightly favored.
}
\label{tab:rho1450}
\end{table}

In Fig.~\ref{fig:rho1450}, we compare our FDR$_{C_N}$ pole positions with the RPP estimate and the values in its $\pi\pi$-mode compilation.  We see that the dispersive result using the Global Fit~I (black full circle) as input is almost compatible with the RPP educated guess, although its width lies more than one deviation away. 
In the figure, we show the dispersive results from Global Fit~II (red full circle) and III (blue full circle), and a rectangular range covering both of them (given in the last line of Table~\ref{tab:rho1450}). 
It is clearly seen that, despite having different masses, the FDR$_{C_N}$ analyses of the three Global Fits yield a narrower resonance than the RPP estimate.

Let us emphasize again that, although we provide two different values for the pole using the FDR$_{C_N}$ method, these correspond to a single resonance when each Global Fit is used as input in the dispersion relations. We never find two nearby poles simultaneously in this region. Namely, our lower pole value cannot be understood as a $\rho(1250)$ resonance besides the $\rho(1450)$. They are different instances of the same pole resulting from two clearly incompatible datasets. Moreover, we have also followed this procedure to cover the rest of the regions of the energy, including segments close to a putative $\rho(1250)$, but we have not found such an additional pole in our dispersive $\pi\pi$ analysis. One may wonder if our method can only find one resonance at a time. However, we will see below how, as previously shown in~\cite{Pelaez:2022qby}, three resonances in the 1.2 to 1.6 GeV region of the $F^{00}$ FDR can be identified from the single [1.15,~1.35]~GeV segment.

\begin{figure}
\centering
\includegraphics[width=0.48\textwidth]{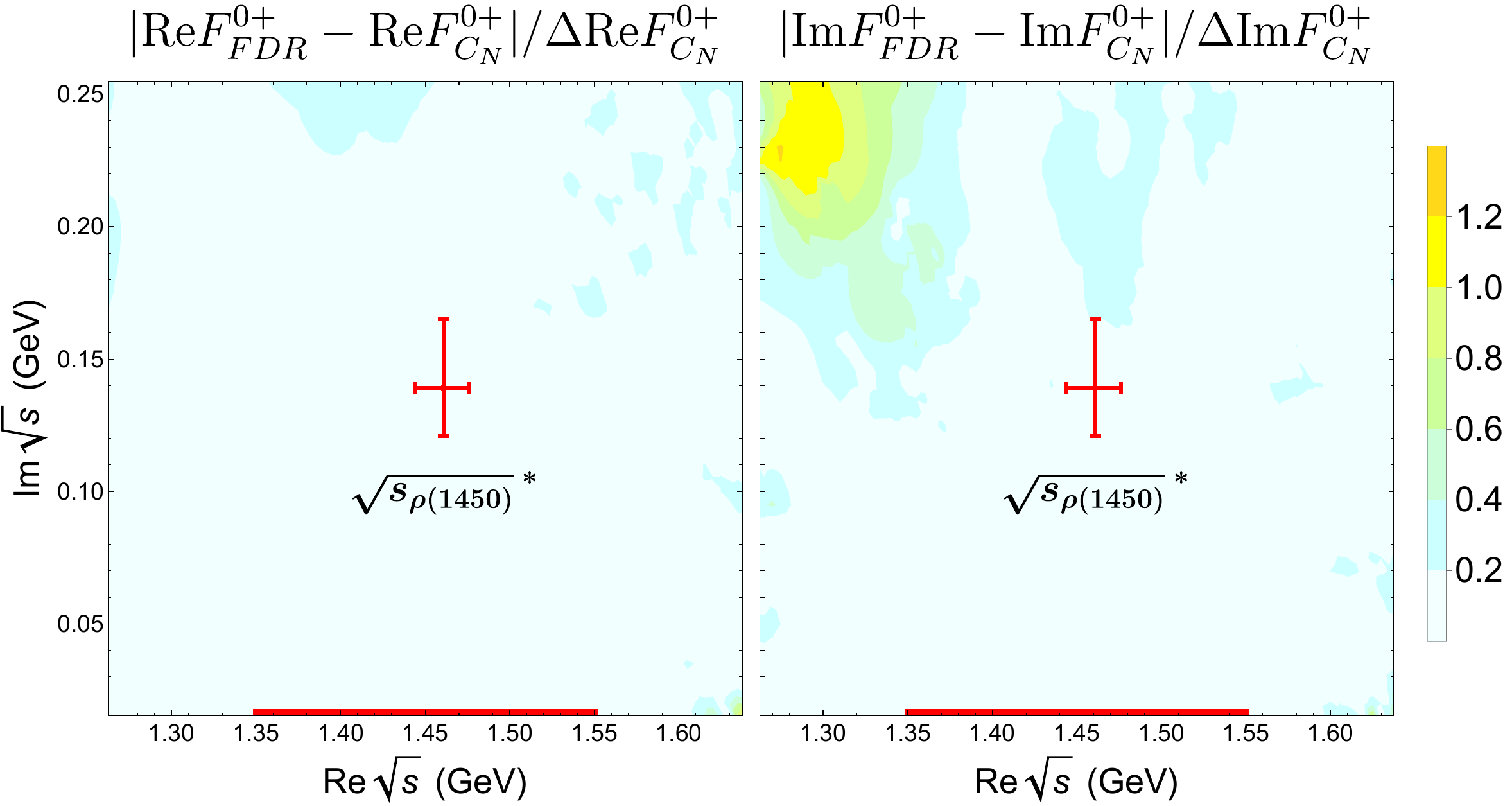}
\caption{ \small \label{fig:rho1450plane} Reliability of the continued fraction analytic continuation method in the upper half $\sqrt s$-plane near the $\rho(1450)$ region. The first Riemann sheet value of the $F^{0+}$ FDR output, $F^{0+}_{FDR}$, lies well within the estimated uncertainty of the continuation made with continued fractions $F^{0+}_{C_N}$. We show the absolute value of the difference for the real parts (left) and for the imaginary parts (right), divided by the uncertainty of the $C_N$ calculation. For reference, we show the conjugate position of the $\rho(1450)$ pole that appears in an unphysical sheet (red cross) and the continued segment (red line).
}
\end{figure}

Finally, as we did in Fig.~\ref{fig:rhoplane} for the $\rho(770)$, Fig.~\ref{fig:rho1450plane} shows that the difference between the analytic continuation in the upper half of the $\sqrt s$-plane using the FDR integrals directly and the $C_N$---using the real segment [1.35,~1.55]~GeV and a representative choice of $N$---lies within the estimated full uncertainties of our method.

\subsubsection{$\rho_3(1690)$}
\label{sssec:rho3}

This is the lightest $J=3$ resonance, which has been observed in different processes, including $\pi\pi\to\pi\pi$ scattering. 
According to the RPP, 14 different decay modes have been seen, 
and its branching fraction is dominated ($\sim$71\%) by 4$\pi$ decays,
followed by $\pi\pi$ ($\sim$24\%). Indeed, from the point of view of $\pi\pi$ scattering, this is a clearly visible resonance. Actually, its shape is very close to a very inelastic BW resonance.
However, although the RPP provides precise averages for its mass and width, $M_{\rho_3(1690)}=1688.8\pm 2.1$ MeV and $\Gamma_{\rho_3(1690)}=161\pm 10$ MeV, it does not list $T$-matrix poles. Moreover, there is a sizable difference between the width of the RPP estimate and that from the $\pi\pi$-mode alone, $186\pm14\,$MeV, or the $K\bar K, K\bar K\pi$ modes, $204\pm18\,$MeV, which come out much wider.

At this point, it is important to note that the mass determination in the $\pi\pi$ mode is dominated by $\pi N\to \pi\pi N'$ scattering analyses, where $\pi\pi$ scattering plays an obviously relevant role. In particular, the RPP includes in this average the CERN-Munich Grayer et al. analysis~\cite{Grayer:1974cr}. Let us remark that, contrary to what is done with other works, the RPP chooses only one result from~\cite{Grayer:1974cr}, $M_{\rho_3}=1693\pm8\,$ MeV, $\Gamma_{\rho_3}=200\pm18\,$MeV, which is obtained from a Breit-Wigner fit to the $\ell=3$ partial-wave intensity. 
However, the authors of~\cite{Grayer:1974cr} explicitly state that \textit{``In principle the masses and widths of resonances are best determined by a phase-shift analysis,... such a procedure gives more reliable results than a fit to the mass spectrum,..."}. The result they quote with such a phase-shift analysis is $M_{\rho_3}=1713\pm4\,$ MeV, $\Gamma=228\pm10\,$MeV, much heavier and wider than the RPP estimates and almost all other determinations from other processes and channels.

Our Global Fits were built with basically the same parametrization as~\cite{Grayer:1974cr} in the $\rho_3(1690)$ region, but modified to be elastic in the elastic region, and thus describe the known threshold parameters. Then, it is not surprising that all three of our Global Fit parametrizations contain poles whose masses and widths are similar to those from the phase-shift analysis of Grayer et al.~\cite{Grayer:1974cr}.
Of course, these are parametrization-dependent determinations. Here we will remove such a model dependence using our FDR$_{C_N}$ method.

It could come as a surprise that we can obtain dispersive information about the $\rho_3(1690)$, since our $F^{0+}$ FDRs have only been imposed and contain input from the Global-Fit partial waves up to 1.6 GeV. Above this energy, we use an ``averaged" Regge description of total cross sections.
Thus, we do not have dispersive output in the region around 1.7 GeV. Still, the naive analytic continuation of the $F^{0+}$ FDR output in the [1.35,~1.55]~GeV segment, gives rise to an unstable pole at $\sqrt{s_{p}}=(1695^{+76}_{-73})-i(125^{+90}_{-83})$ MeV for Global Fit~I (and similar for the other two).  Let us recall that 
in the $F^{0+}$ amplitude, resonances with all spins coexist, and around 
1.7 GeV there is also the $\rho(1700)$. In the RPP, its mass and width are 
$1720\pm20\,$MeV and $250\pm100\,$MeV. Most likely, the pole we find from Global Fit~I with this naive approach is a superposition of the $\rho_3(1690)$ and $\rho(1700)$ resonances.

To disentangle these two poles, we need to continue analytically a segment that reaches the 1.7 GeV region. Unfortunately, in~\cite{Pelaez:2024uav} it was not possible to obtain a constrained Global Fit that simultaneously (i) describes $\pi\pi$ scattering data {\it in all partial waves}, and (ii) satisfies the FDRs beyond 1.6 GeV. This limitation may have two origins: at those higher energies, contributions from partial waves with $J> 4$ could become relevant, but experimental information for such waves is lacking; alternatively, the one-pion-exchange approximation used to extract $\pi\pi$ scattering from $\pi N\to\pi\pi N'$ experiments may be inadequate at these high energies, so the derived $\pi\pi$ input would no longer be reliable. 

\begin{figure}[h]
\centering
\includegraphics[width=0.49\textwidth]{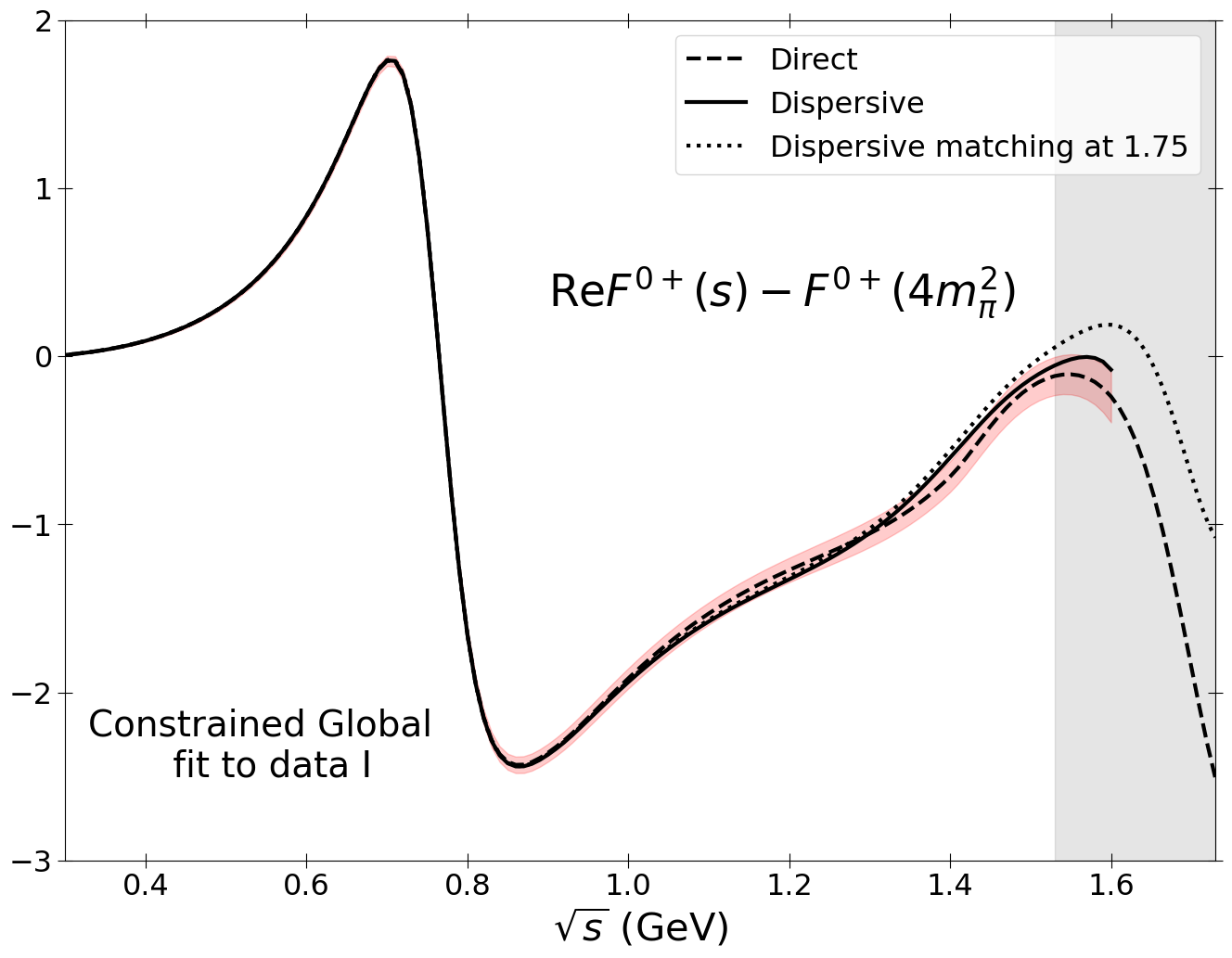}
\caption{ \small \label{fig:F0pextended} 
When the Global Fit parametrization is used as input up to 1.62 GeV, where it is matched to a Regge description, its direct result for $F^{0+}$ is consistent with its dispersive output up to 1.6 GeV. This consistency disappears if the matching is set to 1.75 GeV. We will use the average between the direct and extended-dispersive results to determine the $\rho_3(1690)$ pole and its uncertainty. }
\end{figure}

Nonetheless, we can still calculate the $F^{0+}$ dispersion relation above 1.6 GeV. For that, it is enough to use in the dispersive integrals the $F^{0+}$ amplitude reconstructed from our Global Fits up to a higher energy,
matching the input from the Regge regime at $\sim$1.75~GeV.
In Fig.~\ref{fig:F0pextended}, we show how, when applied up to 1.6 GeV as in~\cite{Pelaez:2024uav}, the dispersive output of the $F^{0+}$ FDR (continuous line) agrees within uncertainties (red band) with the $F^{0+}$ direct calculation used in the input (dashed line). 
In contrast, when the FDR output is extended to 1.75 GeV (dotted line), it no longer agrees with the direct calculation beyond 1.5 GeV.
Thus, neither the direct fit nor the extended-dispersive results can be trusted, since they disagree.
Whatever mechanism corrects the existing data to make them consistent with the dispersive representation, it must bring these two lines together somewhere in the middle. Indeed, up to 1.6 GeV, the original dispersive output, which is consistent with the direct input, lies right between the direct calculation and the extended-dispersive output. 

As expected, the analytic continuation of the extended-dispersive result contains a much more stable $\rho_3(1690)$ pole. Hence, we have determined the pole parameters from the average of the direct and extended-dispersive output of the $F^{0+}$ amplitude in the [1.53,~1.73]~GeV segment (gray area in Fig.~\ref{fig:F0pextended}). As an additional uncertainty, we have considered half the difference between the pole parameters obtained using either one for the analytic continuation. As we will see, our results are completely compatible with the partial-wave poles, but with a larger and less model-dependent error, since it takes into account the fact that the fit with a simple parametrization does not satisfy the dispersive representation.

\begin{table}
\renewcommand{\arraystretch}{1.6}
	      \begin{tabular}{cccc} \hline\hline
		  FDR$_{C_N}$ &$\sqrt{s_{\rho_3(1690)}}$ (MeV)& $|g_{\pi\pi}|$ (GeV$^{-2}$)& $\phi$ ($^\circ$)\\ \hline
		GFI  & $\left(1700^{+21}_{-12}\right)$$\,-\,$$i\left(126^{+7}_{-8}\right)$  & \ $1.46^{+0.17}_{-0.18}$ & $6^{+22}_{-15}$ \\
        GFII  &$\left(1704\pm11\right)$$\,-\,$$i\left(135\pm7\right)$  &   $1.31^{+0.19}_{-0.15}$ &$0^{+12}_{-13}$\\ 
        GFIII  &$\left(1705^{+12}_{-10}\right)$$\,-\,$$i\left(122\pm9\right)$  &   $1.21^{+0.17}_{-0.16}$&$2^{+9}_{-8}$\\ 
         $\mathbf{\overline{\textbf{GF}}}$      & $\mathbf{\left(1704^{+17}_{-16}\right)-\textbf{\textit{i}}\left(129^{+13}_{-16}\right)}$  &   $\mathbf{1.3\pm0.3}$ & $\mathbf{2^{+26}_{-15}}$\\
          \hline
      \hline
	      \end{tabular}
\caption{Dispersive determination of the $\rho_3(1690)$ pole parameters using as input in the dispersive integrals each one of the three dispersively constrained 
Global Fits (GF in the table) from~\cite{Pelaez:2024uav}. Since they are compatible, we combine them into a single $\overline{GF}$ estimate in boldface. }
\label{tab:rho3}
\end{table}

In Table~\ref{tab:rho3}, we present our dispersive pole parameters for the $\rho_3(1690)$, which are compatible within uncertainties for the three Global Fits. For this reason, we provide an estimate $\overline{\text{GF}}$, whose uncertainty covers them all. 

\begin{figure*}
\centering
\includegraphics[width=\textwidth]{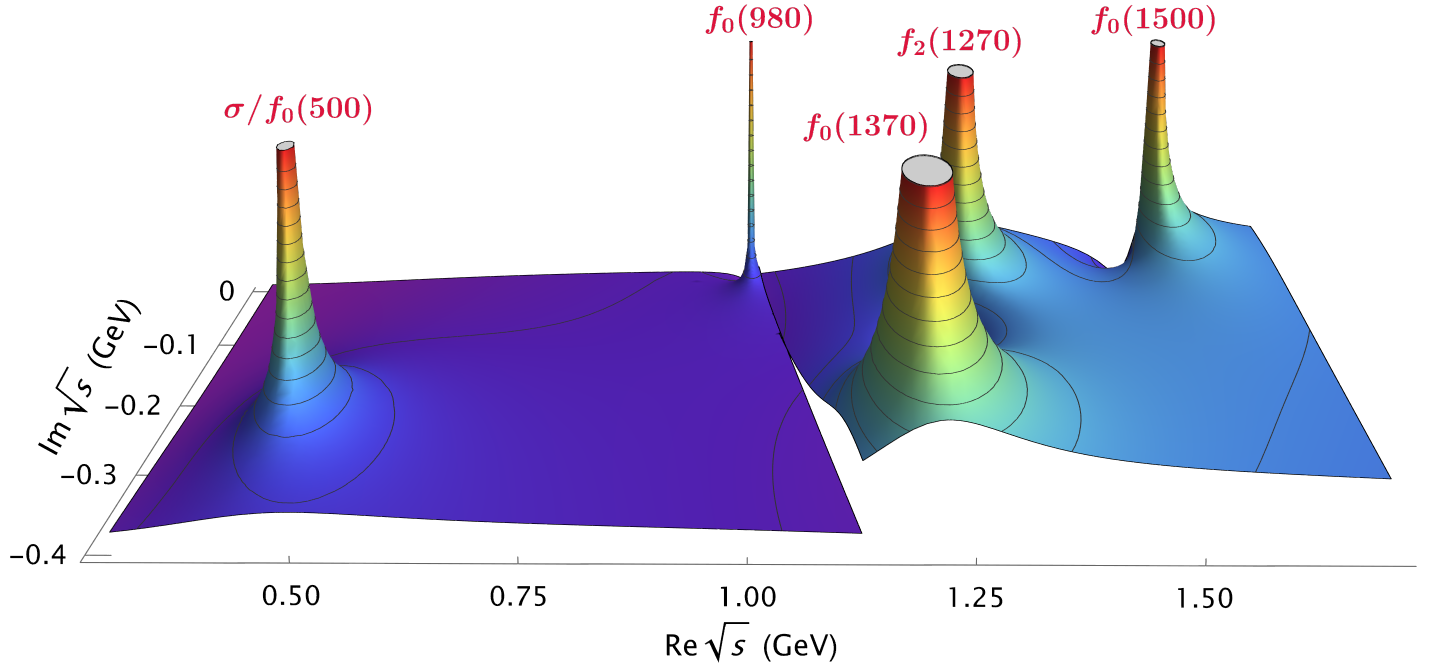}
\caption{ \small \label{fig:f00-rep} 
Analytic continuation of the $F^{00}$ FDR output, for the central values of the Global Fit~I input, and a representative $N$ value. The amplitude is continued from the intervals [0.3,~0.95]~GeV and [1.2,~1.35]~GeV in the elastic and inelastic regions, respectively.
Each interval gives access to different contiguous sheets, which is illustrated with a discontinuity attached to the $K \bar K$ threshold at $\sim$0.991 GeV.
Note the presence of the $f_0(500)$, $f_0(980)$, $f_2(1270)$, $f_0(1370)$, and $f_0(1500)$ resonance poles.}
\end{figure*}

Note that the mass we provide is about 10 MeV lighter than the parametrization-dependent ones obtained directly from each Global Fit or the Grayer et al.~\cite{Grayer:1974cr} original phase-shift analysis. As a consequence, it is closer and indeed compatible with the RPP mass estimate. In contrast, our width $\sim$$258^{+32}_{-26}\,$MeV is definitely larger than, and incompatible with, the RPP estimate. Nevertheless, it is compatible with some of the most recent results quoted in the RPP and, of course, with the Grayer et al. phase-shift analysis~\cite{Grayer:1974cr}. Our uncertainties, both for the mass and width, are larger. However, we consider them more realistic because they take into account the existence of three incompatible datasets and an uncertainty due to parametrization dependence, whereas the dispersive representation is not fulfilled by simple fits to data.
Let us nevertheless recall that ours is a $T$-matrix pole, which often differs by tens of MeV from Breit-Wigner-like masses and widths.

For this resonance, we cannot provide a comparison between the $F^{0+}$
dispersive output in the upper half plane as we did in Figs.~\ref{fig:rhoplane} and~\ref{fig:rho1450plane}, because this time we are not using a purely dispersive output, but its average with the direct calculation from the parametrizations. In addition, our uncertainty has been enlarged to cover both calculations.

Finally, one might wonder what has happened to the other resonance pole that combined with the $\rho_3(1690)$ into a single pole when using the [1.35,~1.55]~GeV segment.
Actually, when we use the [1.53,~1.73]~GeV segment, we also find another pole for the Global Fit~I at $(1835^{+42}_{-38})-i(55\pm35)$ MeV. 
Within uncertainties, it lies in the straight line that connects the poles of the $\rho(1700)$ and $\rho(1900)$
that we find in the Global Fit~I P-wave parametrization, to be discussed in Sec.~\ref{sec:polesfromparams}.
We thus believe it is a superposition of the $\rho(1700)$ and $\rho(1900)$, which cannot be disentangled from the segment [1.53,~1.73]~GeV. We will also see that the $\rho(1700)$ pole that we obtain from the analytic continuation of the Global Fit~I parametrization of the P wave appears around 1780 MeV, well separated from our $\rho_3(1690)$.
No additional pole is obtained from the $F^{0+}$ FDR output of the other two Global Fits. As we will also see below, this could be expected, since the analytic continuation of the P wave for Global Fit~III does not even contain the $\rho(1700)$ and $\rho(1900)$ poles. The P-wave for Global Fit~II, when continued analytically, shows a pole near 1780 MeV. However, it is extremely wide $\sim$400~MeV; therefore, it is reasonable that its effect in the [1.53,~1.73]~GeV segment is not enough to make it seen from the continuation of the FDR output, averaged with the Global Fit~II.
Thus, we conclude that our dispersive results for the $\rho_3(1690)$ are robust and not contaminated by more massive or wider poles unresolved within the precision by the FDR$_{C_N}$ method.

\subsection{Resonances in $F^{00}$}

In this section, we will study the resonances that can be obtained through the analytic continuation of the dispersive output of the $F^{00}$ FDR. 
In Fig.~\ref{fig:f00-rep}, we show such an analytic continuation and the poles that can be identified in the contiguous Riemann sheets for a representative choice of parameters. According to 
Eq.~\eqref{eq:Fsym}, and given that no resonances exist in the $I=2$ channel below 2 GeV, all the resonances obtained are thus isoscalars. 

Part of this program was already carried out in~\cite{Pelaez:2022qby}, but only above 1 GeV: the pole parameters of the $f_2(1270)$, $f_0(1370)$, and $f_0(1500)$ resonances were determined with the same method we are employing here. In this work, we confirm these findings, with minor variations in the resonance parameters, due to the use of our recent dispersive analysis~\cite{Pelaez:2024uav}, where the three Global Fits were revisited, refined, extended in energies, and included partial waves with $\ell\geq2$. In more detail, there were no significant differences in the scalar-isoscalar partial wave, and the matching point with the high-energy regime remained at 1.42 GeV. Nevertheless, there was a relevant update in the D0 wave from~\cite{Pelaez:2024uav}, especially for Global Fit~I, where the inelasticity onset was set at 0.9 GeV instead of at $K\bar K$ threshold used in~\cite{GarciaMartin:2011cn} and for Global Fits~II and III. The G0 and $I=2$ partial waves were refined too, and have slightly smaller uncertainties.

As a novelty in this work, we also obtain the isoscalar resonances below 1 GeV from the FDR$_{C_N}$ approach. 
These are the $f_0(500)$ and $f_0(980)$ resonances, also seen in Fig.~\ref{fig:f00-rep}. 

\begin{figure}
\centering
\includegraphics[width=0.49\textwidth]{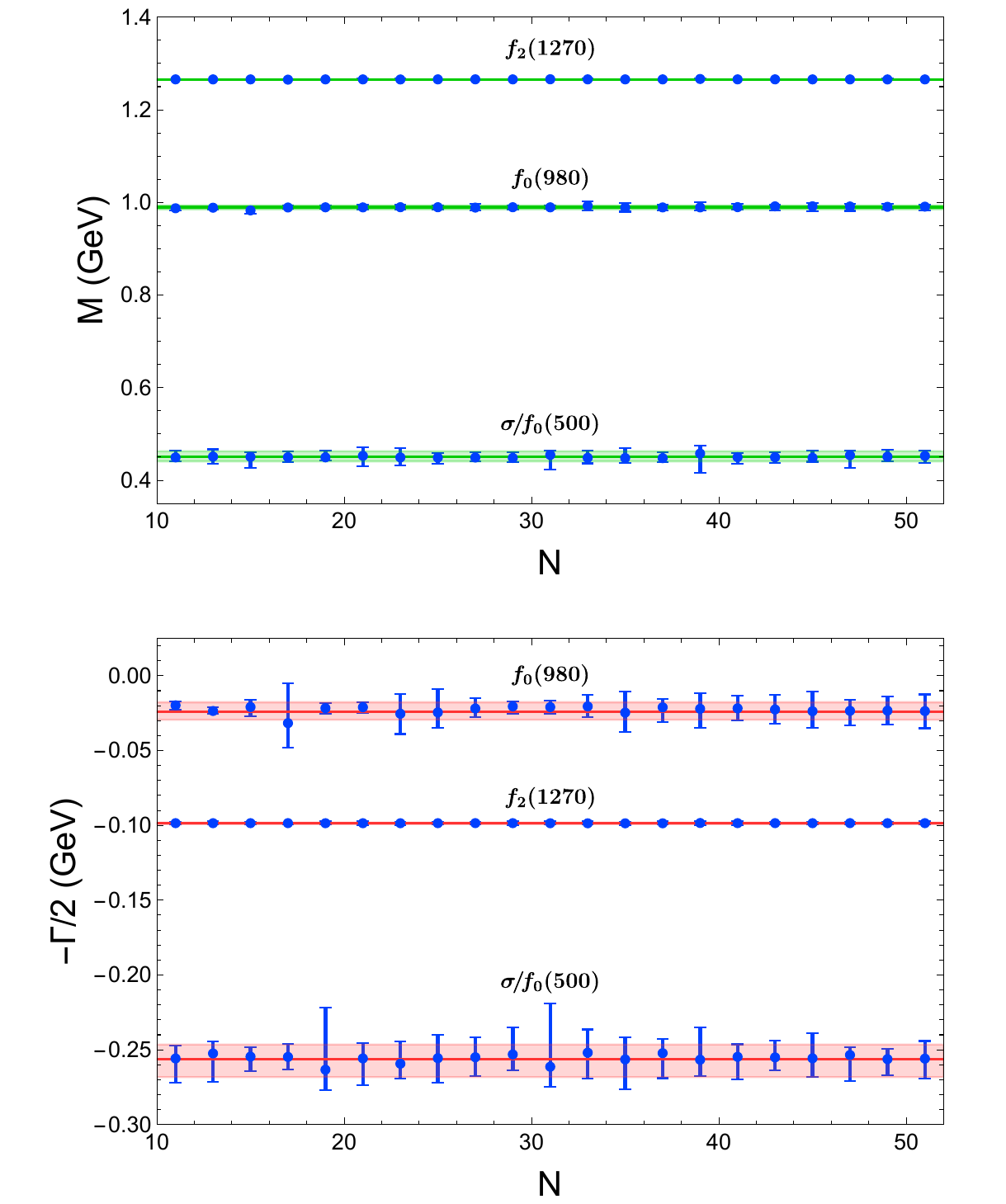}
\caption{ \small \label{fig:resonances-f00-1} 
Pole masses (top) and half-widths (bottom) for the $f_0(500)$, $f_0(980)$, and $f_2(1270)$ resonances. They are obtained from analytic continuations with continued fractions of the output of the $F^{00}$ dispersion relation. The results are stable when varying the number of interpolated points $N$.}
\end{figure}

\begin{figure}
\centering
\includegraphics[width=0.49\textwidth]{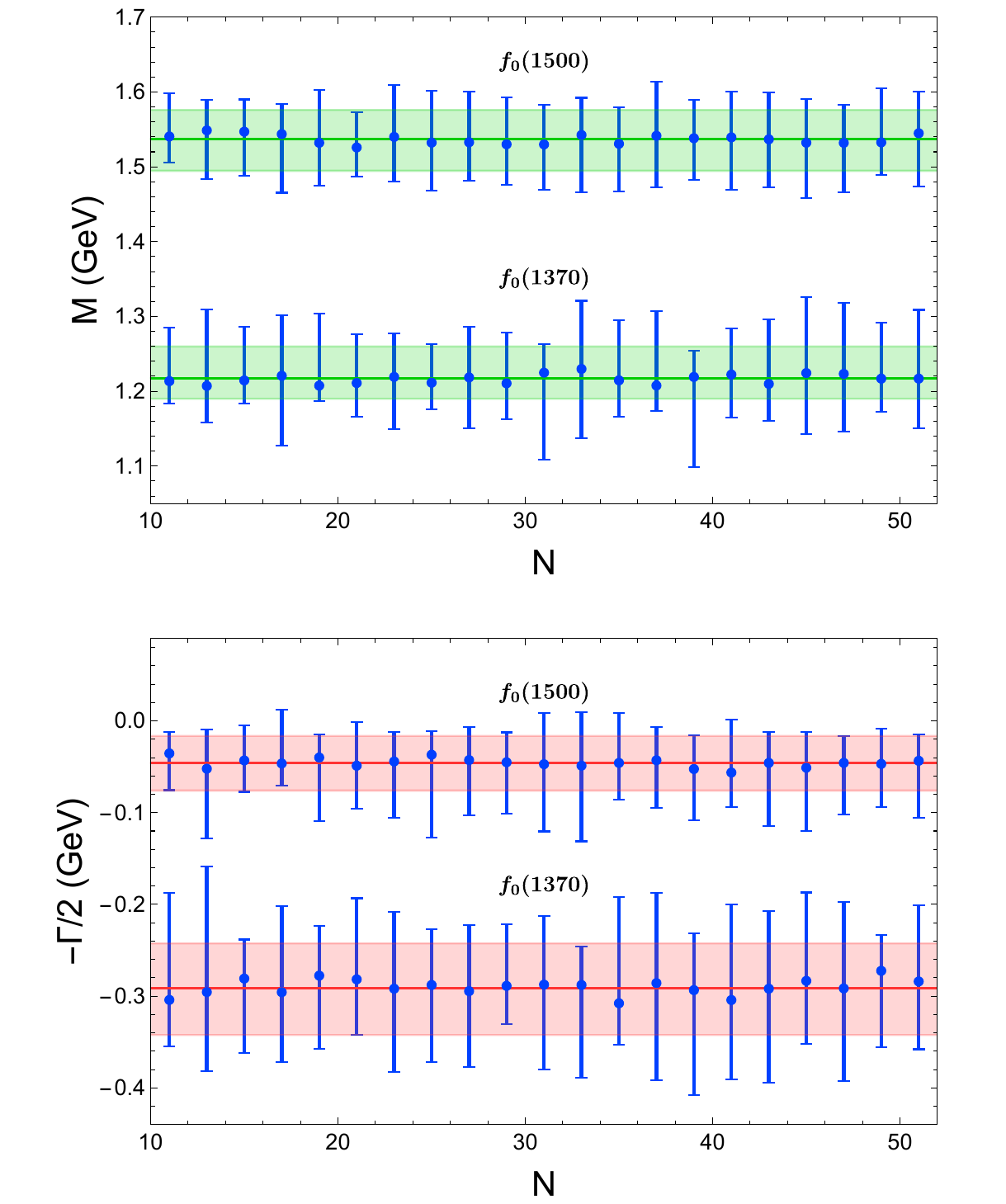}
\caption{ \small \label{fig:resonances-f00-2} 
Pole masses (top) and half-widths (bottom) for the $f_0(1370)$ and $f_0(1500)$ resonances. They are obtained from analytic continuations with continued fractions of the output of the $F^{00}$ dispersion relation. The results are stable when varying the number of interpolated points $N$.}
\end{figure}

As in the case of the $F^{0+}$ FDR, the analytic continuation is made with continued fractions with $N$ as large as 51, and also varying the ends of the intervals that are analytically continued from the real axis. In addition, we vary all parameters of the Global Fit and Regge parametrizations within uncertainties. We collect tens of thousands of samples once again and identify as resonances only those poles that are stable under the variation of all these parameters, rejecting numerical artifacts that appear in just a few percent of the cases.
In Figs.~\ref{fig:resonances-f00-1} and~\ref{fig:resonances-f00-2}, we show the stability of these poles under changes in $N$, which effectively removes the parametrization dependence of our results.

Finally, concerning $J=2$ resonances, we do not find poles for the well-established $f'_2(1525)$ nor the $f_2(1565)$. This was to be expected since, according to the RPP, the $f'_2(1525)$ branching ratio to $\pi\pi$ is $\sim$$10^{-2}$.
Similarly, the $f_2(1565)$ is mostly seen in antinucleon-nucleon annihilation, and its $\pi\pi$ branching ratio is not even determined.
In addition, in the RPP particle listings, there are two other resonances, which are not well established and are omitted from the summary tables. One of them is the $f_2(1430)$, which needs confirmation and whose $\pi\pi$ branching ratio is not determined. The second one is the $f_2(1640)$, which has been seen to decay to $K\bar K$, $4\pi$, and $\omega\omega$, but not to $\pi\pi$. The fact that we do not find them implies that their coupling to two pions, if they exist, must be very small.

Let us discuss separately each isoscalar resonance pole identified in our dispersive analysis of $\pi\pi$ scattering data.

\subsubsection{$f_0(500)/\sigma$}
\label{sssec:sigma}

This resonance was the subject of a decades-long controversy (see~\cite{Pelaez:2015qba} for a dedicated review containing a historical account).
It is extremely wide and its pole lies very close to the $\pi\pi$ threshold. 
Its extraction is further complicated by the nearby Adler zero---due to the spontaneous chiral-symmetry breaking first discussed in~\cite{Adler:1964um}---and by the presence of the $\pi\pi$ left-hand cut required by crossing symmetry.
It was therefore crucial to have all the analytic structure well implemented before determining its parameters. The controversy about its existence and parameters was settled in the RPP when its $T$-matrix pole was established
in a model-independent way using partial-wave dispersion relations~\cite{Caprini:2005zr,GarciaMartin:2011jx,Moussallam:2011zg}.
Although a BW description is completely inadequate for its description~\cite{Meissner:2003pd,RuizdeElvira:2010cs,Ledwig:2014cla,Pelaez:2015qba,Pelaez:2025wma} (see also the Note on ``Scalar mesons below 1 GeV" in the RPP), it is unfortunately still being used. Furthermore, its BW parameters are still listed in the RPP and are inconsistent with the rigorous $T$-matrix pole description, which is the one we will use here. In the RPP, it is estimated to be $(400-550)-i(200-350)\,$MeV. Nevertheless, these caveats are correctly explained in the RPP review on ``Scalar mesons below 1 GeV" pointing to the ``conservative dispersive estimate" $(449^{+22}_{-16})-i(275\pm12)$~MeV provided in~\cite{Pelaez:2015qba} (the uncertainty in the width was corrected to $\pm15$ in~\cite{Pelaez:2021dak}).

Thus, we list in Table~\ref{tab:f0500} the parameters for the pole associated with the $f_0(500)/\sigma$. Note that the fact that this resonance lies within the elastic regime of $\pi\pi$ scattering is what allowed the use of Roy and GKPY partial-wave dispersion relations to determine its associated pole parameters~\cite{Caprini:2005zr,GarciaMartin:2011jx,Moussallam:2011zg}.
As we did with the $\rho(770)$, in the first line of the table, we present the value obtained in~\cite{GarciaMartin:2011jx} using their CFD as input for the GKPY equations. 
Next, we show the results of employing our FDR$_{C_N}$ method to continue the dispersive output in the [0.37,~0.57]~GeV segment. We provide values when using each of the three Global Fits, together with their GKPY results. All the different determinations are compatible within errors. Hence, for each method, we have combined the results of the Global Fits into a single value $\overline{\text{GF}}$ that encompasses all three.
Note that the results obtained through continued fractions present slightly smaller errors for the mass and a somewhat narrower resonance.
\begin{table}
\footnotesize
\resizebox{.48\textwidth}{!}{
\renewcommand{\arraystretch}{1.8}
	      \begin{tabular}{ccccc} \hline\hline
		 Input & Method &  $\sqrt{s_{f_0(500)}}$ (MeV)& $|g_{\pi\pi}|$ (GeV)& $\phi$ ($^\circ$)\\ 
         \hline
        CFD~\cite{GarciaMartin:2011jx} & GKPY& $\left(457^{+14}_{-13}\right)$$\,-\,$$i\left( 279^{+11}_{-7}\right)$ & $3.59^{+0.11}_{-0.13}$& \\
         \hline
         \hline
		\multirow{ 2}{*}{GFI}  & FDR$_{C_N}$ & $\left(451^{+12}_{-10}\right)$$\,-\,$$i\left(256^{+12}_{-10}\right)$  & $3.23\pm0.23$& $-72\pm7$\\
         &GKPY& $\left(449 \pm 16\right)$$\,-\,$$i\left( 270^{+9}_{-11}\right)$ & $3.47^{+0.13}_{-0.14}$&$-74\pm 3$\\\hline
        \multirow{ 2}{*}{GFII}  & FDR$_{C_N}$ &$\left(449^{+7}_{-9}\right)$$\,-\,$$i\left(274\pm7\right)$ & $3.37^{+0.14}_{-0.22}$&$-77^{+4}_{-3}$\\
         &GKPY&$\left(453^{+18}_{-17}\right)$$\,-\,$$i\left( 281^{+9}_{-11}\right)$& $3.56^{+0.16}_{-0.17}$&$-75\pm3$\\\hline
        \multirow{ 2}{*}{GFIII}  & FDR$_{C_N}$ &$\left(454^{+11}_{-7}\right)$$\,-\,$$i\left(270^{+6}_{-7}\right)$ & $3.34^{+0.16}_{-0.20}$&$-77\pm6$\\
         &GKPY&$\left(451^{+18}_{-17}\right)$$\,-\,$$i\left( 279^{+9}_{-11}\right)$& $3.56\pm0.16$&$-74\pm3$\\\hline
           \multirow{ 2}{*}{$\mathbf{\overline{\textbf{GF}}}$}        &   \textbf{FDR$\mathbf{_{\textbf{\textit{C}}_\textbf{\textit{N}}}}$} & $\mathbf{\left(451^{+14}_{-11}\right)-\textbf{\textit{i}}\left(269^{+12}_{-23}\right)}$  &   $\mathbf{3.33^{+0.18}_{-0.33}}$ & $\mathbf{-76^{+11}_{-7}}$ \\
              &   \textbf{GKPY} & $\mathbf{\left(451^{+20}_{-18}\right)-\textbf{\textit{i}}\left(277^{+13}_{-18}\right)}$  &   $\mathbf{3.52^{+0.20}_{-0.19}}$ & $\mathbf{-74^{+3}_{-4}}$\\
              \hline 
      \hline
	      \end{tabular}}
\caption{Dispersive determinations of $f_0(500)$ pole parameters using as input in the dispersive integrals each of the three dispersively constrained 
Global Fits (GF in the table) from~\cite{Pelaez:2024uav}. We present results for our FDR$_{C_N}$ method as well as from the GKPY equations. Since the results for each method are consistent across the three Global Fits, we provide a single range in boldface for each method, covering the results of the three Global Fits. 
The results are compatible with the values obtained using the GKPY equations with the CFD in~\cite{GarciaMartin:2011jx}, but with somewhat larger uncertainties.}
\label{tab:f0500}
\end{table}

As it happened with the $\rho(770)$, we see with this resonance that the FDR$_{C_N}$ method yields competitive or even slightly more precise results than GKPY equations. Still, employing the Global Fits as input yields somewhat larger uncertainties than using the CFD, since the latter focused on precision in the elastic region, rather than covering a large energy range.

\begin{figure}
\centering
\includegraphics[width=0.48\textwidth]{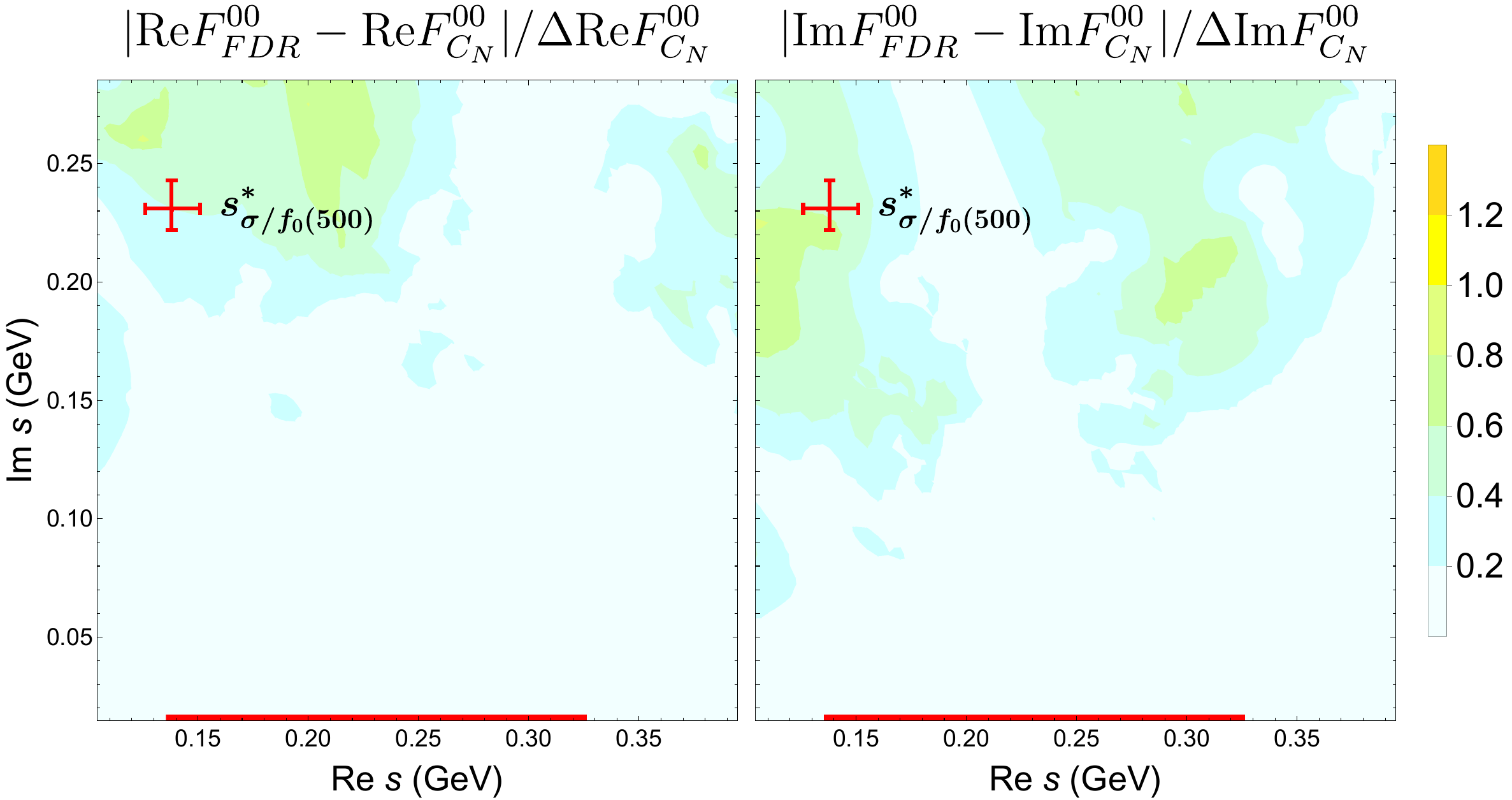}
\caption{ \small \label{fig:f0500plane_s} Reliability of the continued fraction analytic continuation method in the upper half $s$ plane near the $f_0(500)$ region.
The first Riemann sheet value of the $F^{00}$ FDR output, $F^{00}_{FDR}$,
lies well within the estimated uncertainty of the continuation made with continued fractions $F^{00}_{C_N}$. We show the absolute value of the difference for the real parts (left) and for the imaginary parts (right), divided by the uncertainty of the $C_N$ calculation.
For reference, we show the conjugate position of the $f_0(500)$ pole that appears in the second sheet (red cross) and the continued segment (red line).
} 
\end{figure}

Finally, Fig.~\ref{fig:f0500plane_s} presents a plot similar to those in Fig.~\ref{fig:rhoplane} and~\ref{fig:rho1450plane}, showing that the difference between the 
analytic continuation of the $F^{00}$ FDR integral and the continued fraction $C_N$---constructed from the real segment [0.37,~0.57]~GeV and a representative choice of $N$---lies within the estimated full uncertainties of our method. Note that, in this plot, the axes correspond to the real and imaginary parts of $s$ instead of $\sqrt{s}$ as in the other figures. We have displayed most figures in $\sqrt{s}$ because it is the natural variable for quoting resonance masses and widths. Still, the analytic continuation itself is always performed in the variable $s$, for which the analytic properties of the amplitudes are formulated.
For most resonances, their pole width $\Gamma_p$ is much smaller than their pole mass $M_p$, and $\Re\, s_p=M_p^2-\Gamma_p^2/4\simeq M_p^2$ is a good approximation. Hence, choosing an $s$ segment centered around $\Re\,s_p\sim M_p^2$ translates into an $\sqrt{s}$ segment fairly well centered around $\Re\,\sqrt{s_p}\sim M_p$. That is why our $\sqrt s$ segments worked well for most poles.
The $\sig$ is a notable exception, since its pole mass and width are comparable. Thus, Re$\,s_p=M_p^2-\Gamma_p^2/4<  M_p^2$, and the naive expectation $\Re\, s_p\sim M_p^2\sim0.20\,$GeV$^2$ largely overestimates the true value $\Re\, s_p=M_p^2-\Gamma_p^2/4\sim0.13\,$GeV$^2$. 
Consequently, although $M_\sigma$ may look reasonably well centered with respect to our $\sqrt{s}$-segment, $\text{Re}s_\sigma$ lies slightly outside of the $s$-segment. 
Moreover, in this case, we cannot take a lighter segment in the real axis, because we will then get too close to the $\pi\pi$ threshold at $\sqrt{s_{\pi\pi}}\sim 0.28$ GeV. 
It is therefore very relevant that---even when the real part of the $f_0(500)$ pole lies outside the chosen segment, yet not too far from it---we can still determine its pole properties. This is because it is well isolated from other poles, and the continued segment avoids any significant cut singularity. We will see a relatively similar situation with the $f_0(980)$ resonance next, and the $f_0(1500)$ a little later.

\subsubsection{$f_0(980)$}
\label{sssec:f0980}
The $f_0(980)$ resonance is well determined in the RPP, which even provides a $T$-matrix pole estimate at  $(980-1010)-i(20-35)\,$MeV.
As illustrated in Fig.~\ref{fig:f00-rep}, it is a narrow resonance, whose pole lies near the real axis and very close to the $K\bar K$ threshold. The definition of this threshold already presents a subtlety, because the charged $K^+K^-$ threshold is $\sim$987.4~MeV, whereas the neutral threshold lies at $\sim$995.2~MeV.
In the isospin limit, a single cut is considered, located in the average position.

In the dispersive determinations with GKPY equations of~\cite{GarciaMartin:2011jx}, its pole appears in the second Riemann sheet, which is contiguous to the physical sheet when crossing the real axis from {\it below} the $K\bar K$ threshold. This occurs even though the real part of the pole position, 996$\pm$7 MeV, is often nominally larger than two-kaon masses. For these reasons, the $\Im \sqrt{s_{f_0(980)}}$ cannot be interpreted as the total width~\cite{Wang:2022vga,Burkert:2022bqo}. Instead, it approximately corresponds to the difference between the partial width to two pions and the partial width to two kaons~\cite{Burkert:2022bqo}. Consequently, it is seen as a very narrow structure in $\pi\pi$ scattering below the $K\bar K$ threshold.

As the pole lies on the sheet reached by analytic continuation from the elastic region (the second Riemann sheet), it can be computed using the GKPY equations, allowing a direct comparison with the values obtained from our continued-fraction method.
Thus, in Table~\ref{tab:f0980}, as we did with the $\rho(770)$ and $\sig$, we first provide the value obtained in~\cite{GarciaMartin:2011jx} employing GKPY equations with their CFD input.
Next, we show the results using our three Global Fits~in~\cite{Pelaez:2024uav} as input for both the FDR$_{C_N}$ method and GKPY equations. For the former, we have continued the segment [0.77,~0.97]~GeV, slightly below the $K\bar K$ threshold.

All our pole positions are compatible with the RPP $T$-matrix pole estimate. For each Global Fit~input, the GKPY pole masses come slightly heavier and present larger errors than the corresponding results employing the FDR$_{C_N}$ approach. Note also that the FDR$_{C_N}$ results for the different Global Fits are remarkably compatible among themselves---within less than a deviation---and the same happens for the GKPY pole parameters among themselves. 
As usual, in the last rows of the table, we provide a single value $\overline{\text{GF}}$ for each method, covering the results from the three Global Fits.

However, in this case, we have enlarged the final uncertainty
to take into account the existence of the two $K\bar K$ thresholds,
separated by roughly 8 MeV.\footnote{Note that the pion-mass difference effect has been recently proved to be irrelevant for the $f_0(980)$ resonance~\cite{Colangelo:2025iuq}.} The effect that this produces on dispersive pole determinations was studied in~\cite{GarciaMartin:2011jx}, where their unconstrained and constrained fits were refitted to the extreme cases of using
$m_{K^0}$ or $m_{K^+}$ instead of its average in the isospin limit. A significant variation due to this isospin-violating effect was only observed for the $f_0(980)$ half-width, which changes by $\Delta \!\!\!\!\not\! \text{I}=\pm$4.4 MeV for the GKPY equations. We have added such an uncertainty in quadrature to our estimate uncertainties as done in~\cite{GarciaMartin:2011jx}. As a result, our estimated uncertainty is roughly similar to the one found there.
\begin{table}[h]
\footnotesize
\resizebox{.48\textwidth}{!}{
\renewcommand{\arraystretch}{1.8}
	      \begin{tabular}{ccccc} \hline\hline
		Input & Method &  $\sqrt{s_{f_0(980)}}$ (MeV)& $|g_{\pi\pi}|$ (GeV)& $\phi$ ($^\circ$)\\ \hline
         CFD~\cite{GarciaMartin:2011jx} & GKPY& $\left(996\pm7\right)$$\,-\,$$i\left( 25^{+10}_{-6}\right)$ & $2.3\pm0.2$& \\
         \hline
         \hline
		\multirow{ 2}{*}{GFI}  & FDR$_{C_N}$ & $\left(990^{+6}_{-5}\right)$$\,-\,$$i\left(24^{+5}_{-6}\right)$  & $1.6\pm0.3$&$-79\pm13$\\
         &GKPY& $\left(999 \pm 7\right)$$\,-\,$$i\left( 24^{+4}_{-5}\right)$ & $2.31^{+0.21}_{-0.20}$ & $-74^{+10}_{-11}$\\\hline
        \multirow{ 2}{*}{GFII}  & FDR$_{C_N}$ &$\left(991\pm6\right)$$\,-\,$$i\left(27^{+7}_{-6}\right)$ & $1.7\pm0.2$&$-82\pm9$\\
         &GKPY&$\left(997 \pm 7\right)$$\,-\,$$i\left( 25^{+4}_{-5}\right)$& $2.34\pm0.18$ &$-74^{+10}_{-11}$\\\hline
        \multirow{ 2}{*}{GFIII}  & FDR$_{C_N}$ &$\left(992\pm4\right)$$\,-\,$$i\left(27\pm6\right)$ & $1.7^{+0.2}_{-0.3}$&$-78\pm8$\\
         &GKPY&$\left(996 \pm 7\right)$$\,-\,$$i\left( 26^{+4}_{-3}\right)$& $2.28\pm0.14$ &$-76\pm 10$\\\hline
            \multirow{ 2}{*}{$\mathbf{\overline{\textbf{GF}}}+ \mathbf{\Delta \!\!\!\not\! I}$}        &   \textbf{FDR$\mathbf{_{\textbf{\textit{C}}_\textbf{\textit{N}}}}$} & $\mathbf{\left(991^{+7}_{-8}\right)-\textbf{\textit{i}}\left(26\pm9\right)}$  &   $\mathbf{1.7^{+0.2}_{-0.4}}$ & $\mathbf{-80^{+14}_{-12}}$ \\
              &   \textbf{GKPY} & $\mathbf{\left(997^{+10}_{-9}\right)-\textbf{\textit{i}}\left(25^{+7}_{-8}\right)}$  &   $\mathbf{2.3\pm0.2}$ & $\mathbf{-74\pm11}$ \\\hline
      \hline
	      \end{tabular}}
\caption{Dispersive determinations of the $f_0(980)$ pole parameters using as input in the dispersive integrals each of the three dispersively constrained 
Global Fits (GF in the table) from~\cite{Pelaez:2024uav}. We present results for our FDR$_{C_N}$ method as well as from the GKPY equations. Since the results for each method are consistent across the three Global Fits, we provide a single range in boldface for each method, covering the results of the three Global Fits. The results are compatible with the values obtained using the GKPY equations with the CFD in~\cite{GarciaMartin:2011jx}.
Note that in our final result, we have included in the uncertainty an estimate of $\Delta \!\!\!\not\! \text{I}$ of the isospin-breaking effect on the $K\bar K$ threshold, as done in~\cite{GarciaMartin:2011jx} and explained in the main text. 
}
\label{tab:f0980}
\end{table}

\begin{figure}
\centering
\includegraphics[width=0.48\textwidth]{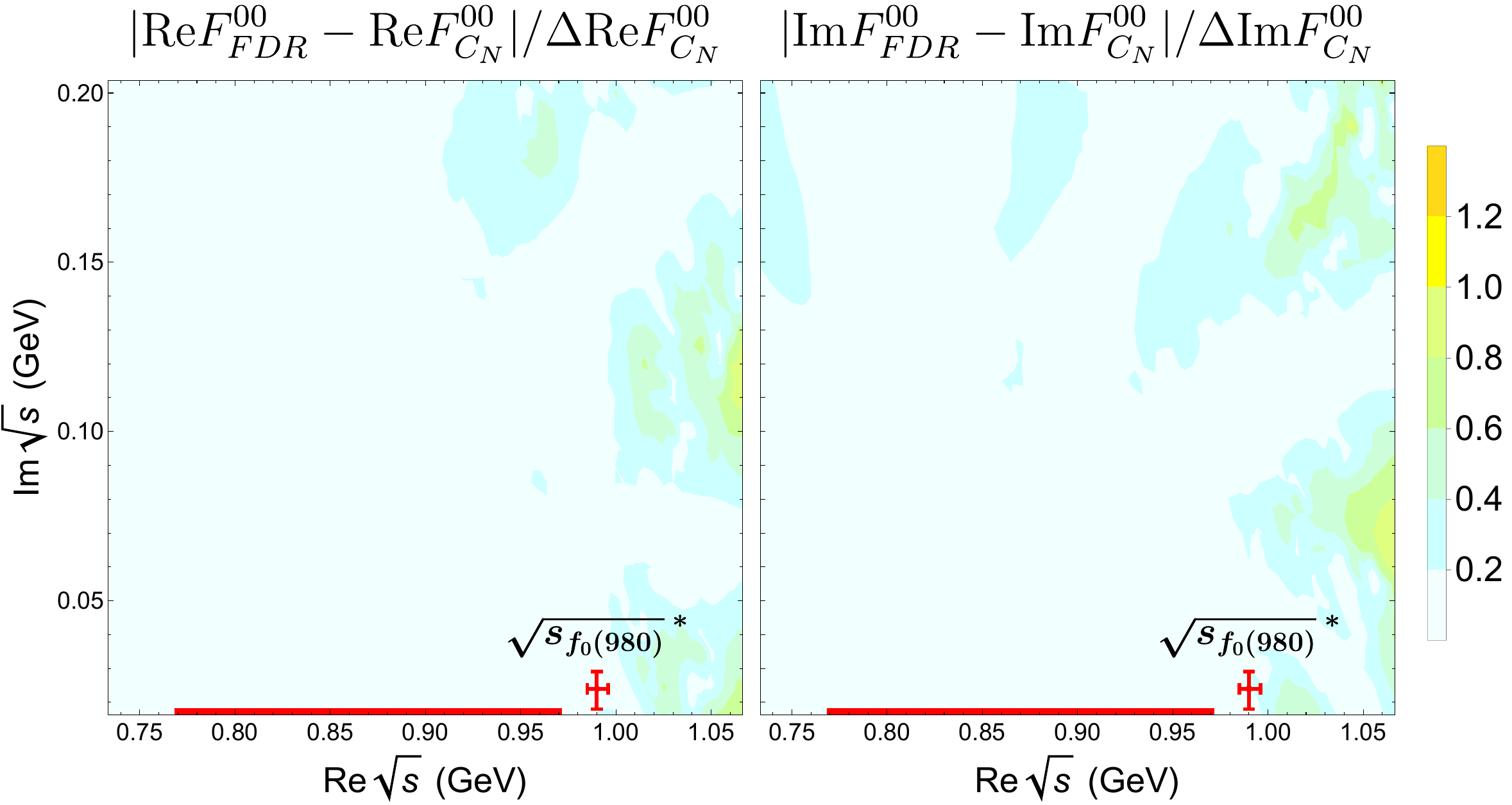}
\caption{ \small \label{fig:f0980plane} Reliability of the continued fraction analytic continuation method in the upper half $\sqrt s$-plane near the $f_0(980)$ region. The first Riemann sheet value of the $F^{00}$ FDR output, $F^{00}_{FDR}$, lies well within the estimated uncertainty of the continuation made with continued fractions $F^{00}_{C_N}$. We show the absolute value of the difference for the real parts (left) and for the imaginary parts (right), divided by the uncertainty of the $C_N$ calculation. For reference, we show the conjugate position of the $f_0(980)$ pole that appears in the second sheet (red cross) and the continued segment (red line).
} 
\end{figure}

Once more, we provide in Fig.~\ref{fig:f0980plane} a plot similar to those in Figs.~\ref{fig:rhoplane},~\ref{fig:rho1450plane}, and~\ref{fig:f0500plane_s} to assess the reliability of the analytic continuation by the continued fractions $C_N$. 
It shows that the difference between the analytic continuation using the $F^{00}$ FDR integral and the $C_N$ analytic continuation---using the real segment [0.77,~0.97]~GeV (in red) and a representative choice of $N$---falls within the estimated full uncertainties of our method.
Note that the segment is not centered around the pole, but lies below the opening of the $K\bar K$ threshold, to ensure that the analytic continuation is to the second Riemann sheet.

\subsubsection{$f_2(1270)$}
\label{sssec:f2}

This isoscalar-tensor resonance is very well established. The RPP provides remarkably precise mass and width averages:
$M_{f_2(1270)}=1275.4\pm0.8$ MeV and $\Gamma_{f_2(1270)}=185.8^{+2.8}_{-2.1}\,$MeV, but also a $T$-matrix pole estimate, $\sqrt{s_{f_2(1270)}}=(1260-1283)-i(90-110)\,$MeV.  Recall that the peak and pole masses are only approximately the same. 
The branching ratio to two pions is $\sim$$84\%$ and its Breit-Wigner-like shape dominates the D0 wave in the 1 to 1.4 GeV region~\cite{ParticleDataGroup:2024cfk}. 

Note that since this resonance lies in the inelastic region, it is far from the applicability range of the GKPY equations. However, in~\cite{Pelaez:2022qby}, it was already shown that this resonance could be obtained very precisely with the FDR$_{C_N}$ method. This was to be expected since the Global Fits~in~\cite{Pelaez:2024uav} were built to have a pole to reproduce the ``peak mass".
It is crucial to have this resonance well described to determine the existence and parameters of the other two nearby poles that appear in this region. These correspond to the $f_0(1370)$ and $f_0(1500)$ and will be discussed later on.

Here, we will briefly revisit the results from the FDR$_{C_N}$ method when using the Global Fits updated in~\cite{Pelaez:2024uav}. 
For this we use our FDR$_{C_N}$ method to continue analytically the output of the $F^{00}$ FDR from the 
[1.15,~1.35]~GeV segment. The results are listed in Table~\ref{tab:f2}.

In the new Global Fits, the peak position of the $f_2(1270)$ was not fixed to the central value of the RPP mass estimate (as done in~\cite{GarciaMartin:2011cn, Pelaez:2022qby}), but it was allowed to vary within the RPP uncertainty. Nonetheless, imposing the dispersive constraints barely changes its position, and the pole is very close to that found in~\cite{Pelaez:2022qby}. 

Note that the functional form of the Global Fits is more flexible than a Breit-Wigner parametrization, which allows for a description of the D0 wave $\pi\pi$ scattering data away from the resonance peak.
This observation is relevant because, although very similar to the eye, the D0 wave of Global Fit~I is incompatible with those of Global Fits~II and III by several standard deviations. 
This is due to the incompatible experimental datasets they fit, and it is only noticeable away from the peak, where the three fits are almost identical. 
In particular, Global Fit~I describes data from the 1973 Hyams et al. analysis~\cite{Hyams:1973zf}, which allowed the D0 inelasticity to open up around 0.9 GeV, well below the $ K {\bar K}$ threshold. This is where the inelasticity sets in for Global Fit~I. 
In contrast, Global Fits~II and III are elastic up to the $K\bar K$ threshold, since they fit two other datasets from the 1975 Hyams et al. analysis~\cite{Hyams:1975mc}, which are elastic in that region.
As a consequence, Global Fits~II and III fit a different inelasticity and prefer a heavier but narrower resonance, with a smaller coupling to two pions.  Note that the two D0-wave datasets in~\cite{Hyams:1975mc} are identical up to 1.4 GeV. This explains why Global Fits~II and III lead to almost identical results, which we have combined into a single estimate $\overline{\text{GF}}_{\text{II,III}}$.
In summary, the value in Table~\ref{tab:f2} obtained using Global Fit~I as input clearly differs from $\overline{\text{GF}}_{\text{II,III}}$.
Compared to the pole position found in~\cite{Pelaez:2022qby} at 1267.5-$i$94~MeV, we see that the effect of lowering the opening of the inelasticity produces a $\sim$4~MeV wider resonance. The small change in the pole mass, by contrast, is driven by allowing the mass to vary within the RPP uncertainties.

We consider our Global Fit~I result to be more reliable. First, because the $f_2(1270)$ branching ratio to $4\pi$ is twice as large as the $K\bar K$ one, which supports the opening of the inelasticity below $K\bar K$ threshold, as it is done in Hyams et al. 1973. Second, because the dispersive analysis~\cite{Pelaez:2024uav} favors the D0-wave inelasticity onset at 0.9 GeV instead of at the $K\bar K$ threshold. Still, all the pole positions from Table~\ref{tab:f2} lie inside the RPP $T$-matrix pole estimate. 
\begin{table}[h]
\footnotesize
\renewcommand{\arraystretch}{1.8}
	      \begin{tabular}{cccc} \hline\hline
		  FDR$_{C_N}$ & $\sqrt{s_{f_2(1270)}}$ (MeV)& $|g_{\pi\pi}|$ (GeV$^{-1}$)& $\phi$ ($^\circ$)\\ \hline
          \textbf{GFI} & $\mathbf{\left(1265.8^{+0.7}_{-0.8}\right)-\textbf{\textit{i}}\left(98.5^{+0.6}_{-0.5}\right)}$  &   $\mathbf{4.47\pm0.12}$ & $\mathbf{3.1^{+1.2}_{-1.5}}$\\ \hline
        GFII &$\left(1273.8^{+0.6}_{-1.0}\right)$$\,-\,$$i\left(91.6^{+0.6}_{-0.8}\right)$  &   $4.08^{+0.10}_{-0.07}$ &$8.2^{+2.0}_{-2.3}$\\ 
        GFIII  &$\left(1273.7^{+0.7}_{-0.8}\right)$$\,-\,$$i\left(91.3^{+0.9}_{-1.0}\right)$  &   $4.04^{+0.08}_{-0.09}$ & $7.6^{+0.9}_{-0.6}$\\ 
        $\mathbf{\overline{\textbf{GF}}_{\textbf{II,III}}}$      & $\mathbf{\left(1273.8^{+0.7}_{-0.9}\right)-\textbf{\textit{i}}\left(91.5^{+0.7}_{-1.2}\right)}$  &   $\mathbf{4.06^{+0.12}_{-0.11}}$ & $\mathbf{7.7^{+2.5}_{-1.8}}$\\
          \hline
      \hline
	      \end{tabular}
\caption{Dispersive determinations of the $f_2(1270)$ pole parameters using as input in the dispersive integrals each one of the three dispersively constrained 
Global Fits (GF in the table) from~\cite{Pelaez:2024uav}. Global Fits~II and III are compatible and thus we have combined them into a single $\overline{\text{GF}}_{\text{II,III}}$ estimate (shown in boldface) that covers both determinations. Since the $GFI$ determination is incompatible with the others, we do not combine them and instead report them separately in boldface.}
\label{tab:f2}
\end{table}

\begin{figure}[t]
\centering
\includegraphics[width=0.48\textwidth]{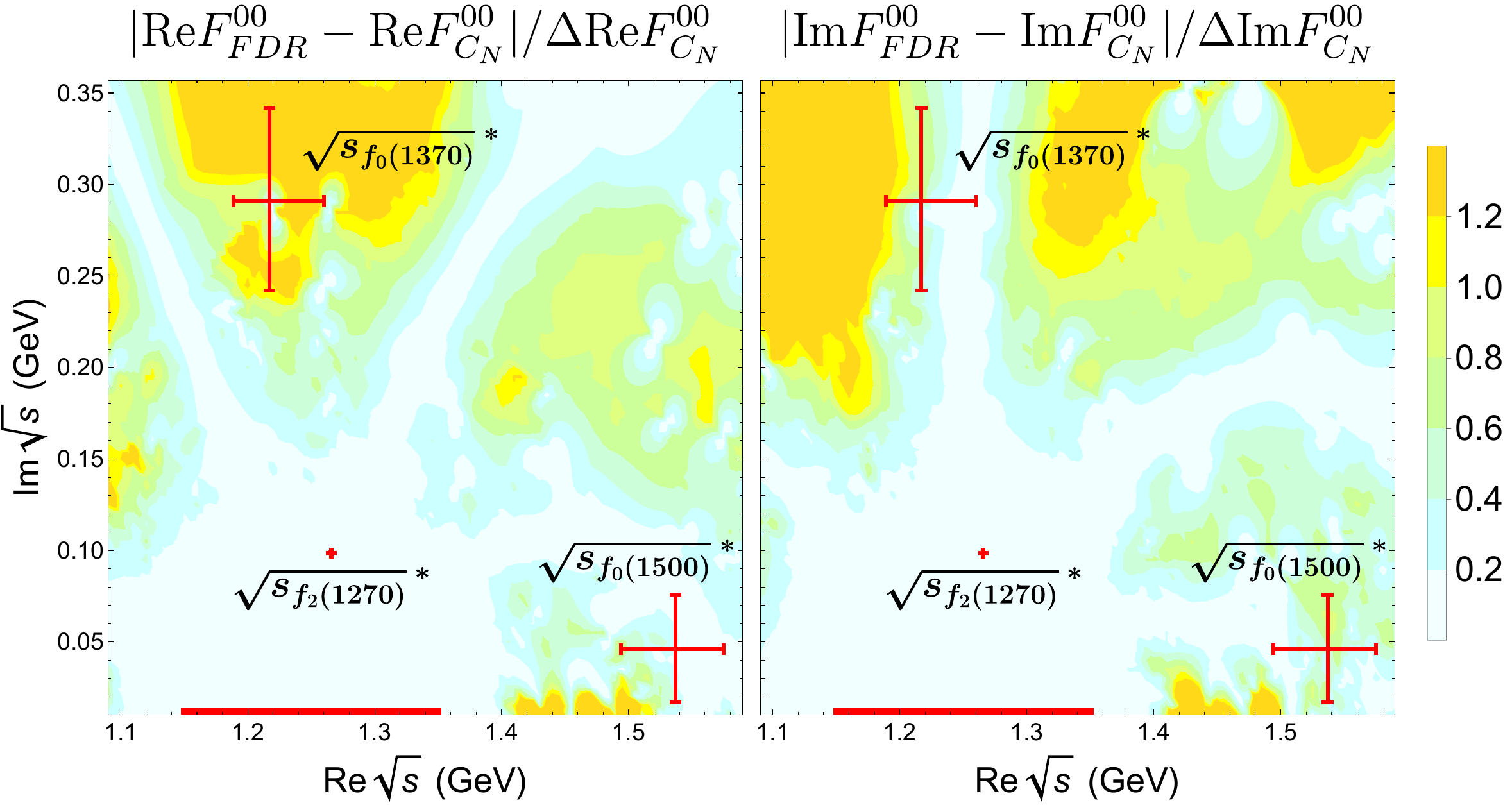}
\caption{ \small \label{fig:fsplane} Reliability of the continued fraction analytic continuation method in the upper half $\sqrt s$-plane near the $f_2(1270)$, $f_0(1370)$, and $f_0(1500)$ region.
The first Riemann sheet value of the $F^{00}$ FDR output, $F^{00}_{FDR}$,
lies well within the estimated uncertainty of the continuation made with continued fractions $F^{00}_{C_N}$. We show the absolute value of the difference for the real parts (left) and for the imaginary parts (right), divided by the uncertainty of the $C_N$ calculation.
For reference, we show the conjugate position of the $f_2(1270)$, $f_0(1370)$, and $f_0(1500)$ poles that appear in unphysical sheets (red crosses) and the continued segment [1.15,~1.35]~GeV (red line). 
For the $f_0(1500)$, the actual calculation is carried out with a slightly different segment closer to the pole.
}
\end{figure}

As usual, we show in Fig.~\ref{fig:fsplane} that the uncertainty in the analytic extrapolation with continued fractions is much smaller than the total uncertainty in our method in the region of the $f_2(1270)$ conjugate pole position, which appears in an unphysical sheet.

\subsubsection{$f_0(1370)$}
\label{sssec:f01370}
Let us now discuss the $f_0(1370)$ resonance, whose existence was not considered well established~\cite{Bugg:2007ja,Klempt:2007cp,Ochs:2013gi,COMPASS:2015gxz}, largely because it was not found in the original analyses of $\pi\pi$ scattering data~\cite{Hyams:1973zf, Grayer:1974cr,Hyams:1975mc,Estabrooks:1974vu}, but also because it was not present in other channels.
The RPP provides a $T$-matrix pole estimate with large uncertainties: $(1250-1440)-i(60-300)\,$MeV, and there is a tendency to find a lighter mass in $\pi\pi$ modes than in the $K\bar K$ ones.

For the above reasons, the resonance was the focus of the application of the FDR$_{C_N}$ method in~\cite{Pelaez:2022qby}. 
In Table~\ref{tab:f01370} we show the results with the updated Global Fits in~\cite{Pelaez:2024uav}, when 
continuing the very same segment [1.15,~1.35]~GeV that we have just used for the $f_2(1270)$.
The results using the three Global Fits as input in the dispersive integrals are compatible with each other within uncertainties. Thus, we provide a single range $\mathbf{\overline{\text{GF}}}$ covering all three of them.
The central value of the mass falls below the RPP $T$-matrix pole estimate (1250-1440)-$i$(60-300)~MeV, but overlaps with it within the uncertainty. The central value of the width falls within the range of the RPP. Our result here is slightly lighter but perfectly compatible with the pole found
in~\cite{Pelaez:2022qby} from $\pipi$ data (shown in the first row of the table\footnote{Note that the coupling provided in~\cite{Pelaez:2022qby} is the one to the neutral $F^{00}$ channel. The coupling to the $\pi\pi$ $I=0$ amplitude requires an additional $\sqrt{3}$ factor.}).
\begin{table}[h]
\renewcommand{\arraystretch}{1.6}
	      \begin{tabular}{cccc} \hline\hline
		   FDR$_{C_N}$& $\sqrt{s_{f_0(1370)}}$ (MeV)& $|g_{\pi\pi}|$ (GeV)& $\phi$ ($^\circ$)\\ \hline
		Ref.~\cite{Pelaez:2022qby} & $\left(1245\pm40\right)$$\,-\,$$i\left(300^{+30}_{-70}\right)$  &   $9.7^{+1.2}_{-2.1}$ &\\
        \hline
        \hline
		GFI & $\left(1217^{+43}_{-28}\right)$$\,-\,$$i\left(291^{+51}_{-49}\right)$  &   $8.8^{+1.1}_{-2.0}$ &$-22^{+9}_{-10}$\\
        GFII  &$\left(1163^{+44}_{-30}\right)$$\,-\,$$i\left(238^{+52}_{-41}\right)$  & $7.7^{+1.6}_{-1.4}$&$-39\pm15$\\ 
        GFIII    &$\left(1189^{+64}_{-40}\right)$$\,-\,$$i\left(260^{+35}_{-40}\right)$  & $9.1^{+1.5}_{-2.2}$&$-36\pm19$\\ 
         $\mathbf{\overline{\textbf{GF}}}$      & $\mathbf{\left(1191^{+69}_{-58}\right)-\textbf{\textit{i}}\left(261^{+81}_{-64}\right)}$  &   $\mathbf{8.5^{+2.1}_{-2.2}}$ & $\mathbf{-28^{+15}_{-27}}$\\
          \hline
      \hline
	      \end{tabular}
\caption{Dispersive determinations of $f_0(1370)$ pole parameters using as input in the dispersive integrals each of the three dispersively constrained 
Global Fits (GF in the table) from~\cite{Pelaez:2024uav}. Since they are compatible, we combine them into a single estimate in boldface. We also compare with the result of~\cite{Pelaez:2022qby} obtained with the FDR$_{C_N}$ method from fits to $\pi\pi$ data. Let us recall that the dispersive analysis of $\pi\pi\to K \bar K$ data in~\cite{Pelaez:2022qby} prefers a heavier resonance.
}
\label{tab:f01370}
\end{table}

At this point, it is important to recall that in~\cite{Pelaez:2022qby} a much heavier pole $1390^{+40}_{-50}-i(220^{+60}_{-40})\,$MeV, was found from $\pi\pi\to K \bar K$ Roy-Steiner dispersion relations. This was interpreted as a conflict between the $\pi\pi$ scattering data and those from 
$\pi\pi\to K \bar K$. Actually, even in the RPP, there is a general tendency to find a heavier $f_0(1370)$ resonance in the $K\bar K$ modes than in $\pi\pi$ modes. From our results here and in~\cite{Pelaez:2022qby}, it seems that $\pi\pi$ scattering data prefer a lighter resonance.

The new results also display larger errors, because the updated dispersive analysis~\cite{Pelaez:2024uav} needed more parameters than in~\cite{GarciaMartin:2011cn} to describe the partial waves up to roughly 1.8 GeV. For the same reasons explained for the $f_2(1270)$, we consider that the Global Fit~I pole parameters are slightly favored over the other two.

In Fig.~\ref{fig:fsplane}, we see that, at the conjugated position of the $f_0(1370)$ pole, the difference between the dispersive calculation and the $C_N$ analytic calculation is as large as the uncertainty of our whole method. Thus, our uncertainties are now dominated by those of the analytic continuation method rather than those from the data itself.

\subsubsection{$f_0(1500)$}
\label{sssec:f01500}
In \cite{Pelaez:2022qby}, a $f_0(1500)$ pole was already found using the FDR$_{C_N}$ method, but since the focus of the analysis was on the controversial $f_0(1370)$ resonance, only the pole for the CFD was calculated and reported. We list it in the first row of Table~\ref{tab:f01500}. It is fairly compatible with the $T$-matrix pole estimate of the RPP: $(1430-1530)-i(40-90)\,$MeV. Remarkably, this pole was obtained without using as input in the $F^{00}$ FDR any partial-wave information above 1.42 GeV. Beyond that, the input of the dispersive integrals comes from the Regge ``average" description of total cross sections.

Here, we will complete the study of the $f_0(1500)$ pole, which is found in the three updated Global Fits
when the dispersive output is analytically continued
from the [1.2,~1.4]~GeV segment. It is almost the same segment as that used to determine the $f_0(1370)$ and $f_2(1270)$ poles. We have, however, moved it closer to the applicability limit of the FDRs to probe this pole more precisely and thereby increase the pole’s stability.
Their results are given in Table~\ref{tab:f01500}.
Global Fit~I describes a heavier resonance, with a larger coupling to two pions, but with also larger uncertainties than in the case of Global Fits~II and III.  All these results are roughly compatible with the pole position given in~\cite{Pelaez:2022qby}, and also lie inside or close to the RPP $T$-matrix pole estimate. We have thus combined the three values in a weighted estimate in the last line of Table~\ref{tab:f01500}.
\begin{table}[h]
\renewcommand{\arraystretch}{1.6}
	      \begin{tabular}{cccc} \hline\hline
		  FDR$_{C_N}$& $\sqrt{s_{f_0(1500)}}$ (MeV)&  $|g_{\pi\pi}|$ (GeV)& $\phi$ ($^\circ$)\\ \hline
          CFD~\cite{Pelaez:2022qby} & $\left(1523^{+16}_{-10}\right)-\textit{i}\left(52^{+16}_{-11}\right)$  &  & \\ 
          \hline     \hline   
          GFI & $\left(1537^{+38}_{-43}\right)-i\left(46^{+30}_{-29}\right)$  &   $5.7^{+1.5}_{-1.7}$ & $-38^{+11}_{-21}$\\
        GFII &$\left(1484^{+13}_{-11}\right)$$\,-\,$$i\left(35^{+11}_{-15}\right)$  &  $4.2^{+0.7}_{-0.6}$ &$-41^{+11}_{-7}$\\ 
        GFIII  &$\left(1491^{+13}_{-15}\right)$$\,-\,$$i\left(32^{+15}_{-14}\right)$  &   $4.6^{+0.7}_{-0.8}$&$-39^{+7}_{-8}$\\ 
           $\mathbf{\overline{\textbf{GF}}}$     & $\mathbf{\left(1489^{+86}_{-16}\right)-\textbf{\textit{i}}\left(35^{+41}_{-18}\right)}$  &   $\mathbf{4.5^{+2.7}_{-0.9}}$ & $\mathbf{-40^{+13}_{-19}}$\\
          \hline
      \hline
	      \end{tabular}
\caption{Dispersive determinations of the $f_0(1500)$ pole parameters using as input in the dispersive integrals each one of the three dispersively constrained 
Global Fits (GF in the table) from~\cite{Pelaez:2024uav}. 
Since they are compatible, we combine them into a single estimate in boldface. We also compare with the result of~\cite{Pelaez:2022qby} obtained with the FDR$_{C_N}$ method using the CFD parametrizations.
}
\label{tab:f01500}
\end{table}

Once more, in Fig.~\ref{fig:fsplane}, we see that, at the conjugated position of the $f_0(1500)$ pole, which appears in the contiguous unphysical sheet, the difference between the dispersive calculation and the $C_N$ analytic calculation is smaller than the total uncertainty in our method. For this plot, we have used the same segment as for the $f_2(1270)$, namely [1.15,~1.35]~GeV. 
Note that the three poles are obtained with the same segment. Nevertheless,
for the actual calculation of the resonance parameters, we have slightly tuned our segment choice. With the actual segment used to determine this resonance with better precision, the uncertainty from the extrapolation is reduced. Our uncertainties are again dominated by those of the analytic continuation method rather than those from the data itself.


\subsection{Resonances in $F^{I_t=1}$}

From the analytic continuation of the $F^{0+}$ and $F^{00}$ amplitudes, we have been able to extract the pole parameters of isoscalar and isovector resonances. We could try the same with the $I_t=1$ amplitude. However, it does not provide any additional information because, as seen in Eq.~\eqref{eq:Fanti}, it encompasses all possible isospin contributions, and also because its input does not have positivity properties. 
As a consequence, the contributions of different resonances overlap very often, and their individual poles cannot be disentangled from this amplitude alone. This is the case of the $\rho(1450)$ and the $f_0(1500)$ resonances, which in most samples appear as a single pole between their respective positions.
Even when poles are clearly isolated, the accuracy in their determination is not competitive with the use of the $F^{00}$ or $F^{0+}$ amplitudes.
For instance,
using Global Fit~I as input for the $F^{I_t=1}$ FDR$_{C_N}$ method, we obtain for the $f_0(500)$ an associated pole at $\sqrt{s_{f_0(500)}}=448^{+50}_{-32}-i\left(273^{+63}_{-90}\right)\,$MeV. Its uncertainties are three to four times larger for the mass and five to nine times larger for the width, compared with the uncertainties provided in Table~\ref{tab:f0500}.

\section{Analytic continuation from the parametrization of resonant partial waves}
\label{sec:parametrizations}

In the previous section, we determined pole parameters from the output of FDRs in the real axis in a model-independent way, using continued fractions as the analytic continuation method. 

However, our application of the FDR$_{C_N}$ method to $\pi\pi$ scattering data has two main limitations.
The first one, intrinsic to the method, is that it does not determine the spin of resonances. Still, taking into account the known absence of $I=2$ resonances below 2 GeV, we have identified as isovector resonances those appearing in $F^{0+}$ and as isoscalar those appearing in $F^{00}$. The second limitation is that we use the Global Fits from~\cite{Pelaez:2024uav} as input for the dispersive integrals, but only up to the energy where they could be made to satisfy dispersive constraints and still provide an acceptable description of $\pi\pi$ data. 
This restricts the resonances that can be extracted with our method to the isoscalar resonances up to $\sim$1.5 GeV, and isovector resonances up to $\sim$1.7 GeV.

In this section, we address these limitations by partially relaxing the parametrization independence. In particular, we will now use continued fractions to analytically extend the partial waves from the Global Fits in~\cite{Pelaez:2024uav} to the complex plane. For that, we use them in a segment of the real axis, hence introducing a parametrization dependence when describing the partial wave on the real axis. 
In contrast, the parametrization dependence is removed in the analytic continuation step, because the identification of stable poles using continued fractions with different values of $N$ does not rely on any specific functional form. Actually, in most cases of interest, the existence of a pole was not even imposed in the Global Fit functional form, which is just a polynomial in a given segment of the real axis, whose naive analytic continuation would be physically meaningless.

The use of a specific parametrization on the real axis generically leads to smaller uncertainties than the use of FDR output. Nevertheless, these parametrizations are dispersively constrained up to 1.4 GeV with the $F^{00}$ FDR, and up to 1.6 GeV with the $F^{0+}$ and $F^{I_t=1}$ FDRs, and thus the parametrization dependence below those energies has been reduced with the dispersive constraints.
Above 1.6 GeV, partial waves are just phenomenological fits to data, and they will give us a fair description of the resonances needed to describe those data, although with error bars that do not incorporate the model dependence.

Finally, we will discuss the Argand diagrams of the partial waves, which provide an intuitive and graphical way to visualize resonant behavior and the effects of inelasticity: elastic partial waves lie on the Argand circumference while inelasticity moves the trajectory inside the circle. As we will see, although visually compelling, these diagrams are not a substitute for a rigorous analytic continuation because neither the presence nor the absence of a complete loop is, by itself, a reliable criterion for a resonance (many light scalar states, for instance, do not produce full circles despite having well-defined poles).
\vspace{-2mm}

\subsection{Poles and their parameters from partial waves}

\vspace{-2mm}

We list the resonances, with their parameters, that can be identified in the analytic continuation of the Global Fit~I, II, and III partial waves in Tables~\ref{tab:resonances_pw_I},~\ref{tab:resonances_pw_II}, and~\ref{tab:resonances_pw_III}, respectively. The resonance content of the three Global Fits is the same for resonances with $J\geq2$, and for the S0 and P waves below 1.5 GeV, although with differences in their parameters.
However, for the S0 and P partial waves, the qualitative behavior of the three Global Fits differs widely beyond 1.4-1.5 GeV, because they fit very different datasets. We will see that S0- and P-wave resonance content is rather different above that energy, depending on the Global Fit.

\begin{table}[h]
\footnotesize
\resizebox{.48\textwidth}{!}{
\renewcommand{\arraystretch}{1.6}
	      \begin{tabular}{cccc} \hline\hline
		\multicolumn{4}{c}{Global Fit~I}  \\ \hline
		Resonance &  $\sqrt{s_{pole}}$ (MeV)& $|g_{\pi\pi}|$ (GeV$^{1-\ell}$)& $\phi$ ($^\circ$)\\ \hline
		\text{$f_0(500)$} &$(463^{+12}_{-14})$ $-$ $i(274^{+20}_{-14})$& $3.4^{+0.4}_{-0.3}$  &$-68\pm7$\\
        \text{$f_0(980)$} &$(995\pm3)$ $-$ $i(23\pm3)$& $2.1\pm0.3$ &$-69^{+10}_{-9}$\\
        \text{ $f_0(1370)$} &$(1197^{+16}_{-13})$ $-$ $i(204^{+11}_{-22})$& $4.0^{+0.4}_{-0.3}$  &$-25^{+7}_{-8}$\\
        \text{$f_0(1500)$} & & &\\
        \text{ $f_0(1710)$} & &  &\\
       
        \text{ $f_0(2020)$} &$(1908^{+40}_{-50})$ $-$ $i(307^{+43}_{-60})$& $6.2^{+1.9}_{-2.6}$ &$5^{+15}_{-18}$\\\hline
        \text{$\rho(770)$} &$(758.1^{+1.4}_{-1.6})$ $-$ $i(74.4^{+3.9}_{-3.5})$ & $6.06^{+0.08}_{-0.07}$&$-6.3^{+0.5}_{-0.3}$\\
		\text{$\rho(1450)$} &$(1410^{+12}_{-15})$ $-$ $i(146^{+16}_{-14})$ & $2.9^{+0.6}_{-0.7}$ &$-90^{+9}_{-10}$\\
         \text{$\rho(1700)$} & $(1776\pm8)$ $-$ $i(132\pm9)$ & $3.1^{+0.4}_{-0.3}$ & $63^{+9}_{-8}$ \\
        \text{$\rho(1900)$} & $(1897^{+27}_{-24})$ $-$ $i(27^{+26}_{-33})$ & $1.0\pm0.4$ & $25^{+27}_{-37}$\\\hline
		\text{$f_2(1270)$} &$(1265.7^{+0.4}_{-0.5})$ $-$ $i(98.6^{+0.5}_{-0.4})$ &$4.50\pm0.05$ &$2.5^{+1.5}_{-1.2}$\\
		\text{$f_2(1950)$} &$(1870^{+25}_{-30})$ $-$ $i(233\pm28)$ &$1.45^{+0.25}_{-0.27}$  &$-18\pm10$\\\hline
     
  		\text{$\rho_3(1690)$} &  $(1697.5^{+0.7}_{-1.2})$ $-$ $i(125.2^{+1.5}_{-1.4})$ & $1.36\pm0.02$ &$5.9^{+1.8}_{-2.0}$\\\hline \hline

	      \end{tabular}}
\caption{Resonance parameters found from the analytic continuation, using continued fractions, of the Global Fit~I partial-wave parametrizations from~\cite{Pelaez:2024uav}. Note the absence of the $f_0(1500)$ in the parametrization, which is nevertheless present in the dispersive FDR$_{C_N}$ analysis.}
\label{tab:resonances_pw_I}
\end{table}

\begin{table}[h]
\footnotesize
\resizebox{.48\textwidth}{!}{
\renewcommand{\arraystretch}{1.6}
	      \begin{tabular}{cccc} \hline\hline
		\multicolumn{4}{c}{Global Fit~II} \\ \hline
		Resonance &  $\sqrt{s_{pole}}$ (MeV)& $|g_{\pi\pi}|$ (GeV$^{1-\ell}$)& $\phi$ ($^\circ$)\\ \hline
		\text{$f_0(500)$} &$(466^{+8}_{-7})$ $-$ $i(286^{+10}_{-9})$& $3.7\pm0.3$ &$-70\pm3$\\
        \text{$f_0(980)$} &$(996\pm4)$ $-$ $i(24^{+3}_{-4})$& $2.3\pm0.4$&$-71^{+15}_{-11}$\\
        \text{ $f_0(1370)$} &$(1177^{+8}_{-6})$ $-$ $i(180^{+7}_{-6})$& $3.8\pm0.5$ &$-22^{+6}_{-5}$\\
        \text{$f_0(1500)$} &$(1533^{+14}_{-16})$ $-$ $i(172\pm14)$& $3.9^{+0.7}_{-0.9}$  &$-75\pm11$\\
        \text{ $f_0(1710)$} & &  &\\
       
        \text{ $f_0(2020)$} &&\\\hline
        \text{$\rho(770)$} &$(758.4^{+1.5}_{-1.6})$ $-$ $i(71.4^{+4.1}_{-3.8})$ & $5.94^{+0.07}_{-0.13}$&$-5.65^{+0.13}_{-0.14}$\\
		\text{$\rho(1450)$} &$(1396^{+28}_{-20})$ $-$ $i(96^{+26}_{-35})$ & $0.54\pm0.14$ &$-89^{+17}_{-18}$\\
         \text{$\rho(1700)$} & $(1783^{+69}_{-74})$ $-$ $i(431^{+58}_{-80})$ & $4.1^{+0.8}_{-1.0}$ & $18^{+15}_{-14}$ \\
        \text{$\rho(1900)$} & & & \\\hline
		\text{$f_2(1270)$} &$(1274.0^{+0.9}_{-1.0})$ $-$ $i(91.2\pm1.2)$ &$4.04\pm0.15$  &$7.5^{+2.7}_{-2.4}$\\
		\text{$f_2(1950)$} &$(1761^{+26}_{-24})$ $-$ $i(186^{+22}_{-25})$ &$1.6 \pm 0.4$  &$-93^{+15}_{-16}$\\\hline
     
  		\text{$\rho_3(1690)$} &  $(1706.5^{+0.9}_{-1.0})$ $-$ $i(137.9^{+1.8}_{-1.1})$ & $1.44\pm0.02$ &$4.8\pm0.5$\\\hline \hline        
	      \end{tabular}}
\caption{Resonance parameters found from the analytic continuation, using continued fractions, of the Global Fit~II partial-wave parametrizations from~\cite{Pelaez:2024uav}.}
\label{tab:resonances_pw_II}
\end{table}

\begin{table}[h]
\footnotesize
\resizebox{.48\textwidth}{!}{
\renewcommand{\arraystretch}{1.6}
	      \begin{tabular}{cccc} \hline\hline
		\multicolumn{4}{c}{Global Fit~III}  \\ \hline
		Resonance&  $\sqrt{s_{pole}}$ (MeV)& $|g_{\pi\pi}|$ (GeV$^{1-\ell}$)& $\phi$ ($^\circ$)\\ \hline
		\text{$f_0(500)$} &$(468^{+7}_{-9})$ $-$ $i(285\pm9)$& $3.6\pm0.3$ &$-68\pm4$ \\
        \text{$f_0(980)$} &$(999^{+7}_{-5})$ $-$ $i(25^{+4}_{-6})$& $2.3^{+0.4}_{-0.5}$ &$-75^{+16}_{-15}$\\
        \text{ $f_0(1370)$} &$(1207^{+8}_{-9})$ $-$ $i(199^{+8}_{-9})$& $4.1\pm0.4$  &$-19^{+5}_{-8}$\\
        \text{$f_0(1500)$} &$(1431^{+17}_{-21})$ $-$ $i(121^{+22}_{-14})$& $2.2\pm0.5$ &$-120^{+16}_{-20}$ \\
        \text{ $f_0(1710)$} & $(1755^{+13}_{-9})$ $-$ $i(117^{+7}_{-10})$& $3.2^{+0.6}_{-0.7}$ &$8^{+10}_{-12}$  \\
        \text{ $f_0(2020)$} &$(1845^{+13}_{-15})$ $-$ $i(77^{+15}_{-11})$& $3.7^{+0.7}_{-0.6}$ &$-70\pm13$\\\hline
        
       \text{$\rho(770)$} &$(758.0^{+1.5}_{-1.7})$ $-$ $i(72.3^{+4.4}_{-4.0})$ & $5.98^{+0.09}_{-0.06}$&$-5.85^{+0.07}_{-0.09}$\\
		\text{$\rho(1450)$} &$(1393^{+24}_{-18})$ $-$ $i(102^{+25}_{-22})$ & $0.66^{+0.11}_{-0.12}$ &$-11^{+22}_{-23}$\\
         \text{$\rho(1700)$} & &   &\\
        \text{$\rho(1900)$} & &  &\\\hline
        
		\text{$f_2(1270)$} &$(1273.7^{+1.1}_{-1.0})$ $-$ $i(91.1^{+0.9}_{-1.1})$ &$3.9\pm0.2$  &$7.2^{+2.5}_{-2.4}$\\
		\text{$f_2(1950)$} &$(1709^{+16}_{-15})$ $-$ $i(180^{+11}_{-13})$ &$2.5^{+0.3}_{-0.4}$  &$-62^{+11}_{-8}$\\\hline
     
  		\text{$\rho_3(1690)$} &  $(1708.2^{+0.9}_{-1.0})$ $-$ $i(120.8^{+0.6}_{-0.8})$ & $1.22\pm0.02$ &$5.63^{+0.20}_{-0.16}$\\\hline \hline        
	      \end{tabular}}
\caption{Resonance parameters found from the analytic continuation, using continued fractions, of the Global Fit~III partial-wave parametrizations from~\cite{Pelaez:2024uav}.}
\label{tab:resonances_pw_III}
\end{table}

\subsubsection{The $f_0$ resonances}

Let us first start discussing the resonances extracted from the S0 wave. The four lightest resonances correspond to the scalar resonances identified from the $F^{00}$ with the FDR$_{C_N}$ method: $f_0(500)$, $f_0(980)$, $f_0(1370)$, and $f_0(1500)$, and confirm the spin assignment.
For the first two, the partial-wave pole parameters are slightly more compatible with the GKPY extractions in Tables~\ref{tab:f0500} and~\ref{tab:f0980} than those from the FDR. Let us recall that GKPY equations were imposed on the S0 wave in~\cite{Pelaez:2019eqa,Pelaez:2024uav} not only in the real axis up to $\sim$$1.1$ GeV, but also in the complex plane, hence including the regions where these two poles appear. Therefore, it is not surprising that these two poles are present in the parametrizations and that, in general, all results agree within one standard deviation. 

The appearance of a $f_0(1370)$ resonance pole in the partial wave is very remarkable, since its existence was not imposed in the functional form. The pole for each Global Fit falls within the uncertainty of its respective FDR$_{C_N}$ value, whereas the widths are clearly smaller, more than one deviation away. However, the couplings to two pions are clearly incompatible with those from the FDRs, which are roughly twice as large as those in Tables~\ref{tab:resonances_pw_I},~\ref{tab:resonances_pw_II}, and~\ref{tab:resonances_pw_III}. 
Such a large disagreement was already found in~\cite{Pelaez:2022qby}.

The appearance of a $f_0(1500)$ is also not imposed in the functional form. Actually, above 1.4 GeV, Global Fit parametrizations are simply Chebyshev polynomials matched continuously and with a continuous derivative to the lower-energy description~\cite{Pelaez:2019eqa}.
Still, the $f_0(1500)$ pole is found from the analytic continuation of Global Fits~II and III. 
However, the values obtained from each Global Fit are incompatible with each other, lying $\sim$100 MeV apart from one another, both for the mass and the width. Moreover, the widths are much larger than the RPP estimated range.
As observed in~\cite{Pelaez:2022qby}, no $f_0(1500)$ is found for Global Fit~I, although it appears when using the FDR$_{C_N}$ method on a segment below 1.4 GeV within the dispersively constrained region. 
The significant instability of the poles obtained directly from unconstrained data parametrization is in stark contrast to the three fairly compatible values we found when the continuation, and its uncertainty, are obtained from a segment where constrained amplitudes can be used. 
This is a nice illustration of the dangers of taking analytic continuations of data fits that do not fulfill the dispersive constraints in the real axis.

Above the dispersively constrained region, two other poles are found in the analytic continuation in at least one of the three Global-Fit S0-wave parametrizations. Once again, these are not imposed in the functional form, but are generated from the fit and its analytic continuation.
Recall that, above 1.5 GeV, the three datasets~\cite{Hyams:1973zf,Hyams:1975mc,Grayer:1974cr,Kaminski:1996da} that lead to these Global Fits are very incompatible with each other. A stable pole compatible with the $f_0(1710)$ $T$-matrix pole RPP estimate, $(1680-1820)-i(50-180)\,$MeV, is only found for Global Fit~III. Let us also recall that, very recently, the RPP also lists a $f_0(1770)$ resonance, which is omitted from the summary tables. Its $\pi\pi$ decay has been seen, but the observations primarily come from other decays. For this reason and also because the pole observed in some of the $\pi\pi$ scattering datasets has traditionally been identified with the $f_0(1710)$, we have kept this identification in the tables. However, it may be a mixture of the two.

Finally, poles close to the RPP $f_0(2020)$ resonance estimate $(1870-2080)-i(120-240)\,$MeV, are found from Global Fits~I and III. However, while the masses are compatible with the RPP range, the widths are either too large or too small, respectively. We should nevertheless recall that our Global Fit~I only fits data up to 1.9 GeV, whereas Global Fits~II and III only up to 1.8 GeV. Therefore, this resonance sits at the very edge of the validity of the Global Fit parametrizations.
Moreover, according to the RPP, this resonance still needs confirmation. The S0 wave from Global Fit~III above 1.4 GeV seems to be the most complete in terms of the resonance poles listed in the RPP.

\subsubsection{The $\rho$ resonances}

The $\rho(770)$ and $\rho(1450)$ resonances were already found in a model-independent way with the $F^{0+}$ FDR$_{C_N}$ in Secs.~\ref{sssec:rho} and~\ref{sssec:rho1450}. In Tables~\ref{tab:resonances_pw_I},~\ref{tab:resonances_pw_II}, and~\ref{tab:resonances_pw_III}, we now list their poles obtained by a direct analytic continuation of the Global Fit partial-wave parametrization, by means of continued fractions. 

In the $\rho(770)$ region, Global Fit parametrizations are constrained not only with FDRs but also with GKPY equations up to $\sim$1.1~GeV.
Moreover, the observed $\rho(770)$ shape is close to a Breit-Wigner form, and the very functional form of the Global Fits is devised to fix its peak mass accurately. Hence, it is not surprising that the results from the analytic continuation of the partial-wave parametrizations are almost identical to those from the FDR$_{C_N}$ method in Table~\ref{tab:rho770}. 

The segment used for the $\rho(1450)$ determination with the FDR$_{C_N}$ method falls within the region constrained by the $F^{0+}$ FDR. However, Global Fit functional forms were not explicitly built to have such a pole. Nevertheless, its pole is found when these parametrizations are continued analytically with continued fractions, and the results for the three Global Fits are quite compatible among themselves. However, Global Fit~I prefers a somewhat heavier and wider resonance, with smaller uncertainties, than the other two Global Fits. The reason for this is that they describe different datasets above 1.4 GeV.
We already saw that the pole from Global Fit~I, using the FDR$_{C_N}$, is even heavier and incompatible with those from Global Fits~II and III, and the tension with them rises to three deviations.

Let us remark that, as it happened with the FDR$_{C_N}$ method with the direct analytic continuation of the Global Fits, there are no other poles in the P-wave $\pi\pi$ scattering data within the region constrained by the FDRs. In particular,  we do not find a stable pole for the $\rho(1570)$. However, as commented above, its $\pi\pi$ decay has not been observed and is omitted from the RPP summary tables. Once again, we do not find a pole for the $\rho(1250)$ resonance, removed from the RPP many years ago.

Our Global Fits remain unconstrained above 1.6 GeV, and beyond that energy, they are merely phenomenological fits that differ widely following different, and highly incompatible, datasets. It is nevertheless interesting to study the analytical continuation of the Global Fits in this region. The reason is that, above 1.6 GeV, their functional form is given by Chebyshev polynomials; therefore, these parametrizations have not been explicitly devised to contain any given number of poles, thus removing such a model dependence.

Still, some stable poles are found when Global Fits are analytically continued with continued fractions. Tentatively, they can be identified with the $\rho(1700)$ and $\rho(1900)$ resonances, both listed in the RPP, although the latter is omitted from the summary tables. The former appears only in Global Fits~I and II. They also fall within the range of the many values compiled by the RPP~\cite{ParticleDataGroup:2024cfk}. The result from Global Fit~II is consistent with the RPP mass and large width estimate, thanks to its large uncertainties.
Contrary to the $\rho(1250)$ pole, our results for Global Fit~I, much narrower and with very small uncertainties, are very 
consistent with the pole found in~\cite{Hammoud:2020aqi}, which they call $\rho(1800)$. This could be expected because the $\pi\pi$ scattering data they fit mostly corresponds to Global Fit~I. 

An additional stable pole is also found in the Global Fit~I, which could be identified with the $\rho(1900)$. This pole is compatible with most of the data collected by the RPP. Nevertheless, this resonance remains omitted from the summary tables. 

\subsubsection{ The $f_2$ resonances}

In the case of the D0 wave, we confirm that the pole extracted from the $F^{00}$ FDR was indeed the $f_2(1270)$ resonance. The results from the partial waves, which we provide in Tables~\ref{tab:resonances_pw_I},~\ref{tab:resonances_pw_II}, and~\ref{tab:resonances_pw_III}, are almost identical to those from Table~\ref{tab:f2}. 

We also find an additional stable pole for the analytic continuation of the D0 wave in the 1.7 to 1.9 GeV range, with a $\sim$400~MeV width, but rather different masses for the three Global Fits. The identification of this second pole is, unfortunately, purely tentative.
There are three $f_2$ resonances listed in the RPP summary tables below 2 GeV, but heavier than the $f_2(1270)$. Apart from being too light for the masses we find, we discard identifying our pole with the $f_2'(1525)$ and $f_2(1565)$ because the first has a $\pi\pi$ branching ratio smaller than $10^{-2}$ and the second is mostly seen in antinucleon-nucleon annihilation. We are left with the $f_2(1950)$, whose $T$-matrix pole is estimated by the RPP to lie in the range $(1830-2020)-i(110-220)\,$MeV, which is compatible with the pole we find from Global Fit~I. The widths found for Global Fits~II and III are consistent with the RPP estimate, but their masses are clearly too low. Still, the $f_2(1950)$ seems the most plausible assignment, and we have used it in the tables.

The RPP lists several other nearby $J=2$ resonances, but omits them from the summary tables. The $f_2(1430)$, which also needs confirmation, is too light and definitely too narrow for our values. The $f_2(1640)$ and $f_2(1910)$ decays to $\pi\pi$ have not been observed, and they are either too light or too heavy for Fits~II and III. 
The $f_2(1810)$, also omitted from the RPP summary table, is too narrow $\Gamma$200~MeV, and too heavy for Global Fits~II and III or too light for Global Fit~I. 

\subsubsection{The $\rho_3$ resonance}
Finally, from the analytic continuation of the F wave, we extract a robust pole in all three Global Fits, whose values we provide in Tables~\ref{tab:resonances_pw_I},~\ref{tab:resonances_pw_II}, and~\ref{tab:resonances_pw_III}. These results confirm that the pole parameters from Table~\ref{tab:rho3} correspond to the $\rho_3(1690)$. The partial-wave central values are almost identical to those obtained with the FDR$_{C_N}$ method. However, we consider that the tiny uncertainties in Tables~\ref{tab:resonances_pw_I},~\ref{tab:resonances_pw_II}, and~\ref{tab:resonances_pw_III} are unrealistic, because of their parametrization dependence and the fact that the FDRs are not well satisfied by any Global Fit above 1.6 GeV. A more realistic uncertainty due to those caveats was estimated in Sec.~\ref{sssec:rho3}, by extracting this resonance from the continuation of the average between the ``direct'' Global Fit and the ``extended-FDR" output, and considering the difference as a systematic source of error.

In this section, we have applied the analytic continuation through continued fractions to the partial-wave amplitudes from~\cite{Pelaez:2024uav}, instead of the FDR output. This has allowed us to identify the resonances obtained from the FDR outputs that we showed in the previous section. Note that the pole parameters in this section carry some model dependence, leading generically to too small uncertainties.
In the region where these parametrizations had been constrained dispersively, their poles agree remarkably well with those obtained from the FDR$_{C_N}$ method.
Above 1.6 GeV, these Global Fits are not dispersively constrained, and the partial waves can only be considered phenomenological fits. However, except for the $\rho_3(1690)$, their functional forms were not devised to have any given number of poles. Therefore, their analytic continuation reduces the model dependence of the usual phenomenological fits. 
Actually, their analytic continuation yields several stable poles in a region beyond the dispersive analysis from~\cite{Pelaez:2024uav}, which can be identified with resonances gathered by the RPP particle listings. We do not find any additional poles that cannot be identified with a nearby resonance in the present RPP particle listings.

\subsection{Argand diagrams for resonant partial waves}\label{sec:polesfromparams}

Argand diagrams provide a widely used graphical characterization for easily identifying certain types of resonances in partial waves (see~\cite{Martin:1970}). For simplicity, let us suppress momentarily the isospin $I$ and angular momentum $\ell$ indices in $t^{(I)}_\ell$ and other quantities. Then, partial waves are cast in terms of their phase shift $\delta(s)$ and elasticity $0\leq\eta(s)\leq1$, as follows:
\begin{equation}
    t(s)=\frac{\eta(s)\, e^{2i\delta(s)}-1}{2i\sigma(s)}; \quad \sigma(s) t(s)=\frac{i}{2}\left( 1-\eta(s) e^{2i\delta(s)}\right),
\end{equation}
where $\sigma(s)=2 k(s)/\sqrt{s}$, and $k(s)$ is the center-of-mass momentum.
In the elastic regime $\eta(s)=1$, and partial waves fall right at the boundary of the so-called Argand circle centered at $i/2$ with radius 1/2, whereas in the inelastic regime, $0\leq\eta(s)<1$, and they fall inside it.

Assuming a Breit-Wigner (multichannel or not) resonance, where the phase shift increases rapidly through $\delta(s_R)=\pi/2$ with increasing energy, and neglecting backgrounds and other singularities, it is easy to show\footnote{See chapter 8.3.1 in~\cite{Martin:1970}; note the different normalization.} that, as the phase shift increases by $\pi$ through $s_R$, the partial wave traces a circular path of radius $\Gamma_p/2\Gamma_{tot}$ inside the Argand circle. Here, $\Gamma_{tot},\,\Gamma_p$ are the BW total width of the resonance and its partial decay width into the channel of interest, respectively. Unfortunately, the absence of these circular paths has often been used to argue against the existence of resonances. However, it only implies the absence of a resonance that satisfies the very restrictive conditions to be described as a BW resonance.

In what follows, we discuss the Argand plots and the related resonance content of the Global Fits~I, II, and III, presented in Figs.~\ref{fig:argand1},~\ref{fig:argand2}, and~\ref{fig:argand3}, respectively. We will find several cases of resonances that trace a full circle in the Argand plot, but many other cases when they do not.
We will thus illustrate with specific examples that {\it the naive assumption that a resonance must trace a complete circle in the Argand plot does not hold in general}.
The rigorous way to determine the existence of a resonance is from the dispersive determination of its associated pole in the contiguous Riemann sheet.

\subsubsection{Argand plot of the S0 wave}

\begin{figure}[h]
\centering
\includegraphics[width=0.49\textwidth]{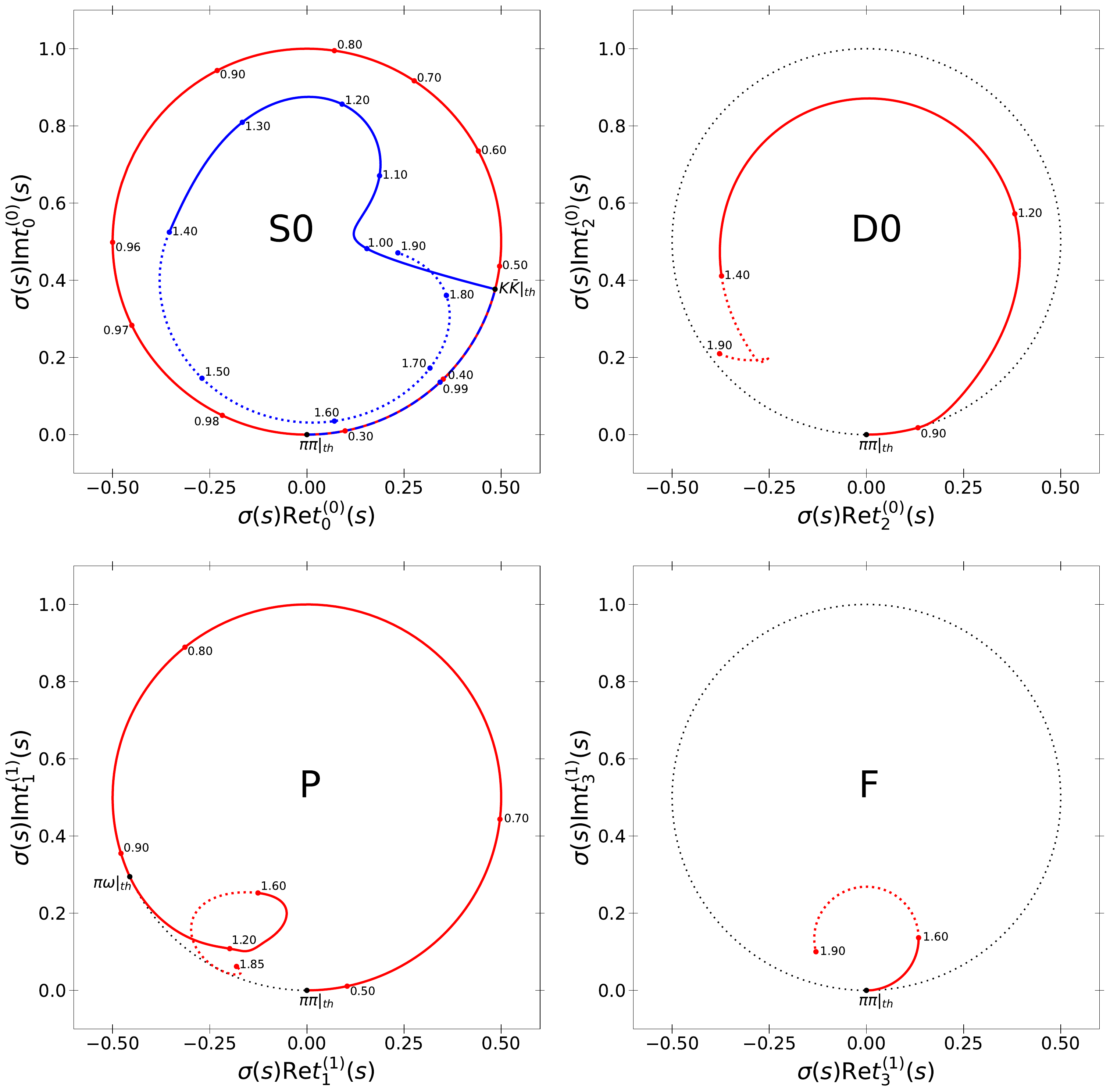}
\caption{ \small \label{fig:argand1} 
Argand plots for the S0, P, D0, and F partial waves of Global Fit~I.  The thin black-dotted curve is the Argand circumference. For the S0, there is an overlap between the first and successive turns (in blue). Thus, in the overlapping region, the S0 curve becomes dashed in red and blue. For all waves, curves are dotted where partial waves are no longer constrained with either the $F^{00}$ FDR (S0 and D0 waves, above 1.4 GeV) or the $F^{0+}$ FDR (P and F waves, above 1.6 GeV). Relevant two-particle thresholds are marked as $\pi\pi\vert_{th}$, $K\bar K\vert_{th}$, $\pi \omega\vert_{th}$.}
\end{figure}

\begin{figure}[h]
\centering
\includegraphics[width=0.49\textwidth]{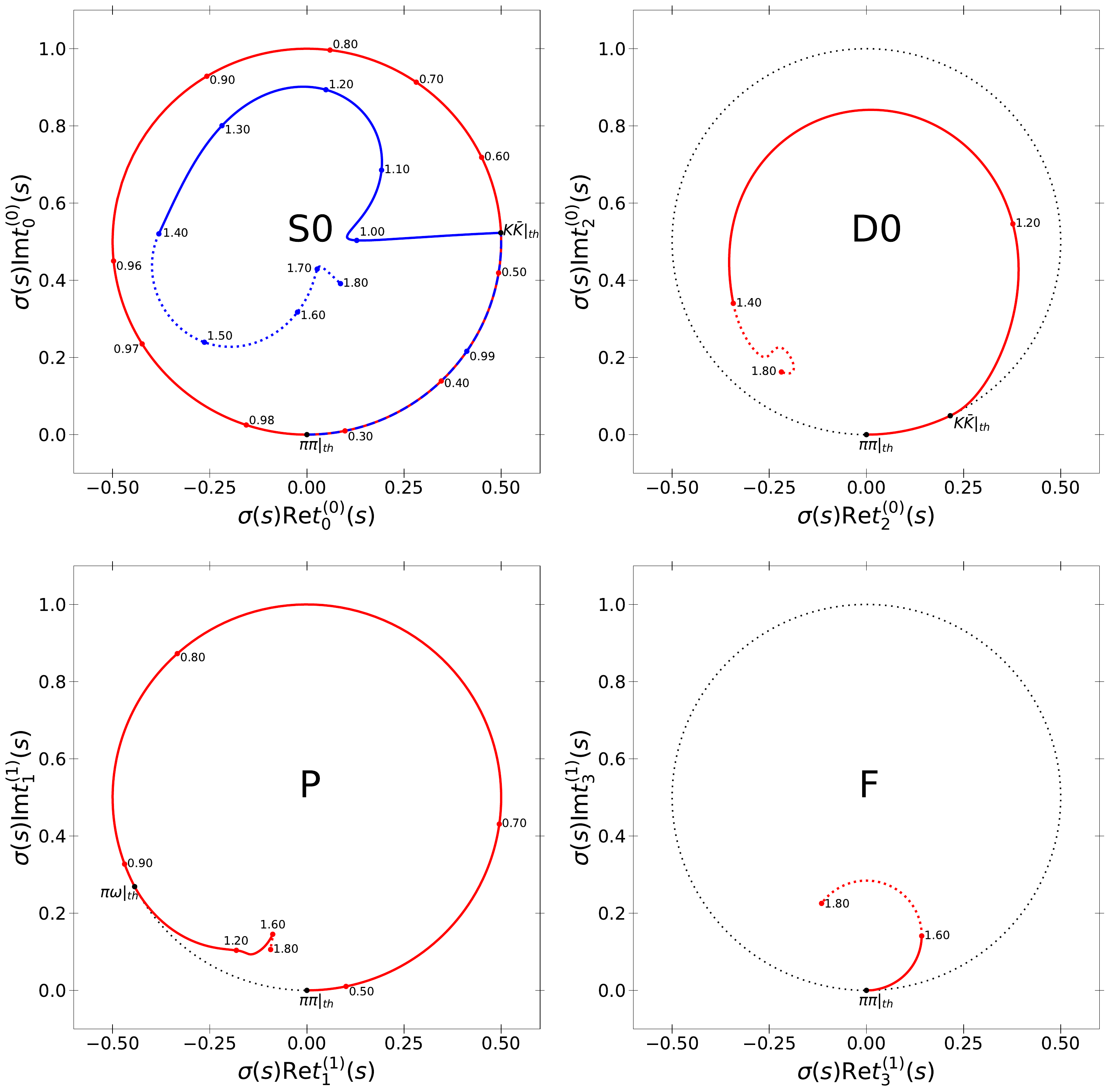}
\caption{ \small \label{fig:argand2} 
As in Fig.~\ref{fig:argand1}, but for Global Fit~II.}
\end{figure}

\begin{figure}[h]
\centering
\includegraphics[width=0.49\textwidth]{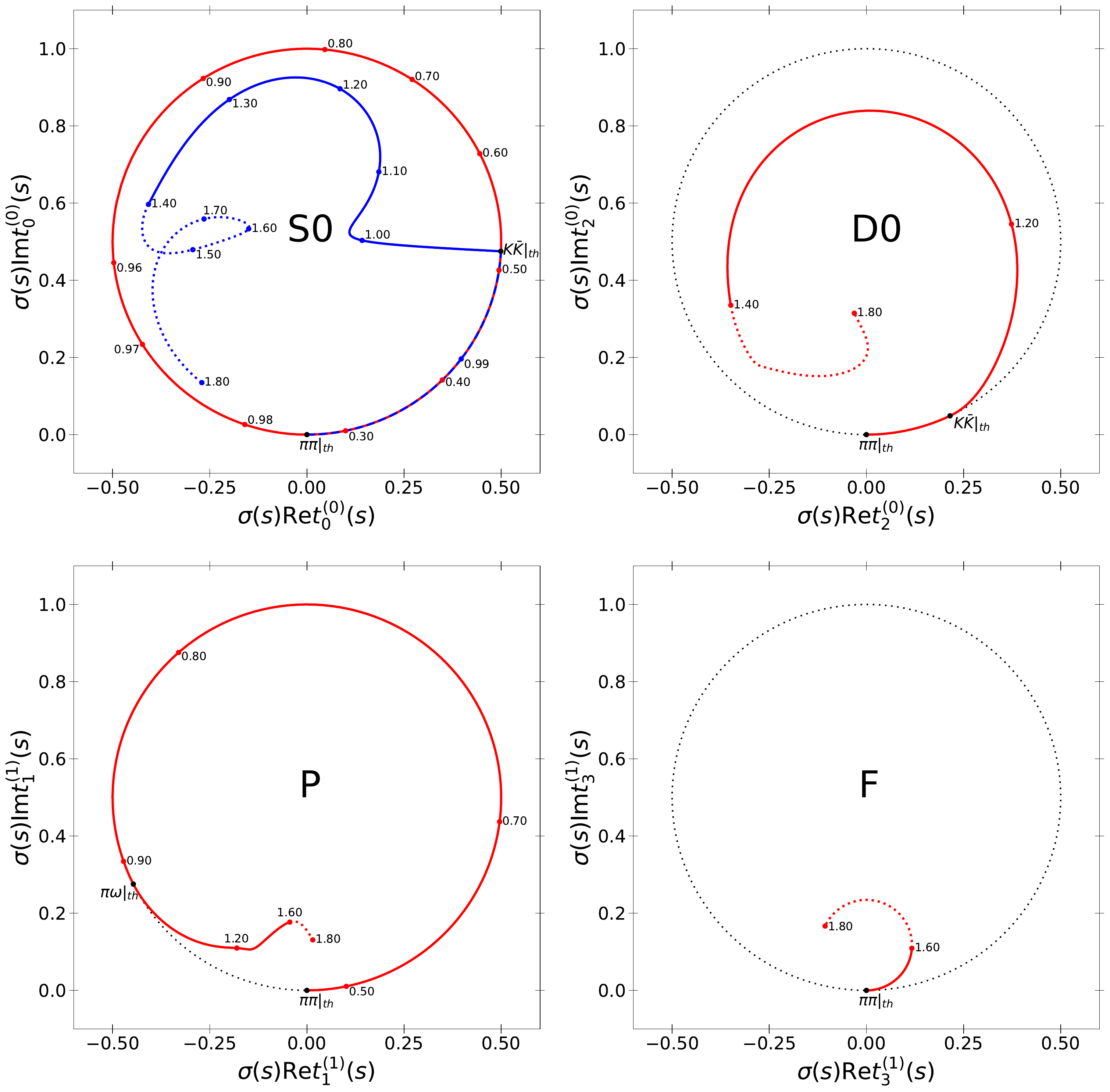}
\caption{ \small \label{fig:argand3} 
As in Fig.~\ref{fig:argand1}, but for Global Fit~III.}
\end{figure}

In the upper left panel of Figs.~\ref{fig:argand1},~\ref{fig:argand2}, and~\ref{fig:argand3}, we show the Argand diagrams of this partial wave for Global Fits~I, II, and III, respectively. Since the three fits are constrained with GKPY and Roy partial-wave dispersion relations up to $\sim$1.1 GeV, their Argand trajectories are almost identical up to that energy. They start at the $\pi\pi$ threshold at the lowest point of the Argand circumference, and they slowly turn counterclockwise to complete half a turn near $\sqrt{s}\simeq0.80-0.85\,$GeV. This first movement roughly corresponds to the $\sigma/f_0(500)$ pole. The partial wave keeps turning, but now at a faster speed, completing a full turn slightly after $\sqrt{s}=0.98\,$GeV. This very fast movement is due to the presence of the narrow $f_0(980)$, and extends up to the $K\bar K$ threshold, where it slows down and the partial wave leaves the circumference, since it becomes inelastic. Note that the first complete loop around the Argand circle is due to the presence of two poles, not just one. As promised, this is an example contrary to the common belief that each resonance must trace a full circle. Actually, neither the $f_0(500)$ nor the $f_0(980)$ resonances meet the criteria explained above for such a behavior.
In particular, we have already discussed that the real part of the $\sigma/f_0(500)$ pole position lies almost at the $\pi\pi$ threshold, very near the S0-wave Adler zero, and its shape is definitely not that of a Breit-Wigner resonance. 
Since the Argand plot movement is plotted only above threshold, there is no reason to expect a full turn due to this resonance alone, but more likely just half of it, as is indeed the case. Similarly, the $f_0(980)$ pole lies on the second Riemann sheet but extremely close to the $K\bar K$ threshold. Again, there is no reason to expect it to trace a full circle, since right above the $K\bar K$ threshold, the real axis is contiguous to a different sheet. Let us recall that it has been shown in~\cite{Burkert:2022bqo} that its apparent width, seen from $\pi\pi$ scattering, roughly corresponds to the difference between the $\pi\pi$ and $K\bar K$ partial widths, not to the naive Breit-Wigner shape with the actual total width.

Finally, we should recall that these two poles exist not only in the analytic continuation of partial-wave parametrizations with continued fractions, but in the very functional forms and in the partial-wave GKPY and Roy dispersion relations (as zeros of the S matrix in the first Riemann sheet). Our result is therefore very robust, and the Argand diagrams have a qualitative behavior similar to those provided in the CERN-Munich 1973 analysis~\cite{Hyams:1973zf} (necessarily so, since Global Fit~I fits those data, although with dispersive constraints).

Above 1 GeV, the S0 partial wave has only been constrained with the $F^{00}$ FDR up to 1.4 GeV (it is constrained by the $F^{I_t=1}$ FDR up to 1.6 GeV, but this is much less stringent). This is why, up to that energy (continuous blue curve), the three Global Fits display a qualitatively similar behavior. Up to 1.4 GeV, the $f_0(1370)$ pole does not trace a full circle, but rather half of it. Let us remark that the $f_0(1370)$ shape is definitely not that of a BW resonance and its width is $\sim$400~MeV. There is no rapid phase variation associated with this pole, which is what made it controversial and led to considering it absent from $\pi\pi$ scattering data, when studied with simple models.

Concerning the $f_0(1500)$, let us recall that it is found with the FDR$_{C_N}$ method for all three Global Fits, by analytically continuing a segment below 1.4 GeV, where the output of a consistent FDR exists. However, to draw the Argand movement, we need the partial wave, which is not constrained dispersively in the $f_0(1500)$
region (dotted blue line), and where the three Global Fits behave differently. In particular, there is no $f_0(1500)$ pole in the S0 partial wave of Global Fit~I.
Only in Global Fit~III do we see a small full loop in the Argand plot that can be partially associated with this resonance. This is not due to its BW shape, but to the fact that the Global Fit~III phase shift does not grow monotonously, but it goes up and down in the 1.4 to 1.6 GeV region  (see figures in~\cite{Pelaez:2019eqa} and~\cite{Pelaez:2024uav}). 

Neither the $f_0(1710)$ nor the $f_0(2020)$ poles, when they exist, trace a full circle in the S0 wave Argand plot, but, of course, they also do not display a BW behavior and do not have an associated fast phase-shift movement.

Therefore, we have seen that in almost all cases, light scalar-isoscalar resonances do not each trace a circle in the Argand plot, despite having a well-identified associated pole in the contiguous Riemann sheet. This is consistent with the well-established fact that they are not Breit-Wigner resonances.

\subsubsection{Argand plot of the P wave}

The Argand diagrams of this partial wave for Global Fits~I, II, and III 
are shown in the lower-left panels of Figs.~\ref{fig:argand1},~\ref{fig:argand2}, and~\ref{fig:argand3}.
Once again, the three Global Fits are almost identical up to $\sim$1.1 GeV, because they are dispersively constrained with GKPY and Roy equations.

With the $\rho(770)$, we see the first case of a resonance that behaves roughly as a BW. Being seen in the elastic region, it almost completes a full turn in the Argand circumference. In its Global Fit description, it becomes inelastic at the $\pi\omega$ threshold, and the three Argand trajectories of the three Global Fits are indistinguishable to the eye up to 1.2 GeV. The P waves exhibit a faster movement up to 1.6 GeV, where they are still constrained with the $F^{0+}$ and $F^{I_t=1}$ FDRs, but each movement is different. In particular, although all of them have a $\rho(1450)$ pole, only for Global Fit~I the $\rho(1450)$ seems to trace a circle, although most likely it is completed with the aid of the $\rho(1700)$, since it closes around 1.75 GeV.

The absence of an Argand loop near the $\rho(1450)$ in Global Fits~II and III, even when they do have an associated pole for that resonance, might be due to the tiny $\pi\pi$ coupling found in these two fits, which is a factor of 4 to 6 smaller than for Global Fit~I. With such a small coupling, the conditions to trace a loop, neglecting other effects, seem not to be met.

Beyond 1.6 GeV (dotted red curves), the partial waves are no longer constrained dispersively, and they behave very differently. While Global Fit~I completes a small half-circle before 1.85 GeV, the other Global Fits do not move much. This is consistent with the phase shift of Global Fit~I, which oscillates widely and rapidly in the 1.6 to 1.8 region, whereas it grows slowly and monotonously for Global Fits~II and III.

\subsubsection{Argand plot of the D0 wave}

The D0-wave Argand diagrams for Global Fits~I, II, and III 
are shown in the upper-right panels of Figs.~\ref{fig:argand1},~\ref{fig:argand2}, and~\ref{fig:argand3}.
This wave is dominated by the $f_2(1270)$, whose shape is close to a BW inelastic form.
Its branching ratio to $\pi\pi$ is $\sim$85$\%$, and therefore it almost completes a full loop very close to the Argand circumference. The three fits are practically indistinguishable to the eye up to 1.4 GeV, where the wave is constrained by the $F^{00}$ FDR.
Beyond that energy (dotted red curve), they are just phenomenological fits to rather different sets of data. They keep a similar behavior until they reach the region of the pole that we have very tentatively identified with the $f_2(1950)$. This resonance would be very wide and not Breit-Wigner-like. One would not expect it to trace a full circle within the region covered by our Global Fits; in fact, there are some hints of a different behavior in all three fits.

\vspace{-4mm}

\subsubsection{Argand plot of the F wave}

The $\rho_3(1690)$ resonance completely dominates the F wave of our Global Fits. It very much follows the shape of a very inelastic BW and therefore it traces a nice circumference well inside the Argand circle, although we cannot see it fully closed within the reach of our fits. It can be nicely seen in the three lower-right panels in Figs.~\ref{fig:argand1},~\ref{fig:argand2}, and~\ref{fig:argand3}.

\section{Summary and Discussion}
\label{sec:discussion}

In this work, we have applied the dispersive method recently developed in~\cite{Pelaez:2022qby} to determine, in a model-independent way, the existence and parameters of poles associated with resonances present in $\pi\pi$ scattering amplitudes.
This method, which we call FDR$_{C_N}$, provides an analytic continuation, to the contiguous Riemann sheet, of the output of FDRs, by means of a series of continued fractions $C_N$ for multiple values of interpolated points $N$. This way, model dependencies or specific parametrization choices are avoided both in the real segment that is analytically continued to the complex plane and also in the analytic continuation procedure. 

Roy and GKPY partial-wave dispersion relations had been used before to study resonances in meson-meson scattering in a model-independent way. However, they are limited, in practice, to resonances in the elastic region, like the $f_0(500)$, $\rho(770)$, and $f_0(980)$ in $\pi\pi$ scattering. The interest of the FDR$_{C_N}$ method is that it can also be applied in the inelastic regime, as was recently shown in~\cite{Pelaez:2022qby}, where it was first used to determine the parameters of the controversial scalar-isoscalar $f_0(1370)$ resonance from the analytic continuation of the FDR output for the $F^{00}$ amplitude. 

In this work, we have applied the FDR$_{C_N}$ method to all resonant partial waves relevant below $\sim$1.7~GeV, obtaining model-independent dispersive determinations of the parameters of all resonances significantly coupled to $\pi\pi$ in such a region.  For our purposes, it is crucial to have an input for the dispersive integrals that, by itself, satisfies the dispersive representation on the real axis. Hence, we have used the partial waves from the three Global Fits in our recent dispersive analysis~\cite{Pelaez:2024uav}. In that work, we extended the original approach of~\cite{Pelaez:2019eqa} to the S2, D, F, and G waves, while improving the P wave and updating the S0 wave. These three Global Fits, valid up to 1.8 GeV, describe the three existing datasets identified in~\cite{Pelaez:2019eqa}, which are incompatible with one another in most of the inelastic region. In~\cite{Pelaez:2024uav}, these Global Fits were constrained to satisfy, within uncertainties, nine dispersion relations: three Roy and three GKPY partial-wave dispersion relations for the S0, P, and S2 waves, up to roughly 1.1 GeV, as well as the $F^{00}$ FDR up to 1.4 GeV and the $F^{0+}$ and $F^{I_t=1}$ FDRs up to 1.6 GeV.

The $F^{0+}$ FDR, which has been continued analytically here for the first time, provides pole parameters for the $\rho(770)$, $\rho(1450)$, and $\rho_3(1690)$ resonances. The $\rho(770)$ values are consistent and 
possess comparable uncertainties to those obtained previously with GKPY equations, thus confirming the reliability of the method.
The results for the $\rho(1450)$ are the most interesting ones. Even if it has been measured from many different reactions, the estimate in the RPP~\cite{ParticleDataGroup:2022pth} has larger uncertainties; there is no estimate of its $T$-matrix pole, and there is a huge spread in the values listed under the $\pi\pi$ mode, which are largely incompatible among themselves and with the very RPP estimate. 
Let us note that this resonance was not imposed in our amplitudes; however, it is found in the continuation from the $F^{0+}$ FDR in all three Global Fits. Unfortunately, the pole mass found for Global Fit~I differs significantly from the other two, which are consistent with each other. Hence, we have provided two values for the $T$-matrix pole. We consider that our Global Fit~I result is slightly favored because it satisfies the dispersive constraints somewhat better than the other two fits~\cite{Pelaez:2024uav}.  

The $\rho_3(1690)$, extracted with a different strategy, required an analytic continuation from a nearby real-energy segment that lies outside the region where constrained Global Fits---i.e., fits that both describe the data in all partial waves and satisfy the FDRs---could be obtained. Thus, we have determined it by continuing the average between the partial-wave Global Fit and the output of an extended $F^{0+}$. These two curves are not compatible, and therefore, their difference is used as an estimate of the systematic uncertainty due to the breaking of the dispersive constraint. The central value is consistent with the values in the RPP listings, which, each being obtained with specific models and a single dataset, have smaller uncertainties than our result.

From the $F^{00}$ FDR, we find the $f_0(500)$, $f_0(980)$, $f_2(1270)$, $f_0(1370)$, and $f_0(1500)$ resonances. In the case of the first two, which appear in the elastic region, we obtain results consistent with those previously obtained from GKPY equations and have rather similar uncertainties. Once again, we consider this agreement a validation of our method. 

As in~\cite{Pelaez:2022qby}, we find poles for the $f_2(1270)$, $f_0(1370)$, and $f_0(1500)$ in the inelastic region.
In particular, the appearance of the controversial $f_0(1370)$ resonance in $\pi\pi$ scattering is confirmed again, with parameters consistent with those of~\cite{Pelaez:2022qby}. However, the mass and width we find now are somewhat smaller, and their uncertainties are larger.
As an improvement with respect to~\cite{Pelaez:2022qby}, we now provide uncertainties for the pole positions of the $f_2(1270)$ and
$f_0(1500)$ resonances.
For the $f_2(1270)$, we find values 
that differ by a few MeV from one Global Fit to another, due to the different datasets fitted in~\cite{Pelaez:2024uav}.
We believe the Global Fit~I results for the $f_2(1270)$ and $f_0(1370)$ to be slightly favored since the D0 wave data that it fits were obtained from a more general analysis than in the other cases, and its inelasticity opens below the $K \bar K$ threshold consistently with the observation of a sizable $f_2(1270)$ decay to four pions. 
As already explained in~\cite{Pelaez:2022qby}, it is remarkable that we find a pole for the $f_0(1500)$, since the $F^{00}$ FDR is imposed and has input from the partial waves only up to 1.4 GeV. Beyond that energy, the input comes from a Regge description of $\pi\pi$ total cross sections. It thus appears that the low-energy tail of this resonance around 1.4 GeV provides sufficient information to reconstruct the pole. 

The $F^{I_t=1}$ FDR has much larger uncertainties than the other two amplitudes and does not provide any competitive resonance determination.

For all dispersive determinations, we have verified that the analytic continuation using the $C_N$ function method onto the first Riemann sheet is consistent, within uncertainties, with the direct FDR output in the regions adjacent to the interpolated segments used to locate the poles.

We have not found in our $\pi\pi$ dispersive analysis any hint of the long-debated $\rho(1250)$, which appeared in previous RPP editions, and was recently reexamined in~\cite{Hammoud:2020aqi}. We also do not find hints of other vector-isovector or tensor resonances listed in the RPP, whose couplings to $\pi\pi$ are known to be very small. 

Those are the model- and parametrization-independent resonance determinations obtained with the FDR$_{C_N}$ dispersive method. However, we have also implemented a mixed approach, where the analytic continuation method by continued fractions is directly applied to the individual partial waves of the Global Fits, which extend at least up to 1.8 GeV. These additional resonance determinations exhibit some parametrization dependence, as we are not using any dispersive output, but the analytic continuation method does not. Although not model-independent and consequently with small uncertainties, the results from this mixed approach allow us to confirm the spin assignment of the resonances extracted from the FDRs. Moreover, this method provides access to resonances above 1.7 GeV that cannot be studied from the FDRs. 

Actually, we have extracted stable and robust additional poles above 1.7 GeV for each of the Global Fits, even if none of them were imposed in the parametrizations. Unfortunately, the three Global Fits differ significantly in this region, leading to rather different resonant spectra. Global Fit~I yields poles tentatively identified as the $\rho(1700)$, $\rho(1900)$, $f_2(1950)$ and $f_0(2020)$ resonances, whereas Global Fit~II contains only the $\rho(1700)$, 
and $f_2(1950)$, and Global Fit~III the $f_0(1710)$, $f_2(1950)$ and $f_0(2020)$. Only the pole tentatively identified with the $f_2(1950)$ appears in the three fits, but with rather incompatible values for the three of them. 
We have found that the S0 wave from Global Fit~III, and the P wave from Global Fit~I seem to be the most comprehensive with respect to the RPP listings. Consequently, unless two of the Global Fits may be discarded by other means or arguments, we believe that no robust conclusions about the light-meson spectrum beyond $\sim$1.7 GeV can be drawn from $\pi\pi$ scattering data alone.

Once the resonant content of each fit has been determined, we have presented and discussed the Argand diagrams of the resonant partial waves. This has allowed us to clarify the common misconception that for every resonance, the Argand trajectory must trace a full circle in the Argand diagram. This is only true for narrow resonances far from other singularities, which typically are reasonably well described by  Breit-Wigner formulas (possibly within multichannel formalisms). However, that is not the case for several resonances as seen in $\pi\pi$ scattering, like the $f_0(500)$, $f_0(980)$, $f_0(1370)$, $\rho(1450)$, and
$f_0(1500)$, which are firmly established by dispersive methods. Therefore, the absence of such loops in Argand plots should not be used to claim that a resonance does not exist.

In summary, the rigorous determination of the existence and parameters of a resonance is through its associated pole. In this work, we have provided such determinations from $\pi\pi$ scattering data alone, by means of a recently proposed method to combine the output of forward dispersion relations, and its analytic continuation by means of a series of continued fractions, using our new dispersively constrained Global Fits~\cite{Pelaez:2024uav} as input. We have validated the method from the results it provides for the $f_0(500)$, $\rho(770)$, and $f_0(980)$ resonances in the elastic region, which had previously been determined using Roy-like partial-wave dispersion relations.

This method is particularly noteworthy because it can be applied in the inelastic regime. Indeed, it has allowed us to provide model-independent results for the $T$-matrix pole parameters of the $f_2(1270)$, $f_0(1370)$, $\rho(1450)$, and $f_0(1500)$ resonances, and the uncertainties in the $\rho_3(1690)$.

We believe that this approach, with minor modifications, could have straightforward applications to study the resonance spectrum appearing in $\pi N$ and $\pi K$ scattering data.

\begin{acknowledgments} 
We are grateful to Robert Kamiński and George Rupp for their comments and for providing valuable references concerning the $\rho(1250)$ resonance. This work is part of the Grant PID2022-136510NB-C31 funded by MCIN/AEI/ 10.13039/501100011033. It has also received funding from the European Union’s Horizon 2020 research and innovation program under grant agreement No.824093. P. R. is supported
by the MIU (Ministerio de Universidades, Spain) fellowship FPU21/03878, and J.R.E. by the Ram\'on y Cajal program RYC2019-027605-I of the Spanish MICIU. 
\end{acknowledgments}

\vspace{1.5cm}


\bibliographystyle{apsrev4-2}
\bibliography{largebiblio.bib}

\begin{thebibliography}{86}%
\makeatletter
\providecommand \@ifxundefined [1]{%
 \@ifx{#1\undefined}
}%
\providecommand \@ifnum [1]{%
 \ifnum #1\expandafter \@firstoftwo
 \else \expandafter \@secondoftwo
 \fi
}%
\providecommand \@ifx [1]{%
 \ifx #1\expandafter \@firstoftwo
 \else \expandafter \@secondoftwo
 \fi
}%
\providecommand \natexlab [1]{#1}%
\providecommand \enquote  [1]{``#1''}%
\providecommand \bibnamefont  [1]{#1}%
\providecommand \bibfnamefont [1]{#1}%
\providecommand \citenamefont [1]{#1}%
\providecommand \href@noop [0]{\@secondoftwo}%
\providecommand \href [0]{\begingroup \@sanitize@url \@href}%
\providecommand \@href[1]{\@@startlink{#1}\@@href}%
\providecommand \@@href[1]{\endgroup#1\@@endlink}%
\providecommand \@sanitize@url [0]{\catcode `\\12\catcode `\$12\catcode
  `\&12\catcode `\#12\catcode `\^12\catcode `\_12\catcode `\%12\relax}%
\providecommand \@@startlink[1]{}%
\providecommand \@@endlink[0]{}%
\providecommand \url  [0]{\begingroup\@sanitize@url \@url }%
\providecommand \@url [1]{\endgroup\@href {#1}{\urlprefix }}%
\providecommand \urlprefix  [0]{URL }%
\providecommand \Eprint [0]{\href }%
\providecommand \doibase [0]{https://doi.org/}%
\providecommand \selectlanguage [0]{\@gobble}%
\providecommand \bibinfo  [0]{\@secondoftwo}%
\providecommand \bibfield  [0]{\@secondoftwo}%
\providecommand \translation [1]{[#1]}%
\providecommand \BibitemOpen [0]{}%
\providecommand \bibitemStop [0]{}%
\providecommand \bibitemNoStop [0]{.\EOS\space}%
\providecommand \EOS [0]{\spacefactor3000\relax}%
\providecommand \BibitemShut  [1]{\csname bibitem#1\endcsname}%
\let\auto@bib@innerbib\@empty
\bibitem [{\citenamefont {Navas}\ \emph {et~al.}(2024)\citenamefont {Navas}
  \emph {et~al.}}]{ParticleDataGroup:2024cfk}%
  \BibitemOpen
  \bibfield  {author} {\bibinfo {author} {\bibfnamefont {S.}~\bibnamefont
  {Navas}} \emph {et~al.} (\bibinfo {collaboration} {Particle Data Group}),\
  }\href {https://doi.org/10.1103/PhysRevD.110.030001} {\bibfield  {journal}
  {\bibinfo  {journal} {Phys. Rev. D}\ }\textbf {\bibinfo {volume} {110}},\
  \bibinfo {pages} {030001} (\bibinfo {year} {2024})}\BibitemShut {NoStop}%
\bibitem [{\citenamefont {Pelaez}(2025)}]{Pelaez:2025wma}%
  \BibitemOpen
  \bibfield  {author} {\bibinfo {author} {\bibfnamefont {J.~R.}\ \bibnamefont
  {Pelaez}},\ }\href@noop {} {\  (\bibinfo {year} {2025})},\ \Eprint
  {https://arxiv.org/abs/2509.08648} {arXiv:2509.08648 [hep-ph]} \BibitemShut
  {NoStop}%
\bibitem [{\citenamefont {Hyams}\ \emph {et~al.}(1973)\citenamefont {Hyams}
  \emph {et~al.}}]{Hyams:1973zf}%
  \BibitemOpen
  \bibfield  {author} {\bibinfo {author} {\bibfnamefont {B.}~\bibnamefont
  {Hyams}} \emph {et~al.},\ }\href
  {https://doi.org/10.1016/0550-3213(73)90618-4} {\bibfield  {journal}
  {\bibinfo  {journal} {Nucl. Phys.}\ }\textbf {\bibinfo {volume} {B64}},\
  \bibinfo {pages} {134} (\bibinfo {year} {1973})}\BibitemShut {NoStop}%
\bibitem [{\citenamefont {Durusoy}\ \emph {et~al.}(1973)\citenamefont
  {Durusoy}, \citenamefont {Baubillier}, \citenamefont {George}, \citenamefont
  {Goldberg}, \citenamefont {Touchard}, \citenamefont {Armenise}, \citenamefont
  {Fogli~Muciaccia},\ and\ \citenamefont {Silvestri}}]{Durusoy:1973aj}%
  \BibitemOpen
  \bibfield  {author} {\bibinfo {author} {\bibfnamefont {N.~B.}\ \bibnamefont
  {Durusoy}}, \bibinfo {author} {\bibfnamefont {M.}~\bibnamefont {Baubillier}},
  \bibinfo {author} {\bibfnamefont {R.}~\bibnamefont {George}}, \bibinfo
  {author} {\bibfnamefont {M.}~\bibnamefont {Goldberg}}, \bibinfo {author}
  {\bibfnamefont {A.~M.}\ \bibnamefont {Touchard}}, \bibinfo {author}
  {\bibfnamefont {N.}~\bibnamefont {Armenise}}, \bibinfo {author}
  {\bibfnamefont {M.~T.}\ \bibnamefont {Fogli~Muciaccia}},\ and\ \bibinfo
  {author} {\bibfnamefont {A.}~\bibnamefont {Silvestri}},\ }\href
  {https://doi.org/10.1016/0370-2693(73)90658-8} {\bibfield  {journal}
  {\bibinfo  {journal} {Phys. Lett. B}\ }\textbf {\bibinfo {volume} {45}},\
  \bibinfo {pages} {517} (\bibinfo {year} {1973})}\BibitemShut {NoStop}%
\bibitem [{\citenamefont {Losty}\ \emph {et~al.}(1974)\citenamefont {Losty},
  \citenamefont {Chaloupka}, \citenamefont {Ferrando}, \citenamefont
  {Montanet}, \citenamefont {Paul}, \citenamefont {Yaffe}, \citenamefont
  {Zieminski}, \citenamefont {Alitti}, \citenamefont {Gandois},\ and\
  \citenamefont {Louie}}]{Losty:1973et}%
  \BibitemOpen
  \bibfield  {author} {\bibinfo {author} {\bibfnamefont {M.~J.}\ \bibnamefont
  {Losty}}, \bibinfo {author} {\bibfnamefont {V.}~\bibnamefont {Chaloupka}},
  \bibinfo {author} {\bibfnamefont {A.}~\bibnamefont {Ferrando}}, \bibinfo
  {author} {\bibfnamefont {L.}~\bibnamefont {Montanet}}, \bibinfo {author}
  {\bibfnamefont {E.}~\bibnamefont {Paul}}, \bibinfo {author} {\bibfnamefont
  {D.}~\bibnamefont {Yaffe}}, \bibinfo {author} {\bibfnamefont
  {A.}~\bibnamefont {Zieminski}}, \bibinfo {author} {\bibfnamefont
  {J.}~\bibnamefont {Alitti}}, \bibinfo {author} {\bibfnamefont
  {B.}~\bibnamefont {Gandois}},\ and\ \bibinfo {author} {\bibfnamefont
  {J.}~\bibnamefont {Louie}},\ }\href
  {https://doi.org/10.1016/0550-3213(74)90131-X} {\bibfield  {journal}
  {\bibinfo  {journal} {Nucl. Phys. B}\ }\textbf {\bibinfo {volume} {69}},\
  \bibinfo {pages} {185} (\bibinfo {year} {1974})}\BibitemShut {NoStop}%
\bibitem [{\citenamefont {Cohen}\ \emph {et~al.}(1973)\citenamefont {Cohen},
  \citenamefont {Ferbel}, \citenamefont {Slattery},\ and\ \citenamefont
  {Werner}}]{Cohen:1973yx}%
  \BibitemOpen
  \bibfield  {author} {\bibinfo {author} {\bibfnamefont {D.~H.}\ \bibnamefont
  {Cohen}}, \bibinfo {author} {\bibfnamefont {T.}~\bibnamefont {Ferbel}},
  \bibinfo {author} {\bibfnamefont {P.}~\bibnamefont {Slattery}},\ and\
  \bibinfo {author} {\bibfnamefont {B.}~\bibnamefont {Werner}},\ }\href
  {https://doi.org/10.1103/PhysRevD.7.661} {\bibfield  {journal} {\bibinfo
  {journal} {Phys. Rev.}\ }\textbf {\bibinfo {volume} {D7}},\ \bibinfo {pages}
  {661} (\bibinfo {year} {1973})}\BibitemShut {NoStop}%
\bibitem [{\citenamefont {Protopopescu}\ \emph {et~al.}(1973)\citenamefont
  {Protopopescu}, \citenamefont {Alston-Garnjost}, \citenamefont
  {Barbaro-Galtieri}, \citenamefont {Flatte}, \citenamefont {Friedman},
  \citenamefont {Lasinski}, \citenamefont {Lynch}, \citenamefont {Rabin},\ and\
  \citenamefont {Solmitz}}]{Protopopescu:1973sh}%
  \BibitemOpen
  \bibfield  {author} {\bibinfo {author} {\bibfnamefont {S.~D.}\ \bibnamefont
  {Protopopescu}}, \bibinfo {author} {\bibfnamefont {M.}~\bibnamefont
  {Alston-Garnjost}}, \bibinfo {author} {\bibfnamefont {A.}~\bibnamefont
  {Barbaro-Galtieri}}, \bibinfo {author} {\bibfnamefont {S.~M.}\ \bibnamefont
  {Flatte}}, \bibinfo {author} {\bibfnamefont {J.~H.}\ \bibnamefont
  {Friedman}}, \bibinfo {author} {\bibfnamefont {T.~A.}\ \bibnamefont
  {Lasinski}}, \bibinfo {author} {\bibfnamefont {G.~R.}\ \bibnamefont {Lynch}},
  \bibinfo {author} {\bibfnamefont {M.~S.}\ \bibnamefont {Rabin}},\ and\
  \bibinfo {author} {\bibfnamefont {F.~T.}\ \bibnamefont {Solmitz}},\ }\href
  {https://doi.org/10.1103/PhysRevD.7.1279} {\bibfield  {journal} {\bibinfo
  {journal} {Phys. Rev.}\ }\textbf {\bibinfo {volume} {D7}},\ \bibinfo {pages}
  {1279} (\bibinfo {year} {1973})}\BibitemShut {NoStop}%
\bibitem [{\citenamefont {Grayer}\ \emph {et~al.}(1974)\citenamefont {Grayer}
  \emph {et~al.}}]{Grayer:1974cr}%
  \BibitemOpen
  \bibfield  {author} {\bibinfo {author} {\bibfnamefont {G.}~\bibnamefont
  {Grayer}} \emph {et~al.},\ }\href
  {https://doi.org/10.1016/0550-3213(74)90545-8} {\bibfield  {journal}
  {\bibinfo  {journal} {Nucl. Phys.}\ }\textbf {\bibinfo {volume} {B75}},\
  \bibinfo {pages} {189} (\bibinfo {year} {1974})}\BibitemShut {NoStop}%
\bibitem [{\citenamefont {Hyams}\ \emph {et~al.}(1975)\citenamefont {Hyams}
  \emph {et~al.}}]{Hyams:1975mc}%
  \BibitemOpen
  \bibfield  {author} {\bibinfo {author} {\bibfnamefont {B.}~\bibnamefont
  {Hyams}} \emph {et~al.},\ }\href
  {https://doi.org/10.1016/0550-3213(75)90616-1} {\bibfield  {journal}
  {\bibinfo  {journal} {Nucl. Phys.}\ }\textbf {\bibinfo {volume} {B100}},\
  \bibinfo {pages} {205} (\bibinfo {year} {1975})}\BibitemShut {NoStop}%
\bibitem [{\citenamefont {Hoogland}\ \emph {et~al.}(1977)\citenamefont
  {Hoogland} \emph {et~al.}}]{Hoogland:1977kt}%
  \BibitemOpen
  \bibfield  {author} {\bibinfo {author} {\bibfnamefont {W.}~\bibnamefont
  {Hoogland}} \emph {et~al.},\ }\href
  {https://doi.org/10.1016/0550-3213(77)90154-7} {\bibfield  {journal}
  {\bibinfo  {journal} {Nucl. Phys.}\ }\textbf {\bibinfo {volume} {B126}},\
  \bibinfo {pages} {109} (\bibinfo {year} {1977})}\BibitemShut {NoStop}%
\bibitem [{\citenamefont {Kaminski}\ \emph {et~al.}(1997)\citenamefont
  {Kaminski}, \citenamefont {Lesniak},\ and\ \citenamefont
  {Rybicki}}]{Kaminski:1996da}%
  \BibitemOpen
  \bibfield  {author} {\bibinfo {author} {\bibfnamefont {R.}~\bibnamefont
  {Kaminski}}, \bibinfo {author} {\bibfnamefont {L.}~\bibnamefont {Lesniak}},\
  and\ \bibinfo {author} {\bibfnamefont {K.}~\bibnamefont {Rybicki}},\ }\href
  {https://doi.org/10.1007/s002880050372} {\bibfield  {journal} {\bibinfo
  {journal} {Z. Phys.}\ }\textbf {\bibinfo {volume} {C74}},\ \bibinfo {pages}
  {79} (\bibinfo {year} {1997})},\ \Eprint
  {https://arxiv.org/abs/hep-ph/9606362} {arXiv:hep-ph/9606362 [hep-ph]}
  \BibitemShut {NoStop}%
\bibitem [{\citenamefont {Pelaez}\ and\ \citenamefont
  {Yndurain}(2005)}]{Pelaez:2004vs}%
  \BibitemOpen
  \bibfield  {author} {\bibinfo {author} {\bibfnamefont {J.~R.}\ \bibnamefont
  {Pelaez}}\ and\ \bibinfo {author} {\bibfnamefont {F.~J.}\ \bibnamefont
  {Yndurain}},\ }\href {https://doi.org/10.1103/PhysRevD.71.074016} {\bibfield
  {journal} {\bibinfo  {journal} {Phys. Rev.}\ }\textbf {\bibinfo {volume}
  {D71}},\ \bibinfo {pages} {074016} (\bibinfo {year} {2005})},\ \Eprint
  {https://arxiv.org/abs/hep-ph/0411334} {arXiv:hep-ph/0411334 [hep-ph]}
  \BibitemShut {NoStop}%
\bibitem [{\citenamefont {Pel\'aez}\ \emph {et~al.}(2021)\citenamefont
  {Pel\'aez}, \citenamefont {Rodas},\ and\ \citenamefont {Ruiz~de
  Elvira}}]{Pelaez:2021dak}%
  \BibitemOpen
  \bibfield  {author} {\bibinfo {author} {\bibfnamefont {J.~R.}\ \bibnamefont
  {Pel\'aez}}, \bibinfo {author} {\bibfnamefont {A.}~\bibnamefont {Rodas}},\
  and\ \bibinfo {author} {\bibfnamefont {J.}~\bibnamefont {Ruiz~de Elvira}},\
  }\href {https://doi.org/10.1140/epjs/s11734-021-00142-9} {\bibfield
  {journal} {\bibinfo  {journal} {Eur. Phys. J. ST}\ }\textbf {\bibinfo
  {volume} {230}},\ \bibinfo {pages} {1539} (\bibinfo {year} {2021})},\ \Eprint
  {https://arxiv.org/abs/2101.06506} {arXiv:2101.06506 [hep-ph]} \BibitemShut
  {NoStop}%
\bibitem [{\citenamefont {Pelaez}\ \emph {et~al.}(2023)\citenamefont {Pelaez},
  \citenamefont {Rodas},\ and\ \citenamefont {Ruiz~de
  Elvira}}]{Pelaez:2022qby}%
  \BibitemOpen
  \bibfield  {author} {\bibinfo {author} {\bibfnamefont {J.~R.}\ \bibnamefont
  {Pelaez}}, \bibinfo {author} {\bibfnamefont {A.}~\bibnamefont {Rodas}},\ and\
  \bibinfo {author} {\bibfnamefont {J.}~\bibnamefont {Ruiz~de Elvira}},\ }\href
  {https://doi.org/10.1103/PhysRevLett.130.051902} {\bibfield  {journal}
  {\bibinfo  {journal} {Phys. Rev. Lett.}\ }\textbf {\bibinfo {volume} {130}},\
  \bibinfo {pages} {051902} (\bibinfo {year} {2023})},\ \bibinfo {note}
  {[Erratum: Phys.Rev.Lett. 132, 239901 (2024)]},\ \Eprint
  {https://arxiv.org/abs/2206.14822} {arXiv:2206.14822 [hep-ph]} \BibitemShut
  {NoStop}%
\bibitem [{\citenamefont {Pel\'aez}\ \emph {et~al.}(2025)\citenamefont
  {Pel\'aez}, \citenamefont {Rab\'an},\ and\ \citenamefont {Ruiz~de
  Elvira}}]{Pelaez:2024uav}%
  \BibitemOpen
  \bibfield  {author} {\bibinfo {author} {\bibfnamefont {J.~R.}\ \bibnamefont
  {Pel\'aez}}, \bibinfo {author} {\bibfnamefont {P.}~\bibnamefont {Rab\'an}},\
  and\ \bibinfo {author} {\bibfnamefont {J.}~\bibnamefont {Ruiz~de Elvira}},\
  }\href {https://doi.org/10.1103/PhysRevD.111.074003} {\bibfield  {journal}
  {\bibinfo  {journal} {Phys. Rev. D}\ }\textbf {\bibinfo {volume} {111}},\
  \bibinfo {pages} {074003} (\bibinfo {year} {2025})},\ \Eprint
  {https://arxiv.org/abs/2412.15327} {arXiv:2412.15327 [hep-ph]} \BibitemShut
  {NoStop}%
\bibitem [{\citenamefont {Roy}(1971)}]{Roy:1971tc}%
  \BibitemOpen
  \bibfield  {author} {\bibinfo {author} {\bibfnamefont {S.~M.}\ \bibnamefont
  {Roy}},\ }\href {https://doi.org/10.1016/0370-2693(71)90724-6} {\bibfield
  {journal} {\bibinfo  {journal} {Phys.Lett.}\ }\textbf {\bibinfo {volume}
  {36B}},\ \bibinfo {pages} {353} (\bibinfo {year} {1971})}\BibitemShut
  {NoStop}%
\bibitem [{\citenamefont {Pelaez}\ \emph {et~al.}(2008)\citenamefont {Pelaez},
  \citenamefont {Garcia-Martin}, \citenamefont {Kaminski},\ and\ \citenamefont
  {Yndurain}}]{Pelaez:2008ry}%
  \BibitemOpen
  \bibfield  {author} {\bibinfo {author} {\bibfnamefont {J.~R.}\ \bibnamefont
  {Pelaez}}, \bibinfo {author} {\bibfnamefont {R.}~\bibnamefont
  {Garcia-Martin}}, \bibinfo {author} {\bibfnamefont {R.}~\bibnamefont
  {Kaminski}},\ and\ \bibinfo {author} {\bibfnamefont {F.~J.}\ \bibnamefont
  {Yndurain}},\ }\href {https://doi.org/10.1063/1.2973509} {\bibfield
  {journal} {\bibinfo  {journal} {AIP Conf. Proc.}\ }\textbf {\bibinfo {volume}
  {1030}},\ \bibinfo {pages} {257} (\bibinfo {year} {2008})},\ \Eprint
  {https://arxiv.org/abs/0804.2632} {arXiv:0804.2632 [hep-ph]} \BibitemShut
  {NoStop}%
\bibitem [{\citenamefont {Kaminski}\ \emph {et~al.}(2009)\citenamefont
  {Kaminski}, \citenamefont {Garcia-Martin}, \citenamefont {Grynkiewicz},
  \citenamefont {Pelaez},\ and\ \citenamefont {Yndurain}}]{Kaminski:2008fu}%
  \BibitemOpen
  \bibfield  {author} {\bibinfo {author} {\bibfnamefont {R.}~\bibnamefont
  {Kaminski}}, \bibinfo {author} {\bibfnamefont {R.}~\bibnamefont
  {Garcia-Martin}}, \bibinfo {author} {\bibfnamefont {P.}~\bibnamefont
  {Grynkiewicz}}, \bibinfo {author} {\bibfnamefont {J.~R.}\ \bibnamefont
  {Pelaez}},\ and\ \bibinfo {author} {\bibfnamefont {F.~J.}\ \bibnamefont
  {Yndurain}},\ }\bibfield  {booktitle} {\emph {\bibinfo {booktitle}
  {{Proceedings, 10th International Workshop on Meson Production, Properties
  and Interaction (MESON 2008): Cracow, Poland, June 6-10, 2008}}},\ }\href
  {https://doi.org/10.1142/S0217751X09043730} {\bibfield  {journal} {\bibinfo
  {journal} {Int. J. Mod. Phys.}\ }\textbf {\bibinfo {volume} {A24}},\ \bibinfo
  {pages} {402} (\bibinfo {year} {2009})},\ \Eprint
  {https://arxiv.org/abs/0809.4766} {arXiv:0809.4766 [hep-ph]} \BibitemShut
  {NoStop}%
\bibitem [{\citenamefont {Garc\'ia-Mart\'in}\ \emph
  {et~al.}(2011{\natexlab{a}})\citenamefont {Garc\'ia-Mart\'in}, \citenamefont
  {Kami\'nski}, \citenamefont {Pel\'aez}, \citenamefont {Ruiz~de Elvira},\ and\
  \citenamefont {Yndur\'ain}}]{GarciaMartin:2011cn}%
  \BibitemOpen
  \bibfield  {author} {\bibinfo {author} {\bibfnamefont {R.}~\bibnamefont
  {Garc\'ia-Mart\'in}}, \bibinfo {author} {\bibfnamefont {R.}~\bibnamefont
  {Kami\'nski}}, \bibinfo {author} {\bibfnamefont {J.~R.}\ \bibnamefont
  {Pel\'aez}}, \bibinfo {author} {\bibfnamefont {J.}~\bibnamefont {Ruiz~de
  Elvira}},\ and\ \bibinfo {author} {\bibfnamefont {F.~J.}\ \bibnamefont
  {Yndur\'ain}},\ }\href {https://doi.org/10.1103/PhysRevD.83.074004}
  {\bibfield  {journal} {\bibinfo  {journal} {Phys.Rev.}\ }\textbf {\bibinfo
  {volume} {D83}},\ \bibinfo {pages} {074004} (\bibinfo {year}
  {2011}{\natexlab{a}})},\ \Eprint {https://arxiv.org/abs/1102.2183}
  {arXiv:1102.2183 [hep-ph]} \BibitemShut {NoStop}%
\bibitem [{\citenamefont {Mahoux}\ \emph {et~al.}(1974)\citenamefont {Mahoux},
  \citenamefont {Roy},\ and\ \citenamefont {Wanders}}]{Mahoux:1974ej}%
  \BibitemOpen
  \bibfield  {author} {\bibinfo {author} {\bibfnamefont {G.}~\bibnamefont
  {Mahoux}}, \bibinfo {author} {\bibfnamefont {S.~M.}\ \bibnamefont {Roy}},\
  and\ \bibinfo {author} {\bibfnamefont {G.}~\bibnamefont {Wanders}},\ }\href
  {https://doi.org/10.1016/0550-3213(74)90480-5} {\bibfield  {journal}
  {\bibinfo  {journal} {Nucl.Phys.}\ }\textbf {\bibinfo {volume} {B70}},\
  \bibinfo {pages} {297} (\bibinfo {year} {1974})}\BibitemShut {NoStop}%
\bibitem [{\citenamefont {Auberson}\ and\ \citenamefont
  {Epele}(1975)}]{Auberson:1974in}%
  \BibitemOpen
  \bibfield  {author} {\bibinfo {author} {\bibfnamefont {G.}~\bibnamefont
  {Auberson}}\ and\ \bibinfo {author} {\bibfnamefont {L.}~\bibnamefont
  {Epele}},\ }\href {https://doi.org/10.1007/BF02820858} {\bibfield  {journal}
  {\bibinfo  {journal} {Nuovo Cim. A}\ }\textbf {\bibinfo {volume} {25}},\
  \bibinfo {pages} {453} (\bibinfo {year} {1975})}\BibitemShut {NoStop}%
\bibitem [{\citenamefont {Roy}\ and\ \citenamefont {Singh}(1975)}]{Roy:1975mq}%
  \BibitemOpen
  \bibfield  {author} {\bibinfo {author} {\bibfnamefont {S.~M.}\ \bibnamefont
  {Roy}}\ and\ \bibinfo {author} {\bibfnamefont {V.}~\bibnamefont {Singh}},\
  }\href {https://doi.org/10.1016/0370-2693(75)90529-8} {\bibfield  {journal}
  {\bibinfo  {journal} {Phys. Lett. B}\ }\textbf {\bibinfo {volume} {60}},\
  \bibinfo {pages} {67} (\bibinfo {year} {1975})}\BibitemShut {NoStop}%
\bibitem [{\citenamefont {Ananthanarayan}(1998)}]{Ananthanarayan:1998hj}%
  \BibitemOpen
  \bibfield  {author} {\bibinfo {author} {\bibfnamefont {B.}~\bibnamefont
  {Ananthanarayan}},\ }\href {https://doi.org/10.1103/PhysRevD.58.036002}
  {\bibfield  {journal} {\bibinfo  {journal} {Phys. Rev. D}\ }\textbf {\bibinfo
  {volume} {58}},\ \bibinfo {pages} {036002} (\bibinfo {year} {1998})},\
  \Eprint {https://arxiv.org/abs/hep-ph/9802338} {arXiv:hep-ph/9802338}
  \BibitemShut {NoStop}%
\bibitem [{\citenamefont {Elias~Mir{\'o}}\ \emph {et~al.}(2025)\citenamefont
  {Elias~Mir{\'o}}, \citenamefont {Guerrieri}, \citenamefont {G{\"u}m{\"u}s},\
  and\ \citenamefont {Zahed}}]{EliasMiro:2025rqo}%
  \BibitemOpen
  \bibfield  {author} {\bibinfo {author} {\bibfnamefont {J.}~\bibnamefont
  {Elias~Mir{\'o}}}, \bibinfo {author} {\bibfnamefont {A.}~\bibnamefont
  {Guerrieri}}, \bibinfo {author} {\bibfnamefont {M.~A.}\ \bibnamefont
  {G{\"u}m{\"u}s}},\ and\ \bibinfo {author} {\bibfnamefont {A.}~\bibnamefont
  {Zahed}},\ }\href@noop {} {\  (\bibinfo {year} {2025})},\ \Eprint
  {https://arxiv.org/abs/2509.14170} {arXiv:2509.14170 [hep-th]} \BibitemShut
  {NoStop}%
\bibitem [{\citenamefont {Kaminski}\ \emph {et~al.}(2006)\citenamefont
  {Kaminski}, \citenamefont {Pelaez},\ and\ \citenamefont
  {Yndurain}}]{Kaminski:2006yv}%
  \BibitemOpen
  \bibfield  {author} {\bibinfo {author} {\bibfnamefont {R.}~\bibnamefont
  {Kaminski}}, \bibinfo {author} {\bibfnamefont {J.~R.}\ \bibnamefont
  {Pelaez}},\ and\ \bibinfo {author} {\bibfnamefont {F.~J.}\ \bibnamefont
  {Yndurain}},\ }\href {https://doi.org/10.1103/PhysRevD.74.014001} {\bibfield
  {journal} {\bibinfo  {journal} {Phys. Rev. D}\ }\textbf {\bibinfo {volume}
  {74}},\ \bibinfo {pages} {014001} (\bibinfo {year} {2006})},\ \bibinfo {note}
  {[Erratum: Phys.Rev.D 74, 079903 (2006)]},\ \Eprint
  {https://arxiv.org/abs/hep-ph/0603170} {arXiv:hep-ph/0603170} \BibitemShut
  {NoStop}%
\bibitem [{\citenamefont {Kaminski}\ \emph {et~al.}(2008)\citenamefont
  {Kaminski}, \citenamefont {Pelaez},\ and\ \citenamefont
  {Yndurain}}]{Kaminski:2006qe}%
  \BibitemOpen
  \bibfield  {author} {\bibinfo {author} {\bibfnamefont {R.}~\bibnamefont
  {Kaminski}}, \bibinfo {author} {\bibfnamefont {J.~R.}\ \bibnamefont
  {Pelaez}},\ and\ \bibinfo {author} {\bibfnamefont {F.~J.}\ \bibnamefont
  {Yndurain}},\ }\href {https://doi.org/10.1103/PhysRevD.77.054015} {\bibfield
  {journal} {\bibinfo  {journal} {Phys. Rev. D}\ }\textbf {\bibinfo {volume}
  {77}},\ \bibinfo {pages} {054015} (\bibinfo {year} {2008})},\ \Eprint
  {https://arxiv.org/abs/0710.1150} {arXiv:0710.1150 [hep-ph]} \BibitemShut
  {NoStop}%
\bibitem [{\citenamefont {Navarro~P\'erez}\ \emph {et~al.}(2015)\citenamefont
  {Navarro~P\'erez}, \citenamefont {Ruiz~Arriola},\ and\ \citenamefont {Ruiz~de
  Elvira}}]{Perez:2015pea}%
  \BibitemOpen
  \bibfield  {author} {\bibinfo {author} {\bibfnamefont {R.}~\bibnamefont
  {Navarro~P\'erez}}, \bibinfo {author} {\bibfnamefont {E.}~\bibnamefont
  {Ruiz~Arriola}},\ and\ \bibinfo {author} {\bibfnamefont {J.}~\bibnamefont
  {Ruiz~de Elvira}},\ }\href {https://doi.org/10.1103/PhysRevD.91.074014}
  {\bibfield  {journal} {\bibinfo  {journal} {Phys.Rev.}\ }\textbf {\bibinfo
  {volume} {D91}},\ \bibinfo {pages} {074014} (\bibinfo {year} {2015})},\
  \Eprint {https://arxiv.org/abs/1502.03361} {arXiv:1502.03361 [hep-ph]}
  \BibitemShut {NoStop}%
\bibitem [{\citenamefont {Ananthanarayan}\ \emph {et~al.}(2001)\citenamefont
  {Ananthanarayan}, \citenamefont {Colangelo}, \citenamefont {Gasser},\ and\
  \citenamefont {Leutwyler}}]{Ananthanarayan:2000ht}%
  \BibitemOpen
  \bibfield  {author} {\bibinfo {author} {\bibfnamefont {B.}~\bibnamefont
  {Ananthanarayan}}, \bibinfo {author} {\bibfnamefont {G.}~\bibnamefont
  {Colangelo}}, \bibinfo {author} {\bibfnamefont {J.}~\bibnamefont {Gasser}},\
  and\ \bibinfo {author} {\bibfnamefont {H.}~\bibnamefont {Leutwyler}},\ }\href
  {https://doi.org/10.1016/S0370-1573(01)00009-6} {\bibfield  {journal}
  {\bibinfo  {journal} {Phys.Rept.}\ }\textbf {\bibinfo {volume} {353}},\
  \bibinfo {pages} {207} (\bibinfo {year} {2001})},\ \Eprint
  {https://arxiv.org/abs/hep-ph/0005297} {arXiv:hep-ph/0005297 [hep-ph]}
  \BibitemShut {NoStop}%
\bibitem [{\citenamefont {Colangelo}\ \emph {et~al.}(2001)\citenamefont
  {Colangelo}, \citenamefont {Gasser},\ and\ \citenamefont
  {Leutwyler}}]{Colangelo:2001df}%
  \BibitemOpen
  \bibfield  {author} {\bibinfo {author} {\bibfnamefont {G.}~\bibnamefont
  {Colangelo}}, \bibinfo {author} {\bibfnamefont {J.}~\bibnamefont {Gasser}},\
  and\ \bibinfo {author} {\bibfnamefont {H.}~\bibnamefont {Leutwyler}},\ }\href
  {https://doi.org/10.1016/S0550-3213(01)00147-X} {\bibfield  {journal}
  {\bibinfo  {journal} {Nucl. Phys.}\ }\textbf {\bibinfo {volume} {B603}},\
  \bibinfo {pages} {125} (\bibinfo {year} {2001})},\ \Eprint
  {https://arxiv.org/abs/hep-ph/0103088} {arXiv:hep-ph/0103088 [hep-ph]}
  \BibitemShut {NoStop}%
\bibitem [{\citenamefont {Caprini}\ \emph {et~al.}(2012)\citenamefont
  {Caprini}, \citenamefont {Colangelo},\ and\ \citenamefont
  {Leutwyler}}]{Caprini:2011ky}%
  \BibitemOpen
  \bibfield  {author} {\bibinfo {author} {\bibfnamefont {I.}~\bibnamefont
  {Caprini}}, \bibinfo {author} {\bibfnamefont {G.}~\bibnamefont {Colangelo}},\
  and\ \bibinfo {author} {\bibfnamefont {H.}~\bibnamefont {Leutwyler}},\ }\href
  {https://doi.org/10.1140/epjc/s10052-012-1860-1} {\bibfield  {journal}
  {\bibinfo  {journal} {Eur. Phys. J.}\ }\textbf {\bibinfo {volume} {C72}},\
  \bibinfo {pages} {1860} (\bibinfo {year} {2012})},\ \Eprint
  {https://arxiv.org/abs/1111.7160} {arXiv:1111.7160 [hep-ph]} \BibitemShut
  {NoStop}%
\bibitem [{\citenamefont {Moussallam}(2011)}]{Moussallam:2011zg}%
  \BibitemOpen
  \bibfield  {author} {\bibinfo {author} {\bibfnamefont {B.}~\bibnamefont
  {Moussallam}},\ }\href {https://doi.org/10.1140/epjc/s10052-011-1814-z}
  {\bibfield  {journal} {\bibinfo  {journal} {Eur. Phys. J.}\ }\textbf
  {\bibinfo {volume} {C71}},\ \bibinfo {pages} {1814} (\bibinfo {year}
  {2011})},\ \Eprint {https://arxiv.org/abs/1110.6074} {arXiv:1110.6074
  [hep-ph]} \BibitemShut {NoStop}%
\bibitem [{\citenamefont {Pelaez}\ \emph {et~al.}(2019)\citenamefont {Pelaez},
  \citenamefont {Rodas},\ and\ \citenamefont {Ruiz~de
  Elvira}}]{Pelaez:2019eqa}%
  \BibitemOpen
  \bibfield  {author} {\bibinfo {author} {\bibfnamefont {J.~R.}\ \bibnamefont
  {Pelaez}}, \bibinfo {author} {\bibfnamefont {A.}~\bibnamefont {Rodas}},\ and\
  \bibinfo {author} {\bibfnamefont {J.}~\bibnamefont {Ruiz~de Elvira}},\ }\href
  {https://doi.org/10.1140/epjc/s10052-019-7509-6} {\bibfield  {journal}
  {\bibinfo  {journal} {Eur. Phys. J. C}\ }\textbf {\bibinfo {volume} {79}},\
  \bibinfo {pages} {1008} (\bibinfo {year} {2019})},\ \Eprint
  {https://arxiv.org/abs/1907.13162} {arXiv:1907.13162 [hep-ph]} \BibitemShut
  {NoStop}%
\bibitem [{\citenamefont {Lehmann}(1958)}]{Lehmann:1958}%
  \BibitemOpen
  \bibfield  {author} {\bibinfo {author} {\bibfnamefont {H.}~\bibnamefont
  {Lehmann}},\ }\href@noop {} {\bibfield  {journal} {\bibinfo  {journal}
  {Nuovo\ Cim.}\ }\textbf {\bibinfo {volume} {10}},\ \bibinfo {pages} {579}
  (\bibinfo {year} {1958})}\BibitemShut {NoStop}%
\bibitem [{\citenamefont {Colangelo}\ \emph {et~al.}(2019)\citenamefont
  {Colangelo}, \citenamefont {Hoferichter},\ and\ \citenamefont
  {Stoffer}}]{Colangelo:2018mtw}%
  \BibitemOpen
  \bibfield  {author} {\bibinfo {author} {\bibfnamefont {G.}~\bibnamefont
  {Colangelo}}, \bibinfo {author} {\bibfnamefont {M.}~\bibnamefont
  {Hoferichter}},\ and\ \bibinfo {author} {\bibfnamefont {P.}~\bibnamefont
  {Stoffer}},\ }\href {https://doi.org/10.1007/JHEP02(2019)006} {\bibfield
  {journal} {\bibinfo  {journal} {JHEP}\ }\textbf {\bibinfo {volume} {02}},\
  \bibinfo {pages} {006}},\ \Eprint {https://arxiv.org/abs/1810.00007}
  {arXiv:1810.00007 [hep-ph]} \BibitemShut {NoStop}%
\bibitem [{\citenamefont {Caprini}\ \emph {et~al.}(2006)\citenamefont
  {Caprini}, \citenamefont {Colangelo},\ and\ \citenamefont
  {Leutwyler}}]{Caprini:2005zr}%
  \BibitemOpen
  \bibfield  {author} {\bibinfo {author} {\bibfnamefont {I.}~\bibnamefont
  {Caprini}}, \bibinfo {author} {\bibfnamefont {G.}~\bibnamefont {Colangelo}},\
  and\ \bibinfo {author} {\bibfnamefont {H.}~\bibnamefont {Leutwyler}},\ }\href
  {https://doi.org/10.1103/PhysRevLett.96.132001} {\bibfield  {journal}
  {\bibinfo  {journal} {Phys.Rev.Lett.}\ }\textbf {\bibinfo {volume} {96}},\
  \bibinfo {pages} {132001} (\bibinfo {year} {2006})},\ \Eprint
  {https://arxiv.org/abs/hep-ph/0512364} {arXiv:hep-ph/0512364 [hep-ph]}
  \BibitemShut {NoStop}%
\bibitem [{\citenamefont {Garc\'ia-Mart\'in}\ \emph
  {et~al.}(2011{\natexlab{b}})\citenamefont {Garc\'ia-Mart\'in}, \citenamefont
  {Kaminski}, \citenamefont {Pel\'aez},\ and\ \citenamefont {Ruiz~de
  Elvira}}]{GarciaMartin:2011jx}%
  \BibitemOpen
  \bibfield  {author} {\bibinfo {author} {\bibfnamefont {R.}~\bibnamefont
  {Garc\'ia-Mart\'in}}, \bibinfo {author} {\bibfnamefont {R.}~\bibnamefont
  {Kaminski}}, \bibinfo {author} {\bibfnamefont {J.~R.}\ \bibnamefont
  {Pel\'aez}},\ and\ \bibinfo {author} {\bibfnamefont {J.}~\bibnamefont
  {Ruiz~de Elvira}},\ }\href {https://doi.org/10.1103/PhysRevLett.107.072001}
  {\bibfield  {journal} {\bibinfo  {journal} {Phys.Rev.Lett.}\ }\textbf
  {\bibinfo {volume} {107}},\ \bibinfo {pages} {072001} (\bibinfo {year}
  {2011}{\natexlab{b}})},\ \Eprint {https://arxiv.org/abs/1107.1635}
  {arXiv:1107.1635 [hep-ph]} \BibitemShut {NoStop}%
\bibitem [{\citenamefont {Beringer}\ \emph {et~al.}(2012)\citenamefont
  {Beringer} \emph {et~al.}}]{Beringer:2012zz}%
  \BibitemOpen
  \bibfield  {author} {\bibinfo {author} {\bibfnamefont {J.}~\bibnamefont
  {Beringer}} \emph {et~al.} (\bibinfo {collaboration} {Particle Data Group}),\
  }\href {https://doi.org/10.1103/PhysRevD.86.010001} {\bibfield  {journal}
  {\bibinfo  {journal} {Phys. Rev. D}\ }\textbf {\bibinfo {volume} {86}},\
  \bibinfo {pages} {010001} (\bibinfo {year} {2012})}\BibitemShut {NoStop}%
\bibitem [{\citenamefont {Pel\'aez}(2016)}]{Pelaez:2015qba}%
  \BibitemOpen
  \bibfield  {author} {\bibinfo {author} {\bibfnamefont {J.~R.}\ \bibnamefont
  {Pel\'aez}},\ }\href {https://doi.org/10.1016/j.physrep.2016.09.001}
  {\bibfield  {journal} {\bibinfo  {journal} {Phys.Rept.}\ }\textbf {\bibinfo
  {volume} {658}},\ \bibinfo {pages} {1} (\bibinfo {year} {2016})},\ \Eprint
  {https://arxiv.org/abs/1510.00653} {arXiv:1510.00653 [hep-ph]} \BibitemShut
  {NoStop}%
\bibitem [{\citenamefont {Pelaez}(2004)}]{Pelaez:2003dy}%
  \BibitemOpen
  \bibfield  {author} {\bibinfo {author} {\bibfnamefont {J.~R.}\ \bibnamefont
  {Pelaez}},\ }\href {https://doi.org/10.1103/PhysRevLett.92.102001} {\bibfield
   {journal} {\bibinfo  {journal} {Phys. Rev. Lett.}\ }\textbf {\bibinfo
  {volume} {92}},\ \bibinfo {pages} {102001} (\bibinfo {year} {2004})},\
  \Eprint {https://arxiv.org/abs/hep-ph/0309292} {arXiv:hep-ph/0309292
  [hep-ph]} \BibitemShut {NoStop}%
\bibitem [{\citenamefont {Hoferichter}\ \emph {et~al.}(2024)\citenamefont
  {Hoferichter}, \citenamefont {Ruiz~de Elvira}, \citenamefont {Kubis},\ and\
  \citenamefont {Mei{\ss}ner}}]{Hoferichter:2023mgy}%
  \BibitemOpen
  \bibfield  {author} {\bibinfo {author} {\bibfnamefont {M.}~\bibnamefont
  {Hoferichter}}, \bibinfo {author} {\bibfnamefont {J.}~\bibnamefont {Ruiz~de
  Elvira}}, \bibinfo {author} {\bibfnamefont {B.}~\bibnamefont {Kubis}},\ and\
  \bibinfo {author} {\bibfnamefont {U.-G.}\ \bibnamefont {Mei{\ss}ner}},\
  }\href {https://doi.org/10.1016/j.physletb.2024.138698} {\bibfield  {journal}
  {\bibinfo  {journal} {Phys. Lett. B}\ }\textbf {\bibinfo {volume} {853}},\
  \bibinfo {pages} {138698} (\bibinfo {year} {2024})},\ \Eprint
  {https://arxiv.org/abs/2312.15015} {arXiv:2312.15015 [hep-ph]} \BibitemShut
  {NoStop}%
\bibitem [{\citenamefont {Garcia-Martin}\ \emph {et~al.}(2007)\citenamefont
  {Garcia-Martin}, \citenamefont {Pelaez},\ and\ \citenamefont
  {Yndurain}}]{Yndurain:2007qm}%
  \BibitemOpen
  \bibfield  {author} {\bibinfo {author} {\bibfnamefont {R.}~\bibnamefont
  {Garcia-Martin}}, \bibinfo {author} {\bibfnamefont {J.~R.}\ \bibnamefont
  {Pelaez}},\ and\ \bibinfo {author} {\bibfnamefont {F.~J.}\ \bibnamefont
  {Yndurain}},\ }\href {https://doi.org/10.1103/PhysRevD.76.074034} {\bibfield
  {journal} {\bibinfo  {journal} {Phys. Rev.}\ }\textbf {\bibinfo {volume}
  {D76}},\ \bibinfo {pages} {074034} (\bibinfo {year} {2007})},\ \Eprint
  {https://arxiv.org/abs/hep-ph/0701025} {arXiv:hep-ph/0701025 [hep-ph]}
  \BibitemShut {NoStop}%
\bibitem [{\citenamefont {Caprini}(2008)}]{Caprini:2008fc}%
  \BibitemOpen
  \bibfield  {author} {\bibinfo {author} {\bibfnamefont {I.}~\bibnamefont
  {Caprini}},\ }\href {https://doi.org/10.1103/PhysRevD.77.114019} {\bibfield
  {journal} {\bibinfo  {journal} {Phys. Rev.}\ }\textbf {\bibinfo {volume}
  {D77}},\ \bibinfo {pages} {114019} (\bibinfo {year} {2008})},\ \Eprint
  {https://arxiv.org/abs/0804.3504} {arXiv:0804.3504 [hep-ph]} \BibitemShut
  {NoStop}%
\bibitem [{\citenamefont {Švarc}\ \emph {et~al.}(2013)\citenamefont {Švarc},
  \citenamefont {Hadzimehmedovic}, \citenamefont {Osmanovic}, \citenamefont
  {Stahov}, \citenamefont {Tiator},\ and\ \citenamefont
  {Workman}}]{Svarc:2013laa}%
  \BibitemOpen
  \bibfield  {author} {\bibinfo {author} {\bibfnamefont {A.}~\bibnamefont
  {Švarc}}, \bibinfo {author} {\bibfnamefont {M.}~\bibnamefont
  {Hadzimehmedovic}}, \bibinfo {author} {\bibfnamefont {H.}~\bibnamefont
  {Osmanovic}}, \bibinfo {author} {\bibfnamefont {J.}~\bibnamefont {Stahov}},
  \bibinfo {author} {\bibfnamefont {L.}~\bibnamefont {Tiator}},\ and\ \bibinfo
  {author} {\bibfnamefont {R.~L.}\ \bibnamefont {Workman}},\ }\href
  {https://doi.org/10.1103/PhysRevC.88.035206} {\bibfield  {journal} {\bibinfo
  {journal} {Phys. Rev.}\ }\textbf {\bibinfo {volume} {C88}},\ \bibinfo {pages}
  {035206} (\bibinfo {year} {2013})},\ \Eprint
  {https://arxiv.org/abs/1307.4613} {arXiv:1307.4613 [hep-ph]} \BibitemShut
  {NoStop}%
\bibitem [{\citenamefont {Švarc}\ \emph {et~al.}(2014)\citenamefont {Švarc},
  \citenamefont {Hadžimehmedović}, \citenamefont {Osmanović}, \citenamefont
  {Stahov}, \citenamefont {Tiator},\ and\ \citenamefont
  {Workman}}]{Svarc:2014sqa}%
  \BibitemOpen
  \bibfield  {author} {\bibinfo {author} {\bibfnamefont {A.}~\bibnamefont
  {Švarc}}, \bibinfo {author} {\bibfnamefont {M.}~\bibnamefont
  {Hadžimehmedović}}, \bibinfo {author} {\bibfnamefont {H.}~\bibnamefont
  {Osmanović}}, \bibinfo {author} {\bibfnamefont {J.}~\bibnamefont {Stahov}},
  \bibinfo {author} {\bibfnamefont {L.}~\bibnamefont {Tiator}},\ and\ \bibinfo
  {author} {\bibfnamefont {R.~L.}\ \bibnamefont {Workman}},\ }\href
  {https://doi.org/10.1103/PhysRevC.89.065208} {\bibfield  {journal} {\bibinfo
  {journal} {Phys. Rev.}\ }\textbf {\bibinfo {volume} {C89}},\ \bibinfo {pages}
  {065208} (\bibinfo {year} {2014})},\ \Eprint
  {https://arxiv.org/abs/1404.1544} {arXiv:1404.1544 [nucl-th]} \BibitemShut
  {NoStop}%
\bibitem [{\citenamefont {\v{S}varc}\ \emph {et~al.}(2015)\citenamefont
  {\v{S}varc}, \citenamefont {Had\v{z}imehmedovi{\'c}}, \citenamefont
  {Osmanovi{\'c}}, \citenamefont {Stahov},\ and\ \citenamefont
  {Workman}}]{Svarc:2014aga}%
  \BibitemOpen
  \bibfield  {author} {\bibinfo {author} {\bibfnamefont {A.}~\bibnamefont
  {\v{S}varc}}, \bibinfo {author} {\bibfnamefont {M.}~\bibnamefont
  {Had\v{z}imehmedovi{\'c}}}, \bibinfo {author} {\bibfnamefont
  {H.}~\bibnamefont {Osmanovi{\'c}}}, \bibinfo {author} {\bibfnamefont
  {J.}~\bibnamefont {Stahov}},\ and\ \bibinfo {author} {\bibfnamefont {R.~L.}\
  \bibnamefont {Workman}},\ }\href {https://doi.org/10.1103/PhysRevC.91.015207}
  {\bibfield  {journal} {\bibinfo  {journal} {Phys.Rev.}\ }\textbf {\bibinfo
  {volume} {C91}},\ \bibinfo {pages} {015207} (\bibinfo {year} {2015})},\
  \Eprint {https://arxiv.org/abs/1405.6474} {arXiv:1405.6474 [nucl-th]}
  \BibitemShut {NoStop}%
\bibitem [{\citenamefont {Masjuan}\ and\ \citenamefont
  {Sanz-Cillero}(2013)}]{Masjuan:2013jha}%
  \BibitemOpen
  \bibfield  {author} {\bibinfo {author} {\bibfnamefont {P.}~\bibnamefont
  {Masjuan}}\ and\ \bibinfo {author} {\bibfnamefont {J.~J.}\ \bibnamefont
  {Sanz-Cillero}},\ }\href {https://doi.org/10.1140/epjc/s10052-013-2594-4}
  {\bibfield  {journal} {\bibinfo  {journal} {Eur. Phys. J.}\ }\textbf
  {\bibinfo {volume} {C73}},\ \bibinfo {pages} {2594} (\bibinfo {year}
  {2013})},\ \Eprint {https://arxiv.org/abs/1306.6308} {arXiv:1306.6308
  [hep-ph]} \BibitemShut {NoStop}%
\bibitem [{\citenamefont {Masjuan}\ \emph {et~al.}(2014)\citenamefont
  {Masjuan}, \citenamefont {Ruiz~de Elvira},\ and\ \citenamefont
  {Sanz-Cillero}}]{Masjuan:2014psa}%
  \BibitemOpen
  \bibfield  {author} {\bibinfo {author} {\bibfnamefont {P.}~\bibnamefont
  {Masjuan}}, \bibinfo {author} {\bibfnamefont {J.}~\bibnamefont {Ruiz~de
  Elvira}},\ and\ \bibinfo {author} {\bibfnamefont {J.~J.}\ \bibnamefont
  {Sanz-Cillero}},\ }\href {https://doi.org/10.1103/PhysRevD.90.097901}
  {\bibfield  {journal} {\bibinfo  {journal} {Phys. Rev.}\ }\textbf {\bibinfo
  {volume} {D90}},\ \bibinfo {pages} {097901} (\bibinfo {year} {2014})},\
  \Eprint {https://arxiv.org/abs/1410.2397} {arXiv:1410.2397 [hep-ph]}
  \BibitemShut {NoStop}%
\bibitem [{\citenamefont {Caprini}\ \emph {et~al.}(2016)\citenamefont
  {Caprini}, \citenamefont {Masjuan}, \citenamefont {Ruiz~de Elvira},\ and\
  \citenamefont {Sanz-Cillero}}]{Caprini:2016uxy}%
  \BibitemOpen
  \bibfield  {author} {\bibinfo {author} {\bibfnamefont {I.}~\bibnamefont
  {Caprini}}, \bibinfo {author} {\bibfnamefont {P.}~\bibnamefont {Masjuan}},
  \bibinfo {author} {\bibfnamefont {J.}~\bibnamefont {Ruiz~de Elvira}},\ and\
  \bibinfo {author} {\bibfnamefont {J.~J.}\ \bibnamefont {Sanz-Cillero}},\
  }\href {https://doi.org/10.1103/PhysRevD.93.076004} {\bibfield  {journal}
  {\bibinfo  {journal} {Phys. Rev.}\ }\textbf {\bibinfo {volume} {D93}},\
  \bibinfo {pages} {076004} (\bibinfo {year} {2016})},\ \Eprint
  {https://arxiv.org/abs/1602.02062} {arXiv:1602.02062 [hep-ph]} \BibitemShut
  {NoStop}%
\bibitem [{\citenamefont {Pel\'aez}\ \emph {et~al.}(2017)\citenamefont
  {Pel\'aez}, \citenamefont {Rodas},\ and\ \citenamefont {Ruiz~de
  Elvira}}]{Pelaez:2016klv}%
  \BibitemOpen
  \bibfield  {author} {\bibinfo {author} {\bibfnamefont {J.~R.}\ \bibnamefont
  {Pel\'aez}}, \bibinfo {author} {\bibfnamefont {A.}~\bibnamefont {Rodas}},\
  and\ \bibinfo {author} {\bibfnamefont {J.}~\bibnamefont {Ruiz~de Elvira}},\
  }\href {https://doi.org/10.1140/epjc/s10052-017-4668-1} {\bibfield  {journal}
  {\bibinfo  {journal} {Eur. Phys. J.}\ }\textbf {\bibinfo {volume} {C77}},\
  \bibinfo {pages} {91} (\bibinfo {year} {2017})},\ \Eprint
  {https://arxiv.org/abs/1612.07966} {arXiv:1612.07966 [hep-ph]} \BibitemShut
  {NoStop}%
\bibitem [{\citenamefont {Schlessinger}(1968)}]{Schlessinger:1968}%
  \BibitemOpen
  \bibfield  {author} {\bibinfo {author} {\bibfnamefont {L.}~\bibnamefont
  {Schlessinger}},\ }\href {https://doi.org/10.1103/PhysRev.167.1411}
  {\bibfield  {journal} {\bibinfo  {journal} {Phys. Rev.}\ }\textbf {\bibinfo
  {volume} {167}},\ \bibinfo {pages} {1411} (\bibinfo {year}
  {1968})}\BibitemShut {NoStop}%
\bibitem [{\citenamefont {Tripolt}\ \emph {et~al.}(2017)\citenamefont
  {Tripolt}, \citenamefont {Haritan}, \citenamefont {Wambach},\ and\
  \citenamefont {Moiseyev}}]{Tripolt:2016cya}%
  \BibitemOpen
  \bibfield  {author} {\bibinfo {author} {\bibfnamefont {R.-A.}\ \bibnamefont
  {Tripolt}}, \bibinfo {author} {\bibfnamefont {I.}~\bibnamefont {Haritan}},
  \bibinfo {author} {\bibfnamefont {J.}~\bibnamefont {Wambach}},\ and\ \bibinfo
  {author} {\bibfnamefont {N.}~\bibnamefont {Moiseyev}},\ }\href
  {https://doi.org/10.1016/j.physletb.2017.10.001} {\bibfield  {journal}
  {\bibinfo  {journal} {Phys. Lett. B}\ }\textbf {\bibinfo {volume} {774}},\
  \bibinfo {pages} {411} (\bibinfo {year} {2017})},\ \Eprint
  {https://arxiv.org/abs/1610.03252} {arXiv:1610.03252 [hep-ph]} \BibitemShut
  {NoStop}%
\bibitem [{\citenamefont {Binosi}\ and\ \citenamefont
  {Tripolt}(2020)}]{Binosi:2019ecz}%
  \BibitemOpen
  \bibfield  {author} {\bibinfo {author} {\bibfnamefont {D.}~\bibnamefont
  {Binosi}}\ and\ \bibinfo {author} {\bibfnamefont {R.-A.}\ \bibnamefont
  {Tripolt}},\ }\href {https://doi.org/10.1016/j.physletb.2019.135171}
  {\bibfield  {journal} {\bibinfo  {journal} {Phys. Lett. B}\ }\textbf
  {\bibinfo {volume} {801}},\ \bibinfo {pages} {135171} (\bibinfo {year}
  {2020})},\ \Eprint {https://arxiv.org/abs/1904.08172} {arXiv:1904.08172
  [hep-ph]} \BibitemShut {NoStop}%
\bibitem [{\citenamefont {Binosi}\ \emph {et~al.}(2023)\citenamefont {Binosi},
  \citenamefont {Pilloni},\ and\ \citenamefont {Tripolt}}]{Binosi:2022ydc}%
  \BibitemOpen
  \bibfield  {author} {\bibinfo {author} {\bibfnamefont {D.}~\bibnamefont
  {Binosi}}, \bibinfo {author} {\bibfnamefont {A.}~\bibnamefont {Pilloni}},\
  and\ \bibinfo {author} {\bibfnamefont {R.-A.}\ \bibnamefont {Tripolt}},\
  }\href {https://doi.org/10.1016/j.physletb.2023.137809} {\bibfield  {journal}
  {\bibinfo  {journal} {Phys. Lett. B}\ }\textbf {\bibinfo {volume} {839}},\
  \bibinfo {pages} {137809} (\bibinfo {year} {2023})},\ \Eprint
  {https://arxiv.org/abs/2205.02690} {arXiv:2205.02690 [hep-ph]} \BibitemShut
  {NoStop}%
\bibitem [{\citenamefont {Pelaez}\ and\ \citenamefont
  {Rodas}(2018)}]{Pelaez:2018qny}%
  \BibitemOpen
  \bibfield  {author} {\bibinfo {author} {\bibfnamefont {J.~R.}\ \bibnamefont
  {Pelaez}}\ and\ \bibinfo {author} {\bibfnamefont {A.}~\bibnamefont {Rodas}},\
  }\href {https://doi.org/10.1140/epjc/s10052-018-6296-9} {\bibfield  {journal}
  {\bibinfo  {journal} {Eur. Phys. J.}\ }\textbf {\bibinfo {volume} {C78}},\
  \bibinfo {pages} {897} (\bibinfo {year} {2018})},\ \Eprint
  {https://arxiv.org/abs/1807.04543} {arXiv:1807.04543 [hep-ph]} \BibitemShut
  {NoStop}%
\bibitem [{\citenamefont {Flatt\'{e}}(1976)}]{Flatte:1976xu}%
  \BibitemOpen
  \bibfield  {author} {\bibinfo {author} {\bibfnamefont {S.~M.}\ \bibnamefont
  {Flatt\'{e}}},\ }\href {https://doi.org/10.1016/0370-2693(76)90654-7}
  {\bibfield  {journal} {\bibinfo  {journal} {Phys. Lett. B}\ }\textbf
  {\bibinfo {volume} {63}},\ \bibinfo {pages} {224} (\bibinfo {year}
  {1976})}\BibitemShut {NoStop}%
\bibitem [{\citenamefont {Burkert}\ \emph {et~al.}(2023)\citenamefont {Burkert}
  \emph {et~al.}}]{Burkert:2022bqo}%
  \BibitemOpen
  \bibfield  {author} {\bibinfo {author} {\bibfnamefont {V.}~\bibnamefont
  {Burkert}} \emph {et~al.},\ }\href
  {https://doi.org/10.1016/j.physletb.2023.138070} {\bibfield  {journal}
  {\bibinfo  {journal} {Phys. Lett. B}\ }\textbf {\bibinfo {volume} {844}},\
  \bibinfo {pages} {138070} (\bibinfo {year} {2023})},\ \Eprint
  {https://arxiv.org/abs/2207.08472} {arXiv:2207.08472 [hep-ph]} \BibitemShut
  {NoStop}%
\bibitem [{\citenamefont {Ceci}\ \emph {et~al.}(2017)\citenamefont {Ceci},
  \citenamefont {Had{\v{z}}imehmedovi{\'c}}, \citenamefont {Osmanovi{\'c}},
  \citenamefont {Percan},\ and\ \citenamefont {Zauner}}]{Ceci:2016pdn}%
  \BibitemOpen
  \bibfield  {author} {\bibinfo {author} {\bibfnamefont {S.}~\bibnamefont
  {Ceci}}, \bibinfo {author} {\bibfnamefont {M.}~\bibnamefont
  {Had{\v{z}}imehmedovi{\'c}}}, \bibinfo {author} {\bibfnamefont
  {H.}~\bibnamefont {Osmanovi{\'c}}}, \bibinfo {author} {\bibfnamefont
  {A.}~\bibnamefont {Percan}},\ and\ \bibinfo {author} {\bibfnamefont
  {B.}~\bibnamefont {Zauner}},\ }\href {https://doi.org/10.1038/srep45246}
  {\bibfield  {journal} {\bibinfo  {journal} {Sci. Rep.}\ }\textbf {\bibinfo
  {volume} {7}},\ \bibinfo {pages} {45246} (\bibinfo {year} {2017})},\ \Eprint
  {https://arxiv.org/abs/1608.06485} {arXiv:1608.06485 [hep-ph]} \BibitemShut
  {NoStop}%
\bibitem [{\citenamefont {Ceci}\ \emph {et~al.}(2026)\citenamefont {Ceci},
  \citenamefont {Omerovi{\'c}}, \citenamefont {Osmanovi{\'c}}, \citenamefont
  {Uroi{\'c}},\ and\ \citenamefont {Zauner}}]{Ceci:2025gsm}%
  \BibitemOpen
  \bibfield  {author} {\bibinfo {author} {\bibfnamefont {S.}~\bibnamefont
  {Ceci}}, \bibinfo {author} {\bibfnamefont {R.}~\bibnamefont {Omerovi{\'c}}},
  \bibinfo {author} {\bibfnamefont {H.}~\bibnamefont {Osmanovi{\'c}}}, \bibinfo
  {author} {\bibfnamefont {M.}~\bibnamefont {Uroi{\'c}}},\ and\ \bibinfo
  {author} {\bibfnamefont {B.}~\bibnamefont {Zauner}},\ }\href
  {https://doi.org/10.1016/j.physletb.2025.140136} {\bibfield  {journal}
  {\bibinfo  {journal} {Phys. Lett. B}\ }\textbf {\bibinfo {volume} {872}},\
  \bibinfo {pages} {140136} (\bibinfo {year} {2026})},\ \Eprint
  {https://arxiv.org/abs/2511.09199} {arXiv:2511.09199 [hep-ph]} \BibitemShut
  {NoStop}%
\bibitem [{\citenamefont {Ceci}\ \emph {et~al.}(2025)\citenamefont {Ceci},
  \citenamefont {Osmanovi{\'c}},\ and\ \citenamefont {Zauner}}]{Ceci:2025yas}%
  \BibitemOpen
  \bibfield  {author} {\bibinfo {author} {\bibfnamefont {S.}~\bibnamefont
  {Ceci}}, \bibinfo {author} {\bibfnamefont {H.}~\bibnamefont
  {Osmanovi{\'c}}},\ and\ \bibinfo {author} {\bibfnamefont {B.}~\bibnamefont
  {Zauner}},\ }\href@noop {} {\  (\bibinfo {year} {2025})},\ \Eprint
  {https://arxiv.org/abs/2505.16880} {arXiv:2505.16880 [hep-ph]} \BibitemShut
  {NoStop}%
\bibitem [{\citenamefont {Anisovich}\ \emph {et~al.}(2012)\citenamefont
  {Anisovich}, \citenamefont {Beck}, \citenamefont {Klempt}, \citenamefont
  {Nikonov}, \citenamefont {Sarantsev},\ and\ \citenamefont
  {Thoma}}]{Anisovich:2011fc}%
  \BibitemOpen
  \bibfield  {author} {\bibinfo {author} {\bibfnamefont {A.~V.}\ \bibnamefont
  {Anisovich}}, \bibinfo {author} {\bibfnamefont {R.}~\bibnamefont {Beck}},
  \bibinfo {author} {\bibfnamefont {E.}~\bibnamefont {Klempt}}, \bibinfo
  {author} {\bibfnamefont {V.~A.}\ \bibnamefont {Nikonov}}, \bibinfo {author}
  {\bibfnamefont {A.~V.}\ \bibnamefont {Sarantsev}},\ and\ \bibinfo {author}
  {\bibfnamefont {U.}~\bibnamefont {Thoma}},\ }\href
  {https://doi.org/10.1140/epja/i2012-12015-8} {\bibfield  {journal} {\bibinfo
  {journal} {Eur.Phys.J.}\ }\textbf {\bibinfo {volume} {A48}},\ \bibinfo
  {pages} {15} (\bibinfo {year} {2012})},\ \Eprint
  {https://arxiv.org/abs/1112.4937} {arXiv:1112.4937 [hep-ph]} \BibitemShut
  {NoStop}%
\bibitem [{\citenamefont {\v{S}varc}\ \emph {et~al.}(2014)\citenamefont
  {\v{S}varc}, \citenamefont {Had\v{z}imehmedovi\'c}, \citenamefont
  {Omerovi\'c}, \citenamefont {Osmanovi\'c},\ and\ \citenamefont
  {Stahov}}]{Svarc:2014zja}%
  \BibitemOpen
  \bibfield  {author} {\bibinfo {author} {\bibfnamefont {A.}~\bibnamefont
  {\v{S}varc}}, \bibinfo {author} {\bibfnamefont {M.}~\bibnamefont
  {Had\v{z}imehmedovi\'c}}, \bibinfo {author} {\bibfnamefont {R.}~\bibnamefont
  {Omerovi\'c}}, \bibinfo {author} {\bibfnamefont {H.}~\bibnamefont
  {Osmanovi\'c}},\ and\ \bibinfo {author} {\bibfnamefont {J.}~\bibnamefont
  {Stahov}},\ }\href {https://doi.org/10.1103/PhysRevC.89.045205} {\bibfield
  {journal} {\bibinfo  {journal} {Phys.Rev.}\ }\textbf {\bibinfo {volume}
  {C89}},\ \bibinfo {pages} {045205} (\bibinfo {year} {2014})},\ \Eprint
  {https://arxiv.org/abs/1401.1947} {arXiv:1401.1947 [nucl-th]} \BibitemShut
  {NoStop}%
\bibitem [{\citenamefont {Arndt}\ \emph {et~al.}(2006)\citenamefont {Arndt},
  \citenamefont {Briscoe}, \citenamefont {Strakovsky},\ and\ \citenamefont
  {Workman}}]{Arndt:2006bf}%
  \BibitemOpen
  \bibfield  {author} {\bibinfo {author} {\bibfnamefont {R.~A.}\ \bibnamefont
  {Arndt}}, \bibinfo {author} {\bibfnamefont {W.~J.}\ \bibnamefont {Briscoe}},
  \bibinfo {author} {\bibfnamefont {I.~I.}\ \bibnamefont {Strakovsky}},\ and\
  \bibinfo {author} {\bibfnamefont {R.~L.}\ \bibnamefont {Workman}},\ }\href
  {https://doi.org/10.1103/PhysRevC.74.045205} {\bibfield  {journal} {\bibinfo
  {journal} {Phys.Rev.}\ }\textbf {\bibinfo {volume} {C74}},\ \bibinfo {pages}
  {045205} (\bibinfo {year} {2006})},\ \Eprint
  {https://arxiv.org/abs/nucl-th/0605082} {arXiv:nucl-th/0605082 [nucl-th]}
  \BibitemShut {NoStop}%
\bibitem [{\citenamefont {R{\"o}nchen}\ \emph {et~al.}(2022)\citenamefont
  {R{\"o}nchen}, \citenamefont {D{\"o}ring}, \citenamefont {Mei{\ss}ner},\ and\
  \citenamefont {Shen}}]{Ronchen:2022hqk}%
  \BibitemOpen
  \bibfield  {author} {\bibinfo {author} {\bibfnamefont {D.}~\bibnamefont
  {R{\"o}nchen}}, \bibinfo {author} {\bibfnamefont {M.}~\bibnamefont
  {D{\"o}ring}}, \bibinfo {author} {\bibfnamefont {U.-G.}\ \bibnamefont
  {Mei{\ss}ner}},\ and\ \bibinfo {author} {\bibfnamefont {C.-W.}\ \bibnamefont
  {Shen}},\ }\href {https://doi.org/10.1140/epja/s10050-022-00852-1} {\bibfield
   {journal} {\bibinfo  {journal} {Eur. Phys. J. A}\ }\textbf {\bibinfo
  {volume} {58}},\ \bibinfo {pages} {229} (\bibinfo {year} {2022})},\ \Eprint
  {https://arxiv.org/abs/2208.00089} {arXiv:2208.00089 [nucl-th]} \BibitemShut
  {NoStop}%
\bibitem [{\citenamefont {Hoferichter}\ \emph {et~al.}(2017)\citenamefont
  {Hoferichter}, \citenamefont {Kubis},\ and\ \citenamefont
  {Zanke}}]{Hoferichter:2017ftn}%
  \BibitemOpen
  \bibfield  {author} {\bibinfo {author} {\bibfnamefont {M.}~\bibnamefont
  {Hoferichter}}, \bibinfo {author} {\bibfnamefont {B.}~\bibnamefont {Kubis}},\
  and\ \bibinfo {author} {\bibfnamefont {M.}~\bibnamefont {Zanke}},\ }\href
  {https://doi.org/10.1103/PhysRevD.96.114016} {\bibfield  {journal} {\bibinfo
  {journal} {Phys. Rev. D}\ }\textbf {\bibinfo {volume} {96}},\ \bibinfo
  {pages} {114016} (\bibinfo {year} {2017})},\ \Eprint
  {https://arxiv.org/abs/1710.00824} {arXiv:1710.00824 [hep-ph]} \BibitemShut
  {NoStop}%
\bibitem [{\citenamefont {Dax}\ \emph {et~al.}(2021)\citenamefont {Dax},
  \citenamefont {Stamen},\ and\ \citenamefont {Kubis}}]{Dax:2020dzg}%
  \BibitemOpen
  \bibfield  {author} {\bibinfo {author} {\bibfnamefont {M.}~\bibnamefont
  {Dax}}, \bibinfo {author} {\bibfnamefont {D.}~\bibnamefont {Stamen}},\ and\
  \bibinfo {author} {\bibfnamefont {B.}~\bibnamefont {Kubis}},\ }\href
  {https://doi.org/10.1140/epjc/s10052-021-08951-x} {\bibfield  {journal}
  {\bibinfo  {journal} {Eur. Phys. J. C}\ }\textbf {\bibinfo {volume} {81}},\
  \bibinfo {pages} {221} (\bibinfo {year} {2021})},\ \Eprint
  {https://arxiv.org/abs/2012.04655} {arXiv:2012.04655 [hep-ph]} \BibitemShut
  {NoStop}%
\bibitem [{\citenamefont {Niehus}\ \emph {et~al.}(2021)\citenamefont {Niehus},
  \citenamefont {Hoferichter},\ and\ \citenamefont {Kubis}}]{Niehus:2021iin}%
  \BibitemOpen
  \bibfield  {author} {\bibinfo {author} {\bibfnamefont {M.}~\bibnamefont
  {Niehus}}, \bibinfo {author} {\bibfnamefont {M.}~\bibnamefont
  {Hoferichter}},\ and\ \bibinfo {author} {\bibfnamefont {B.}~\bibnamefont
  {Kubis}},\ }\href {https://doi.org/10.1007/JHEP12(2021)038} {\bibfield
  {journal} {\bibinfo  {journal} {JHEP}\ }\textbf {\bibinfo {volume} {12}},\
  \bibinfo {pages} {038}},\ \Eprint {https://arxiv.org/abs/2110.11372}
  {arXiv:2110.11372 [hep-ph]} \BibitemShut {NoStop}%
\bibitem [{\citenamefont {Aston}\ \emph {et~al.}(1991)\citenamefont {Aston}
  \emph {et~al.}}]{Aston:1990wg}%
  \BibitemOpen
  \bibfield  {author} {\bibinfo {author} {\bibfnamefont {D.}~\bibnamefont
  {Aston}} \emph {et~al.},\ }\href
  {https://doi.org/10.1016/0920-5632(91)90243-8} {\bibfield  {journal}
  {\bibinfo  {journal} {Nucl. Phys. B Proc. Suppl.}\ }\textbf {\bibinfo
  {volume} {21}},\ \bibinfo {pages} {105} (\bibinfo {year} {1991})}\BibitemShut
  {NoStop}%
\bibitem [{\citenamefont {Bertin}\ \emph {et~al.}(1997)\citenamefont {Bertin}
  \emph {et~al.}}]{OBELIX:1997zla}%
  \BibitemOpen
  \bibfield  {author} {\bibinfo {author} {\bibfnamefont {A.}~\bibnamefont
  {Bertin}} \emph {et~al.} (\bibinfo {collaboration} {OBELIX}),\ }\href
  {https://doi.org/10.1016/S0370-2693(97)01189-1} {\bibfield  {journal}
  {\bibinfo  {journal} {Phys. Lett. B}\ }\textbf {\bibinfo {volume} {414}},\
  \bibinfo {pages} {220} (\bibinfo {year} {1997})}\BibitemShut {NoStop}%
\bibitem [{\citenamefont {Surovtsev}\ and\ \citenamefont
  {Bydzovsky}(2008)}]{Surovtsev:2008zza}%
  \BibitemOpen
  \bibfield  {author} {\bibinfo {author} {\bibfnamefont {Y.~S.}\ \bibnamefont
  {Surovtsev}}\ and\ \bibinfo {author} {\bibfnamefont {P.}~\bibnamefont
  {Bydzovsky}},\ }\href {https://doi.org/10.1016/j.nuclphysa.2008.04.005}
  {\bibfield  {journal} {\bibinfo  {journal} {Nucl. Phys. A}\ }\textbf
  {\bibinfo {volume} {807}},\ \bibinfo {pages} {145} (\bibinfo {year}
  {2008})}\BibitemShut {NoStop}%
\bibitem [{\citenamefont {Surovtsev}\ \emph {et~al.}(2010)\citenamefont
  {Surovtsev}, \citenamefont {Bydzovsky}, \citenamefont {Kaminski},\ and\
  \citenamefont {Nagy}}]{Surovtsev:2010cjf}%
  \BibitemOpen
  \bibfield  {author} {\bibinfo {author} {\bibfnamefont {Y.~S.}\ \bibnamefont
  {Surovtsev}}, \bibinfo {author} {\bibfnamefont {P.}~\bibnamefont
  {Bydzovsky}}, \bibinfo {author} {\bibfnamefont {R.}~\bibnamefont
  {Kaminski}},\ and\ \bibinfo {author} {\bibfnamefont {M.}~\bibnamefont
  {Nagy}},\ }\href {https://doi.org/10.1103/PhysRevD.81.016001} {\bibfield
  {journal} {\bibinfo  {journal} {Phys. Rev. D}\ }\textbf {\bibinfo {volume}
  {81}},\ \bibinfo {pages} {016001} (\bibinfo {year} {2010})}\BibitemShut
  {NoStop}%
\bibitem [{\citenamefont {Hammoud}\ \emph {et~al.}(2020)\citenamefont
  {Hammoud}, \citenamefont {Kami\'nski}, \citenamefont {Nazari},\ and\
  \citenamefont {Rupp}}]{Hammoud:2020aqi}%
  \BibitemOpen
  \bibfield  {author} {\bibinfo {author} {\bibfnamefont {N.}~\bibnamefont
  {Hammoud}}, \bibinfo {author} {\bibfnamefont {R.}~\bibnamefont {Kami\'nski}},
  \bibinfo {author} {\bibfnamefont {V.}~\bibnamefont {Nazari}},\ and\ \bibinfo
  {author} {\bibfnamefont {G.}~\bibnamefont {Rupp}},\ }\href
  {https://doi.org/10.1103/PhysRevD.102.054029} {\bibfield  {journal} {\bibinfo
   {journal} {Phys. Rev. D}\ }\textbf {\bibinfo {volume} {102}},\ \bibinfo
  {pages} {054029} (\bibinfo {year} {2020})},\ \Eprint
  {https://arxiv.org/abs/2009.06317} {arXiv:2009.06317 [hep-ph]} \BibitemShut
  {NoStop}%
\bibitem [{\citenamefont {Henner}(1985)}]{henner:1985}%
  \BibitemOpen
  \bibfield  {author} {\bibinfo {author} {\bibfnamefont {V.}~\bibnamefont
  {Henner}},\ }\href {https://doi.org/10.1007/BF01571389} {\bibfield  {journal}
  {\bibinfo  {journal} {Z. Phys. C}\ }\textbf {\bibinfo {volume} {29}},\
  \bibinfo {pages} {107} (\bibinfo {year} {1985})}\BibitemShut {NoStop}%
\bibitem [{\citenamefont {Workman}\ \emph {et~al.}(2022)\citenamefont {Workman}
  \emph {et~al.}}]{ParticleDataGroup:2022pth}%
  \BibitemOpen
  \bibfield  {author} {\bibinfo {author} {\bibfnamefont {R.~L.}\ \bibnamefont
  {Workman}} \emph {et~al.} (\bibinfo {collaboration} {Particle Data Group}),\
  }\href {https://doi.org/10.1093/ptep/ptac097} {\bibfield  {journal} {\bibinfo
   {journal} {PTEP}\ }\textbf {\bibinfo {volume} {2022}},\ \bibinfo {pages}
  {083C01} (\bibinfo {year} {2022})}\BibitemShut {NoStop}%
\bibitem [{\citenamefont {Barto{\v{s}}}\ \emph {et~al.}(2017)\citenamefont
  {Barto{\v{s}}}, \citenamefont {Dubni{\v{c}}ka}, \citenamefont {Liptaj},
  \citenamefont {Dubni{\v{c}}kov{\'a}},\ and\ \citenamefont
  {Kami{\'n}ski}}]{Bartos:2017ils}%
  \BibitemOpen
  \bibfield  {author} {\bibinfo {author} {\bibfnamefont {E.}~\bibnamefont
  {Barto{\v{s}}}}, \bibinfo {author} {\bibfnamefont {S.}~\bibnamefont
  {Dubni{\v{c}}ka}}, \bibinfo {author} {\bibfnamefont {A.}~\bibnamefont
  {Liptaj}}, \bibinfo {author} {\bibfnamefont {A.~Z.}\ \bibnamefont
  {Dubni{\v{c}}kov{\'a}}},\ and\ \bibinfo {author} {\bibfnamefont
  {R.}~\bibnamefont {Kami{\'n}ski}},\ }\href
  {https://doi.org/10.1103/PhysRevD.96.113004} {\bibfield  {journal} {\bibinfo
  {journal} {Phys. Rev. D}\ }\textbf {\bibinfo {volume} {96}},\ \bibinfo
  {pages} {113004} (\bibinfo {year} {2017})}\BibitemShut {NoStop}%
\bibitem [{\citenamefont {Adler}(1965)}]{Adler:1964um}%
  \BibitemOpen
  \bibfield  {author} {\bibinfo {author} {\bibfnamefont {S.~L.}\ \bibnamefont
  {Adler}},\ }\href {https://doi.org/10.1103/PhysRev.137.B1022} {\bibfield
  {journal} {\bibinfo  {journal} {Phys. Rev.}\ }\textbf {\bibinfo {volume}
  {137}},\ \bibinfo {pages} {B1022} (\bibinfo {year} {1965})},\ \bibinfo {note}
  {[,140(1964)]}\BibitemShut {NoStop}%
\bibitem [{\citenamefont {Meissner}\ and\ \citenamefont
  {Gardner}(2003)}]{Meissner:2003pd}%
  \BibitemOpen
  \bibfield  {author} {\bibinfo {author} {\bibfnamefont {U.-G.}\ \bibnamefont
  {Meissner}}\ and\ \bibinfo {author} {\bibfnamefont {S.}~\bibnamefont
  {Gardner}},\ }\href {https://doi.org/10.1140/epja/i2002-10279-1} {\bibfield
  {journal} {\bibinfo  {journal} {Eur. Phys. J. A}\ }\textbf {\bibinfo {volume}
  {18}},\ \bibinfo {pages} {543} (\bibinfo {year} {2003})}\BibitemShut
  {NoStop}%
\bibitem [{\citenamefont {Ruiz~de Elvira}\ \emph {et~al.}(2011)\citenamefont
  {Ruiz~de Elvira}, \citenamefont {Pelaez}, \citenamefont {Pennington},\ and\
  \citenamefont {Wilson}}]{RuizdeElvira:2010cs}%
  \BibitemOpen
  \bibfield  {author} {\bibinfo {author} {\bibfnamefont {J.}~\bibnamefont
  {Ruiz~de Elvira}}, \bibinfo {author} {\bibfnamefont {J.~R.}\ \bibnamefont
  {Pelaez}}, \bibinfo {author} {\bibfnamefont {M.~R.}\ \bibnamefont
  {Pennington}},\ and\ \bibinfo {author} {\bibfnamefont {D.~J.}\ \bibnamefont
  {Wilson}},\ }\href {https://doi.org/10.1103/PhysRevD.84.096006} {\bibfield
  {journal} {\bibinfo  {journal} {Phys. Rev. D}\ }\textbf {\bibinfo {volume}
  {84}},\ \bibinfo {pages} {096006} (\bibinfo {year} {2011})},\ \Eprint
  {https://arxiv.org/abs/1009.6204} {arXiv:1009.6204 [hep-ph]} \BibitemShut
  {NoStop}%
\bibitem [{\citenamefont {Ledwig}\ \emph {et~al.}(2014)\citenamefont {Ledwig},
  \citenamefont {Nieves}, \citenamefont {Pich}, \citenamefont {Ruiz~Arriola},\
  and\ \citenamefont {Ruiz~de Elvira}}]{Ledwig:2014cla}%
  \BibitemOpen
  \bibfield  {author} {\bibinfo {author} {\bibfnamefont {T.}~\bibnamefont
  {Ledwig}}, \bibinfo {author} {\bibfnamefont {J.}~\bibnamefont {Nieves}},
  \bibinfo {author} {\bibfnamefont {A.}~\bibnamefont {Pich}}, \bibinfo {author}
  {\bibfnamefont {E.}~\bibnamefont {Ruiz~Arriola}},\ and\ \bibinfo {author}
  {\bibfnamefont {J.}~\bibnamefont {Ruiz~de Elvira}},\ }\href
  {https://doi.org/10.1103/PhysRevD.90.114020} {\bibfield  {journal} {\bibinfo
  {journal} {Phys. Rev. D}\ }\textbf {\bibinfo {volume} {90}},\ \bibinfo
  {pages} {114020} (\bibinfo {year} {2014})},\ \Eprint
  {https://arxiv.org/abs/1407.3750} {arXiv:1407.3750 [hep-ph]} \BibitemShut
  {NoStop}%
\bibitem [{\citenamefont {Wang}\ \emph {et~al.}(2022)\citenamefont {Wang},
  \citenamefont {Kang}, \citenamefont {Oller},\ and\ \citenamefont
  {Zhang}}]{Wang:2022vga}%
  \BibitemOpen
  \bibfield  {author} {\bibinfo {author} {\bibfnamefont {Z.-Q.}\ \bibnamefont
  {Wang}}, \bibinfo {author} {\bibfnamefont {X.-W.}\ \bibnamefont {Kang}},
  \bibinfo {author} {\bibfnamefont {J.~A.}\ \bibnamefont {Oller}},\ and\
  \bibinfo {author} {\bibfnamefont {L.}~\bibnamefont {Zhang}},\ }\href
  {https://doi.org/10.1103/PhysRevD.105.074016} {\bibfield  {journal} {\bibinfo
   {journal} {Phys. Rev. D}\ }\textbf {\bibinfo {volume} {105}},\ \bibinfo
  {pages} {074016} (\bibinfo {year} {2022})},\ \Eprint
  {https://arxiv.org/abs/2201.00492} {arXiv:2201.00492 [hep-ph]} \BibitemShut
  {NoStop}%
\bibitem [{\citenamefont {Colangelo}\ \emph {et~al.}(2025)\citenamefont
  {Colangelo}, \citenamefont {Cottini},\ and\ \citenamefont {Ruiz~de
  Elvira}}]{Colangelo:2025iuq}%
  \BibitemOpen
  \bibfield  {author} {\bibinfo {author} {\bibfnamefont {G.}~\bibnamefont
  {Colangelo}}, \bibinfo {author} {\bibfnamefont {M.}~\bibnamefont {Cottini}},\
  and\ \bibinfo {author} {\bibfnamefont {J.}~\bibnamefont {Ruiz~de Elvira}},\
  }\href@noop {} {\  (\bibinfo {year} {2025})},\ \Eprint
  {https://arxiv.org/abs/2511.08680} {arXiv:2511.08680 [hep-ph]} \BibitemShut
  {NoStop}%
\bibitem [{\citenamefont {Bugg}(2007)}]{Bugg:2007ja}%
  \BibitemOpen
  \bibfield  {author} {\bibinfo {author} {\bibfnamefont {D.~V.}\ \bibnamefont
  {Bugg}},\ }\href {https://doi.org/10.1140/epjc/s10052-007-0389-1} {\bibfield
  {journal} {\bibinfo  {journal} {Eur. Phys. J. C}\ }\textbf {\bibinfo {volume}
  {52}},\ \bibinfo {pages} {55} (\bibinfo {year} {2007})},\ \Eprint
  {https://arxiv.org/abs/0706.1341} {arXiv:0706.1341 [hep-ex]} \BibitemShut
  {NoStop}%
\bibitem [{\citenamefont {Klempt}\ and\ \citenamefont
  {Zaitsev}(2007)}]{Klempt:2007cp}%
  \BibitemOpen
  \bibfield  {author} {\bibinfo {author} {\bibfnamefont {E.}~\bibnamefont
  {Klempt}}\ and\ \bibinfo {author} {\bibfnamefont {A.}~\bibnamefont
  {Zaitsev}},\ }\href {https://doi.org/10.1016/j.physrep.2007.07.006}
  {\bibfield  {journal} {\bibinfo  {journal} {Phys.Rept.}\ }\textbf {\bibinfo
  {volume} {454}},\ \bibinfo {pages} {1} (\bibinfo {year} {2007})},\ \Eprint
  {https://arxiv.org/abs/0708.4016} {arXiv:0708.4016 [hep-ph]} \BibitemShut
  {NoStop}%
\bibitem [{\citenamefont {Ochs}(2013)}]{Ochs:2013gi}%
  \BibitemOpen
  \bibfield  {author} {\bibinfo {author} {\bibfnamefont {W.}~\bibnamefont
  {Ochs}},\ }\href {https://doi.org/10.1088/0954-3899/40/4/043001} {\bibfield
  {journal} {\bibinfo  {journal} {J. Phys.}\ }\textbf {\bibinfo {volume}
  {G40}},\ \bibinfo {pages} {043001} (\bibinfo {year} {2013})},\ \Eprint
  {https://arxiv.org/abs/1301.5183} {arXiv:1301.5183 [hep-ph]} \BibitemShut
  {NoStop}%
\bibitem [{\citenamefont {Adolph}\ \emph {et~al.}(2017)\citenamefont {Adolph}
  \emph {et~al.}}]{COMPASS:2015gxz}%
  \BibitemOpen
  \bibfield  {author} {\bibinfo {author} {\bibfnamefont {C.}~\bibnamefont
  {Adolph}} \emph {et~al.} (\bibinfo {collaboration} {COMPASS}),\ }\href
  {https://doi.org/10.1103/PhysRevD.95.032004} {\bibfield  {journal} {\bibinfo
  {journal} {Phys. Rev. D}\ }\textbf {\bibinfo {volume} {95}},\ \bibinfo
  {pages} {032004} (\bibinfo {year} {2017})},\ \Eprint
  {https://arxiv.org/abs/1509.00992} {arXiv:1509.00992 [hep-ex]} \BibitemShut
  {NoStop}%
\bibitem [{\citenamefont {Estabrooks}\ and\ \citenamefont
  {Martin}(1974)}]{Estabrooks:1974vu}%
  \BibitemOpen
  \bibfield  {author} {\bibinfo {author} {\bibfnamefont {P.}~\bibnamefont
  {Estabrooks}}\ and\ \bibinfo {author} {\bibfnamefont {A.~D.}\ \bibnamefont
  {Martin}},\ }\href {https://doi.org/10.1016/0550-3213(74)90488-X} {\bibfield
  {journal} {\bibinfo  {journal} {Nucl. Phys.}\ }\textbf {\bibinfo {volume}
  {B79}},\ \bibinfo {pages} {301} (\bibinfo {year} {1974})}\BibitemShut
  {NoStop}%
\bibitem [{\citenamefont {Martin}\ and\ \citenamefont
  {Spearman}(1970)}]{Martin:1970}%
  \BibitemOpen
  \bibfield  {author} {\bibinfo {author} {\bibfnamefont {A.}~\bibnamefont
  {Martin}}\ and\ \bibinfo {author} {\bibfnamefont {T.}~\bibnamefont
  {Spearman}},\ }\href {https://books.google.com/books?id=sxAzAAAAMAAJ} {\emph
  {\bibinfo {title} {Elementary particle theory}}}\ (\bibinfo  {publisher}
  {North-Holland Pub. Co.},\ \bibinfo {year} {1970})\BibitemShut {NoStop}%
\end{thebibliography}%

\end{document}